\documentclass[a4paper,12pt,twoside]{article}

\usepackage[dvips]{graphicx}
\usepackage{amssymb}
\usepackage{amsmath}
\usepackage{cite}
\usepackage{eepic}
\usepackage{epic}
\usepackage{multirow}

\makeatletter
 \renewcommand\section{\@startsection {section}{1}{\z@}%
                                   {3.5ex \@plus 1ex \@minus .2ex}%
                                   {2.3ex \@plus.2ex}%
                                   {\normalfont\Large\bfseries}}
 \renewcommand\subsection{\@startsection{subsection}{2}{\z@}%
                                     {3.25ex\@plus 1ex \@minus .2ex}%
                                     {1.5ex \@plus .2ex}%
                                 {\normalfont\large\bfseries}}
 \renewcommand\subsubsection{\@startsection{subsubsection}{3}{\z@}%
                                     {3.25ex\@plus 1ex \@minus .2ex}%
                                     {1.5ex \@plus .2ex}%
                                {\normalfont\normalsize\bfseries}}
 \def\ps@myheadings{%
    \let\@oddfoot\@empty\let\@evenfoot\@empty
    \def\@evenhead{\footnotesize\thepage\hfil\slshape\leftmark\hfil}%
    \def\@oddhead{\footnotesize\hfil{\slshape\rightmark}\hfil\thepage}%
    \let\@mkboth\@gobbletwo
    \let\sectionmark\@gobble
    \let\subsectionmark\@gobble
    }
 \def\ps@firstpage{%
    \let\@oddfoot\@empty\let\@evenfoot\@empty
    \def\@evenhead{\thepage\hfil\slshape\leftmark}%
    \def\@oddhead{\hfil
    \parbox[t]{360pt}{\footnotesize Memoirs 
    of the Faculty of Science, Kyoto
    University, Series of Physics, Astrophysics,\par Geophysics and
    Chemistry, Vol.XXXXIV, No.1, Article 2, 2003}\hfil\thepage}%
    \let\@mkboth\@gobbletwo
    \let\sectionmark\@gobble
    \let\subsectionmark\@gobble
    }
\newcommand{\department}[1]{\gdef\@department{#1}}
\newcommand{\adress}[1]{\gdef\@adress{#1}}
\def\@maketitle{%
  \newpage
  \null
  \begin{center}%
  \let \footnote \thanks
    {\LARGE \@title \par}%
    \vskip 2.0em%
   {\footnotesize By \par}%
    \vskip 1.0em%
   {\large
      \lineskip .5em%
      \begin{tabular}[t]{c}%
        \@author
      \end{tabular}\par}%
    \vskip 0.5em%
    {\footnotesize \@department \\}%
    {\footnotesize \@adress }%
    \vskip 0.5em%
    {\footnotesize ( \it \@date )}%
  \end{center}%
  \par
  \vskip 1.5em}
 \renewcommand\maketitle{\par
  \begingroup
    \renewcommand\thefootnote{\@fnsymbol\c@footnote}%
    \def\@makefnmark{\rlap{\@textsuperscript{\normalfont\@thefnmark}}}%
    \long\def\@makefntext##1{\parindent 1em\noindent
            \hb@xt@1.8em{%
                \hss\@textsuperscript{\normalfont\@thefnmark}}##1}%
    \if@twocolumn
      \ifnum \col@number=\@ne
        \@maketitle
      \else
        \twocolumn[\@maketitle]%
      \fi
    \else
      \newpage
      \global\@topnum\z@   
      \@maketitle
    \fi
    \thispagestyle{firstpage}\@thanks
  \endgroup
  \setcounter{footnote}{0}%
  \global\let\thanks\relax
  \global\let\maketitle\relax
  \global\let\@maketitle\relax
  \global\let\@thanks\@empty
  \global\let\@author\@empty
  \global\let\@date\@empty
  \global\let\@title\@empty
  \global\let\title\relax
  \global\let\author\relax
  \global\let\date\relax
  \global\let\and\relax
}
\makeatother
\title{$J/\psi$ Production in p+p Collisions \\
at $\sqrt{s}$ = 200 GeV}
\author{Hiroki Sato}
\department{Department of Physics, Faculty of Science, Kyoto University}
\adress{Kyoto 606-8052, Japan}
\date{Received December 9, 2002}
\begin{document}
 \maketitle
 \begin{abstract}


Total and differential cross sections for inclusive $J/\psi$ production
have been measured 
in $\sqrt{s} =$ 200~GeV p+p collisions 
at Relativistic Heavy Ion Collider in
Brookhaven National Laboratory.
$J/\psi$ particles have been clearly identified 
via $\mu^{+}\mu^{-}$ decays
measured in the forward muon spectrometer covering $1.2 < y < 2.2$ and 
via $e^{+}e^{-}$ decays measured
in the mid-rapidity spectrometers of the PHENIX experiment.
Details of the muon channel measurement are presented in this paper
based upon an 81-nb$^{-1}$ integrated luminosity
obtained in the second run period at RHIC.

Corrected with the detection efficiency,
the branching fraction times 
rapidity-differential cross section at forward rapidity
$Br (J/\psi \rightarrow \mu^{+}\mu^{-})$
$d\sigma_{J/\psi}/dy|_{y=1.7}$ = 37 $\pm$ 7 (stat.) $\pm$ 11 (syst.)~nb
has been obtained.
The average transverse momentum of the $J/\psi$'s, 
$\langle p_{T} \rangle$ = 1.66 $\pm$ 0.18 (stat.) $\pm$ 0.12 (syst.)
GeV/$c$,
is slightly higher than lower-energy results and consistent
with a model expectation based on perturbative QCD.
The total production cross section 
$\sigma_{J/\psi}$ = 3.8 $\pm$ 0.6 (stat.) $\pm$ 1.3 (syst.)~$\mu$b
has been extracted by fitting both
the muon and electron channel measurements, 
which is related with lower-energy results
by gluon distribution functions currently available.
Its normalization is well reproduced by both the color-octet model 
and color-evaporation model.

The results have established the baseline
necessary for the QGP search in Au+Au collisions
as well as gluon polarization measurements in
polarized p+p collisions at RHIC,
where the unpolarized cross sections presented here 
will determine the absolute normalization.



 \end{abstract}
 \tableofcontents
 \newpage

\section{Introduction}
\label{sec:intro}

Since the sensational discovery of the $J/\psi$ particle
\cite{ref:obs_J,ref:obs_psi}
as a triumph of the Quark Model,
heavy-quarkonium production in hadron-hadron collisions and 
in photon-hadron collisions has drawn attentions
because it involves both perturbative and
non-perturbative aspects of Quantum Chromodynamics (QCD),
which describes interactions between quarks and gluons.
A sufficiently large scale $Q$ $\sim M_{Q}$ (heavy quark mass)
enables perturbative treatments of the production of a 
heavy-quark pair.
On the other hand, its hadronization into a quarkonium state, with a scale of 
a typical hadron size ($\sim \Lambda_{QCD}$), 
is considered to be a non-perturbative
phenomenon, where perturbative calculation is not applicable.

Thanks to narrow resonances found 
in the invariant mass of lepton pairs and relatively large cross-sections
especially for the case of $J/\psi$ (1S),
abundant experimental data have been accumulated on heavy quarkonia.
Results of cross sections, polarization,
relative yields such as the $\psi'$ to $J/\psi$ ratio, 
have led to better understanding of
quarkonium production.
However the production mechanism
involving non-perturbative QCD phenomena
is still controversial.
To understand it, more data on various
observables in a wider energy range are needed.

We have added new measurements of $J/\psi$ in p+p collisions 
at unexploited energy ($\sqrt{s}$ = 200 GeV)
with the PHENIX experiment at Relativistic Heavy Ion Collider (RHIC).
The results are expected to 
provide better understanding of non-perturbative aspects of
the hadro-production of $J/\psi$, as well as
further knowledge of perturbative QCD.

In addition, our measurements play crucial roles
for both two innovative measurements realized for the first time at RHIC,
heavy ion physics and spin physics,
which will bring new insights into
both perturbative and non-perturbative aspects of QCD.

\subsection{Heavy ion physics}

Provided nucleons are in high temperature and/or high density
environment, hadronic matter is expected to turn into
the Quark-Gluon Plasma (QGP), a new phase of matter whose
degrees of freedom are quarks and gluons instead of hadrons.
Understanding the behavior of bulk matter governed by
QCD elementary degrees of freedom and interactions,
and studying how it turns into hadronic matter, offers
challenging perspectives and touches fundamental issues in
the study of QCD in its non perturbative regime, such as the nature of
confinement and chiral symmetry breaking.

RHIC is the first heavy-ion collider in the world to
create the highest temperature environment experimentally.
In the most central Au+Au collisions at 
the center-of-mass energy $\sqrt{s}$ = 200 GeV per 
nucleon-nucleon collision,
initial temperature, $T_{0}$, is expected 
to reach as high as $T_{0} \approx$ 0.5 GeV \cite{ref:Vogt}
at which the formation of QGP is expected.

A heavy quarkonium, a bound state of a heavy quark ($Q$) and anti heavy
quark ($\bar{Q}$),
is detected via a leptonic decay with the PHENIX detector at RHIC.
It is considered to be one of the best probes for the earliest stages of
collisions since 
$Q\bar{Q}$ pairs can be produced only in hard parton interactions
(dominantly gluon fusion at RHIC energies)
and final state leptons do not suffer from interactions
with hadronic matter in the later stages
of the collisions.
In deconfined (QGP) matter,
$Q$ and $\bar{Q}$ tend not to form a quarkonium
and diffuse away from each other instead,
because attractive color forces between them
are reduced by a Debye-type screening. 
Therefore, suppression of quarkonium yield
is expected in QGP \cite{ref:Matsui}.
Another theoretical model predicts
enhancement of quarkonium production due to recombination
of $Q$ and $\bar{Q}$ produced in different nucleon-nucleon collisions
because of their higher mobility inside the QGP \cite{ref:Thews}.
If we observed an anomalous suppression
or enhancement of quarkonium production in Au+Au collisions,
it could be interpreted as evidence of QGP.
This study definitely requires the reference of unsuppressed cross
section,
which can be best served by p+p collisions at the same energy.
There has been, however, no data of $J/\psi$ production 
in p+p collisions at $\sqrt{s}$ = 200~GeV,
and such measurements play a crucial role
to establish the basis of the QGP search in heavy ion collisions.

The $J/\psi$ has the largest production cross-section 
among all the heavy quarkonia and 
possible to be measured even with such a small integrated luminosity
as obtained in the RHIC Run-2 period (2001-2002) where 
accelerator performance was in the stage of progress.
With an increased luminosity, measurements and
comparison with other charmonium and bottomonium
states are important,
since different binding radii for each quarkonium 
will result in different degrees of suppressions.


\subsection{Spin physics}

Spin is one of the fundamental properties indigenous to
nuclei and elementary particles.
Studies of spin have been providing deep
understandings of properties of particles and their interactions. 

An example is the 
studies of nucleon spin structure
using deep inelastic scattering (DIS) of
polarized leptons from polarized nucleons.
Since the 1970s, the polarized structure functions
of nucleons, $g_{1}^{p(n)}(x)$ have been measured
using polarized electrons or muons and polarized
nuclear targets \cite{ref:pol-DIS}.
As a result, the net quark polarization inside a proton
$\Delta\Sigma$ has turned out to be as small as 0.1 to 0.3
which is much smaller than the naive expectation ($\sim$~1).
This discrepancy is called the ``proton spin puzzle''.
A significant fraction of the proton spin is 
allegedly carried by gluons in analogy to the case of momentum.
However, 
polarized gluon distribution $\Delta g(x)
= g(x)^{+} - g(x)^{-}$
where $g(x)^{+} (g(x)^{-})$ is distribution of gluons
polarized in the same (opposite) direction as the proton helicity,
suffers from
large theoretical uncertainties in extraction
from the DIS results
since gluons do not couple to photons
at the leading order.

RHIC has been added a new feature of accelerating
polarized protons up to 250 GeV/$c$ momentum
keeping 70\% polarization
with the advent of innovative technologies such as Siberian Snakes
\cite{ref:Siberian_Snake}
to keep proton polarization through acceleration.
Studying polarized phenomena at the perturbative QCD regime is now
accessible using polarized p+p collisions at
the highest energies up to $\sqrt{s}$ = 500 GeV
\cite{ref:RHIC-spin},
which is much higher than the previous polarized p+p experiment
($\sqrt{s}$ = 19 GeV) \cite{ref:E704}.
For example,
we plan the first direct measurement of 
$\Delta g(x)$, 
through double-spin asymmetries ($A_{LL}$)
for cross sections of gluon-initiated subprocesses such as
the direct-photon production \cite{ref:Goto-deltaG}.
It is important to measure $A_{LL}$ for various subprocesses
to elude theoretical uncertainties to extract $\Delta g(x)$.
Quarkonium production is one of them. 
An asymmetry $A_{LL}$ for quarkonium $\psi_{Q}$ production can be written as
\[ A_{LL}^{\psi_{Q}} = \frac{\Delta g(x_{1})}{g(x_{1})} 
	    \frac{\Delta g(x_{2})}{g(x_{2})} 
            a_{LL}^{gg \rightarrow \psi_{Q} X} \]
where $x_{1(2)}$ denotes a momentum fraction of each parton
and $a_{LL}^{gg \rightarrow \psi_{Q} X}$ is a partonic
asymmetry for the subprocess $g+g \rightarrow \psi_{Q} X$\footnote{
represents inclusive production}.
An advantage of this channel is 
an experimental feasibility due to 
a relatively large cross-section and small background
especially for the case of $J/\psi$ as already stated.
Also the subprocess (gluon fusion) can be identified with a small 
($< 10$\%) contamination from others.
Therefore we will be able to measure 
$A_{LL}^{J/\psi}$ for $g+g \rightarrow J/\psi X$ with 
a small uncertainty.

However, the extraction of $\Delta g(x)$ from $A_{LL}^{J/\psi}$
suffers from large theoretical ambiguities
due to the lack of knowledge of 
$a_{LL}^{gg \rightarrow \psi_{Q}}$,
which is sensitive to the production mechanism.
Therefore the elucidation of the production mechanism
is a crucial key
to extract $\Delta g(x)$ from $A_{LL}^{J/\psi}$.

\vspace{1cm}

In summary, measurements of $J/\psi$ in 
p+p collisions at RHIC are vital to:
(1) discuss the formation of QGP in Au+Au collisions
and 
(2) clarify the mechanism of quarkonium production to obtain $\Delta
g(x)$ in polarized p+p collisions.

\subsection{$J/\psi$ production in hadron-hadron collisions}

Current theoretical and experimental knowledge of
hadro-production of charmonia is described
in the following.

\begin{figure}[htbp]
\begin{center}
\includegraphics[width=4.5in,angle=0]{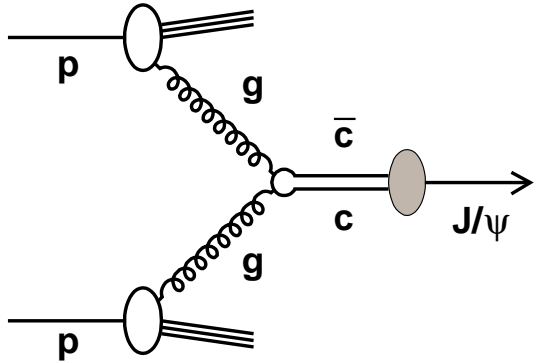}
\end{center}
\caption[An example of hadro-production of $J/\psi$
]{
An example of $J/\psi$ hadro-production. In this example,
a $J/\psi$ meson is produced in the subprocess $g + g \rightarrow J/\psi$ 
in a p+p collision. The oval between the $c\bar{c}$ pair and
the $J/\psi$ represents the formation of the $J/\psi$ from the $c\bar{c}$ pair
which is a non-perturbative phenomenon.
}
\label{fig:feynman_jpsi1}
\end{figure}

Production of $J/\psi$ and other charmonia in hadron-hadron collisions
is understood in the following framework
based upon the QCD factorization theorem 
\cite{ref:factorization_theorem}.
Figure~\ref{fig:feynman_jpsi1} shows an example of
Feynman diagram for hadro-production of $J/\psi$.
According to the factorization theorem,
the cross section to produce a charmonium $\psi$ 
in a collision of hadron $A$ and $B$, 
$\sigma(AB \rightarrow \psi X)$ can be factorized into: 

\begin{itemize}
\item $f_{a/A}(x,Q)$: 
  Probability for a parton $a$ to be found in a hadron $A$,
  called a \b{P}arton \b{D}istribution \b{F}unction (PDF),
  where $x$ is a momentum fraction of $a$ to $A$ 
  and $Q$ is a scale for the parton interaction,  

\item $f_{b/B}(x,Q)$: 
  Probability for a parton $b$ to be found in a hadron $B$ and
\item $\sigma(ab \rightarrow \psi X)$:
  Cross section for the partonic subprocess
  $ab \rightarrow \psi X$, where $\psi$ is produced in the 
  hard scattering of the parton $a$ and $b$,

\end{itemize}

\noindent
and written as 

\[ \sigma(AB \rightarrow \psi X) = \sum\limits_{ab} 
 \int \!\!\! \int dx_{1}dx_{2} f_{a/A}(x_{1},Q)f_{b/B}(x_{2},Q)
 \sigma(ab \rightarrow \psi X).
\]

\noindent
Following 2$\rightarrow$1 subprocesses contribute to
low-$p_{T}$ (transverse momentum) production of $\psi$ ($\lesssim$ 1 GeV/$c$):

\begin{itemize}
\item $g + g \rightarrow \psi$ and
\vspace{-3mm}
\item $q + \bar{q} \rightarrow \psi$,
\end{itemize}

\noindent
where $q$ and $g$ denote a quark and a gluon respectively. 
Figure~\ref{fig:feynman_jpsi1} is an example of the $g + g \rightarrow
\psi$ subprocess.
Medium and high $p_{T}$ $\psi$'s
are produced in the following 2$\rightarrow$2 subprocesses:

\begin{itemize}
\item $g + g \rightarrow \psi + g$,
\vspace{-3mm}
\item $g + q \rightarrow \psi + q$ and
\vspace{-3mm}
\item $q + \bar{q} \rightarrow \psi + g$.
\end{itemize}

\noindent
At RHIC energies ($\sqrt{s}$ = 200 to 500 GeV)
$g + g$ and $g + q$ subprocesses contribute 
to cross sections significantly. 

Production of $\psi$ from initial partons
is further separated into two steps:
production of a $c\bar{c}$ pair in the hard scattering of
the initial partons ({\bf step~1}) and
hadronization of a charmonium $\psi$ from the $c\bar{c}$ pair
({\bf step~2}).
This is possible since energy scales for each step are well separated,
that is, 2$M_{c} \sim$ 3~GeV ($M_{c}$ is the charm quark mass) for 
{\bf step~1} and
$\Lambda_{QCD} \sim$ 0.2~GeV for {\bf step~2} which is the inverse of 
a typical hadron size. 
The cross section for {\bf step~1} can be calculated using perturbative QCD.
On the other hand, the probability for {\bf step~2} is not calculable
with it, because it is a non-perturbative phenomenon. 
There are some
theoretical models for charmonium production
each of which assumes a different assumption 
and treatment on {\bf step~2}.

\begin{figure}[htbp]
\begin{center}
\includegraphics[width=4.5in,angle=0]{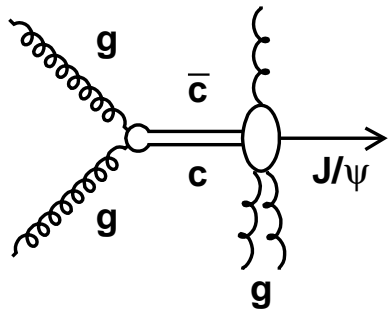}
\end{center}
\caption[$J/\psi$ production with the color-evaporation model
]{
A schematic diagram for $J/\psi$ production from gluon fusion
with the color-evaporation model.
Incoming protons are omitted in the figure.
Multiple soft-gluon emissions destroy the
information on quantum numbers of the $c\bar{c}$ pair.
}
\label{fig:feynman_jpsi_cem}
\end{figure}

The color-evaporation model (CEM)
or the semi-local duality approach~\cite{ref:cem_org1,ref:cem_org2},
born in the 1970s, simply ignores the color
and other quantum numbers of $c\bar{c}$ pairs 
and assumes a certain fraction of them,
which is needed to be determined from experimental data,
forms each charmonium state through multiple 
soft-gluon emissions as illustrated in
Fig.~\ref{fig:feynman_jpsi_cem}.
The CEM describes experimental data on
low-$p_{T}$ or $p_{T}$-integrated results well,
where the soft gluon picture is expected to be valid.
For example, the CEM describes $J/\psi$'s total cross sections
in both hadro-production and photo-production
at lower energies~\cite{ref:cem_update}.
Also the CEM prediction of zero polarization (spin-alignment) of $J/\psi$'s
is consistent with the lower-energy experiments
where low-$p_{T}$ contribution is dominant
\cite{ref:jpsi_pol_fixed_target},
but contradicts to the CDF data at
medium and high $p_{T}$~\cite{ref:cdf_jpsi_pol}.

\begin{figure}[htbp]
\begin{center}
\includegraphics[width=4.5in,angle=0]{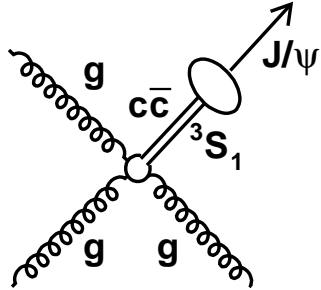}
\end{center}
\caption[$J/\psi$ production with the color-singlet model
]{
An example of the lowest order diagram for direct $J/\psi$ production
from gluon fusion with the color-singlet model.
Incoming protons are omitted in the figure.
The $c\bar{c}$ pair is in the color-singlet state.
}
\label{fig:feynman_jpsi_csm}
\end{figure}

A more sophisticated model born in the 1980s,
the color-singlet model (CSM) \cite{ref:color_singlet_model,ref:Shuler_CSM_1994},
requires a $c\bar{c}$ pair to 
be the color-singlet state and have 
the same quantum numbers as the charmonium to be formed. 
Figure~\ref{fig:feynman_jpsi_csm} shows an example of 
the lowest order production of a $J/\psi$ with the CSM
where the $c\bar{c}$ pair should be in ${}^{2s+1}L_{J}$ = ${}^{3}S_{1}$
and the color-singlet state as the $J/\psi$.
It should be noted that an additional hard-gluon emission 
is necessary to conserve the $C$-parity.
The CSM can unambiguously predict production cross sections for each
charmonium without any free parameters
and has explained $p_{T}$ distributions of $J/\psi$
production at ISR energies 
($\sqrt{s}$ = 30 to 63 GeV) reasonably well \cite{ref:color_singlet_model}.
However the CSM failed to explain $p_{T}$ 
differential cross sections 
of the Tevatron data 
in p+$\bar{\mbox{p}}$ collisions at
$\sqrt{s}$ = 1.8 TeV
by large factors (30 $\sim$ 50) \cite{ref:cdf_jpsi}.
CSM predictions do not agree either with
the total cross sections at lower energies
by a factor of about 20~\cite{ref:COM_fixed_target}.

\begin{figure}[htbp]
\begin{center}
\includegraphics[width=4.5in,angle=0]{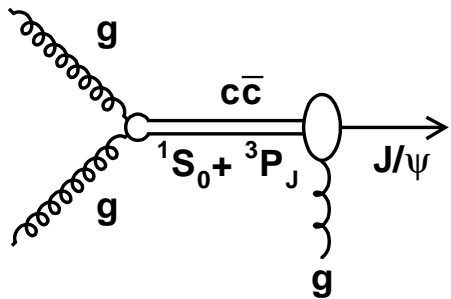}
\end{center}
\caption[$J/\psi$ production with the color-octet model
]{
An example of the lowest order diagram for direct $J/\psi$ production
from gluon fusion with the color-octet model.
Incoming protons are omitted in the figure.
The $c\bar{c}$ pair is in the color-octet state.
}
\label{fig:feynman_jpsi_com}
\end{figure}

\begin{figure}[htbp]
\begin{center}
\includegraphics[width=4.0in,angle=0]{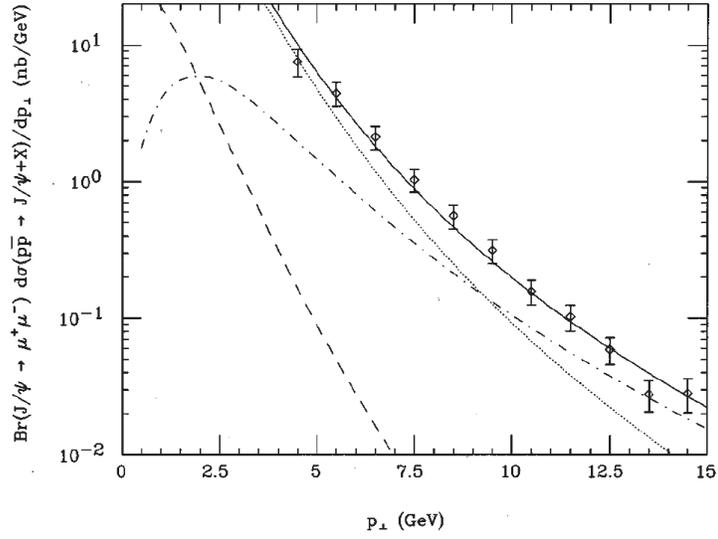}
\end{center}
\caption[Transverse momentum spectrum for $J/\psi$ production
cross section measured by the CDF experiment at $\sqrt{s}$ = 1.8~TeV
]{
Transverse momentum differential cross sections for 
prompt (excluding $b$-quark decays) $J/\psi$ production
measured by the CDF experiment
compared with theoretical predictions.
The dashed curve depicts the direct color-singlet contribution.
The dot-dashed curve illustrates the contribution
of the ${}^{3}S_{1}$ octet state of the $c\bar{c}$ pairs
and the dotted curve denotes the combined contribution
of the ${}^{1}S_{0}$ and ${}^{3}P_{J}$ octet states. 
The solid curve equals the sum of the color-singlet and
color-octet contributions.
All curves are multiplied by the muon branching ratio
$B(J/\psi \rightarrow \mu^{+}\mu^{-})$.
}
\label{fig:cdf_pt}
\end{figure}

To explain these discrepancies,
the color-octet model (COM) \cite{ref:color_octet_model}
has been developed in the 1990s
based upon the non-relativistic QCD framework~\cite{ref:NRQCD}.
The COM allows the formation of a charmonium 
from a color-octet $c\bar{c}$ pair
with one or some soft gluon emissions.
Figure~\ref{fig:feynman_jpsi_com} shows an example of 
the lowest order production of a $J/\psi$ with the COM
from the gluon fusion subprocess.
The $c\bar{c}$ pair, which is in 
${}^{1}S_{0}$ (or ${}^{3}P_{J}$) and the
color-octet state, forms a $J/\psi$ 
with a soft gluon emission.
Using appropriate color-octet matrix elements,
which are additional free parameters 
needed to be extracted from experimental data,
the COM has successfully reproduced $p_{T}$
distributions at CDF 
\cite{ref:COM_with_CDF,ref:com_cano_coloma,ref:sakuma_master_thesis}
as shown in Fig.~\ref{fig:cdf_pt}
and total cross sections at
lower-energy experiments
\cite{ref:COM_fixed_target,ref:Beneke_96,ref:WKT_96}.
COM predictions for the relative
yields for each charmonium state 
(for example, $\chi_{c}$ to direct $J/\psi$ ratio)
are also consistent with the experimental data
\cite{ref:Beneke_96}, which will be discussed in Appendix \ref{chap:feeddown}.
However extraction of these matrix elements
is still controversial and therefore
large ambiguities are left for the
prediction with the COM.

Hadro-production data on $J/\psi$ and
other charmonia are available
in both fixed-target experiments
and collider experiments
in a wide energy range (6.1~GeV $\le$ $\sqrt{s}$ $\le$ 1.8~TeV)
~\cite{ref:exp_sqrt_6.1,ref:exp_sqrt_6.8,ref:exp_sqrt_8.7,ref:exp_sqrt_11.5,ref:exp_sqrt_16.8,ref:exp_sqrt_19.4,ref:exp_sqrt_20.5,ref:exp_sqrt_23.7,ref:E705_chi,ref:exp_sqrt_24.3,ref:exp_sqrt_27.4_2,ref:exp_sqrt_31.5_E672,ref:exp_sqrt_31.5_E789,ref:b_jpsi_E789,ref:E672-706,ref:exp_sqrt_52.0,ref:exp_sqrt_63.0,ref:exp_sqrt_30-63,ref:jpsi_pol_fixed_target,ref:ua1_jpsi,ref:cdf_jpsi,ref:cdf_jpsi_chi,ref:cdf_jpsi_pol,ref:d0_jpsi}. 
Measurements of such as total and 
$p_{T}$-differential cross-sections, polarization,
and relative yields of each charmonium have promoted
better understanding of production mechanism.
However more data on different observables
at different energies are required
since currently none of the theoretical models
can successfully explain all the experimental data 
at all energies. 

\vspace{1cm}

Our measurements of $J/\psi$ in p+p collisions using the PHENIX
detector at RHIC
will result in another critical input for
understanding of hadro-production of $J/\psi$.
The energy region of RHIC ($\sqrt{s}$ = 200 to 500~GeV)
has never been exploited yet by previous experiments.
Thanks to excellent lepton-identification capabilities 
of two independent spectrometers of PHENIX
which cover both the central rapidity region
($|y| < $0.35) 
and the forward region (1.2 $< y < $1.2),
we are able
to extract total cross-section for $J/\psi$ production
with less ambiguity in the extrapolation to 
unmeasured kinematical region, 
unlike higher-energy ($\sqrt{s} \ge$ 630~GeV)
collider experiments with a $p_{T}$ cut ($p_{T} >$ 4 or 5~GeV/$c$)
and limited kinematical coverage in the forward region.
Therefore, we can obtain total cross-section results at the highest
energies, which are sensitive to the production models.

In addition, our measurements of differential
cross sections will provide further information
to understand perturbative QCD.
Our wider kinematical coverage enables us to measure rapidity-differential
cross section which is reflected by 
gluon distribution function $g(x)$ in the proton.
Transverse momentum ($p_{T}$) distribution at lower $p_{T}$,
which dominates our inclusive yield, and the average 
value of $p_{T}$ are sensitive to intrinsic 
transverse-momentum of partons ($k_{T}$).

RHIC has started a physics run from year 2000.
In the 2001-2002 run period (Run-2), 
both Au+Au collision and p+p collision
data have been accumulated, where $J/\psi$ particles
have been successfully detected via both the
$e^{+}e^{-}$ and $\mu^{+}\mu^{-}$ decay channels.
In this paper, 
measurements of total and differential cross sections
for inclusive $J/\psi$ production
in p+p collisions at $\sqrt{s}$ = 200~GeV are presented.
In section~2 and 3, 
details of the experimental setup 
and analysis procedure are described 
respectively, focusing on
the $\mu^{+}\mu^{-}$ channel measurement.
In section~4, results 
are discussed 
together with theoretical
predictions as well as results of other experiments,
followed by the conclusion.

\section{Experimental setup and data acquisition}
\label{chap:exp_setup}

\subsection{The RHIC accelerator complex}

The \b{R}elativistic \b{H}eavy-\b{I}on \b{C}ollider (RHIC), 
located in Brookhaven
National Laboratory, Upton, New York,
is capable of accelerating a wide variety of nuclei and ions from 
protons to Au (gold) nuclei up to 250-GeV energy for protons
(or 100-GeV per nucleon for Au)
using two independent rings and colliding them at six interaction
points. The design luminosities are
2 $\times$ $10^{26}$~cm$^{-2}$~sec$^{-1}$ for Au beams and
2 $\times$ 10$^{31}$~cm$^{-2}$~sec$^{-1}$ for proton beams 
(2 $\times$ 10$^{31}$~cm$^{-2}$~sec$^{-1}$ in an enhanced mode)
at the top energy. 

RHIC also has a capability of accelerating polarized protons.
In the Run-2 period (year 2001-2002), RHIC has successfully
achieved transversely-polarized proton-proton collisions 
at $\sqrt{s}=$ 200~GeV and delivered an integrated luminosity of 
roughly 1 pb$^{-1}$ to each experiment. 
In this paper we will present unpolarized cross sections for the
$J/\psi$ production, that is, 
spin-averaged results. 
Therefore, 
descriptions of the polarized proton acceleration are not described here.


\begin{figure}[htbp]
\begin{center}
\includegraphics[width=3.5in, keepaspectratio]{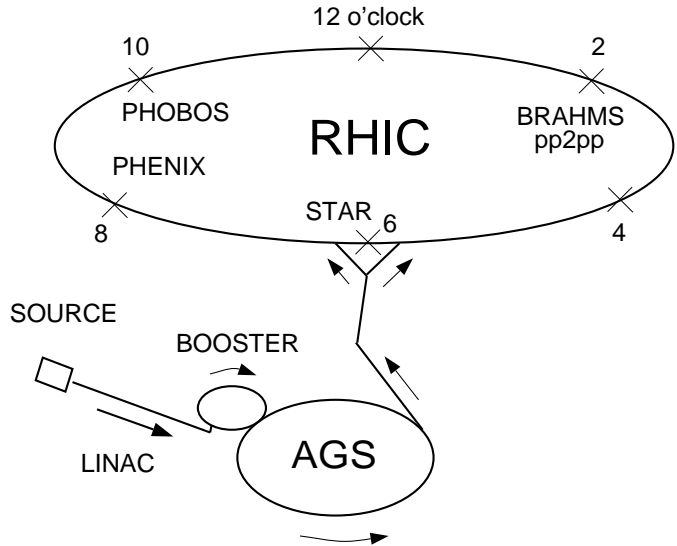}
\end{center}
\caption{The proton acceleration to RHIC}
\label{fig:rhic_line}
\end{figure}

Attached figure~1 shows an aerial view of
the RHIC accelerator complex. 
The route for the proton acceleration
to RHIC is shown in Fig.~\ref{fig:rhic_line}.
Starting from the polarized ion source,
(polarized) protons are accelerated through Linac,
Booster and AGS then injected into both rings of RHIC,
whose circumference is 3.834 km.
The Blue ring runs clockwise and the Yellow ring runs 
counter-clockwise. There are currently 120 bunch buckets
in each ring whose interval is 106 nsec (or 9.4 MHz frequency).
In Run-2, only about a half of them (54 bunch buckets) have been filled.
Typical bunch length was 2 nsec (60 cm) in rms during the p+p run.

Experiments are located at the interaction points in RHIC where
two bunches in each ring collide at an angle of 0$^{\circ}$.
There are six interaction points called
12, 2, 4, 6, 8, and 10 o'clock respectively starting from the north and 
going clockwise.
Table~\ref{tab:RHIC_experiments} shows the experiments at each 
interaction point.
Each experiment demonstrates its unique feature.
STAR and PHENIX are the largest experiments at RHIC each with
more than 400 collaborators.
STAR (\b{S}olenoid \b{T}racker \b{A}t \b{R}HIC) tracks and identifies charged particles
with a time projection chamber covering a large solid angle.
PHENIX (\b{P}ioneering \b{H}igh \b{E}nergy \b{N}uclear
and \b{I}on e\b{X}periment)
was designed to measure hadrons, leptons and photons
in both high multiplicity and high rate environments.
BRAHMS measures hadrons over wide ranges of rapidity
and momentum using two magnetic spectrometers.
PHOBOS consists of a large number of silicon detectors surrounding the
interaction region to measure charged particle multiplicities
even in the most central Au+Au collisions.
The pp2pp experiment is aiming at measuring
p+p total and elastic cross sections.

\begin{table}[htbp]
\begin{center}
\begin{tabular}{|c|c|}\hline
  Interaction Point & Experiment \\
  (o'clock) &   \\ \hline

  12   & - \\
  2    & pp2pp, BRAHMS \\ 
  4    & -  \\ 
  6    & STAR \\ 
  8    & PHENIX \\
  10   & PHOBOS \\ \hline

\end{tabular}
\end{center}
\caption{Experiments at RHIC}
\label{tab:RHIC_experiments}
\end{table}

\subsection{Coordinates and formulae}

In this subsection, coordinate system and formulae
used to describe the experimental setup
and results are introduced.

\subsubsection*{Coordinate system at PHENIX}

The $z$ axis stands for the beam line which 
runs straight in the experimental area,
where positive $z$ points to the north.
The polar angle $\theta$ and azimuthal angle $\phi$ are defined with
respect to the $z$ axis. 
The north (south) direction is usually defined as $\theta$ = 0
(180) degrees and the 
west (east) direction is defined as $\phi$ = 0 (180) degrees.


\subsubsection*{Rapidity}
Rapidity $y$ of a particle is defined as 

\[ y \equiv \frac{1}{2} \ln \frac{E + p_{z}}{E - p_{z}} \]

\noindent
where $E$ is the energy and $p_{z}$ is
the $z$-component of momentum of the particle.

\subsubsection*{Pseudo rapidity}
Pseudo rapidity $\eta$ of a particle is defined as 

\[ \eta \equiv \frac{1}{2} \ln \frac{p_{tot} + p_{z}}{p_{tot} - p_{z}} 
= -\ln \tan \frac{\theta}{2}
\]

\noindent
where $p_{tot}$ is the scalar value of momentum
and $\theta$ is the polar angle of the particle direction.
In the massless limit, $y$ reaches $\eta$. 

In the definition above, $y (\eta$) of a particle
going to the south is negative. However, 
absolute (positive) values of $y (\eta$) will be often used 
for those particles in this paper,
since results with the spectrometer on the south side of PHENIX
will be mainly described and physics should be symmetric with
respect to $y (\eta)$ = 0.

\subsection{The PHENIX experiment overview}


PHENIX
is one of the largest experiments at RHIC, 
located at the 8-o'clock interaction region. 
Attached figure~2 shows a schematic view of the
PHENIX experiment.
PHENIX was designed to measure leptons, photons and
hadrons in both high-multiplicity heavy-ion collisions and 
high event-rate p+p collisions.

There are two independent spectrometers
in PHENIX which cover different pseudo-rapidity regions.
Two Central Arms, East and West Arms,
cover the pseudo-rapidity range of 
$|\eta| < 0.35$ with a quarter azimuth for
each Arm
and measure electrons, photons and hadrons.
Two Muon Arms, North and South Arms,
cover $1.2$ $<\eta<$ $2.4$ and
$-2.2 <\eta< -1.2$ respectively with a full azimuth
and measures muons.

There are three magnets in PHENIX. The Central Magnet provides an
axial magnetic field for the Central Arms while two Muon Magnets
produce a radial field for each Muon Arm.
Attached figure~3 shows magnetic field
lines inside the magnets.
A part of the Central Magnet steel
and a copper spacer mounted on each side of it
(called a copper nosecone)
work as a hadron absorber
of about five interaction-length
for the Muon Arms.
The positions of the nosecones,
$z$ = $\pm$ 40 cm,
determine a useful vertex region
for physics events ($|z| <$ 30 to 40 cm depending on analysis).


The Central Arms consist of three kinds of tracking chambers
(Drift Chambers, Pad Chambers and Time-Expansion
Chambers), Ring-Imaging $\check{\mbox{C}}$erenkov Counters
for the electron identification, Time-Of-Flight detectors
for the particle-identification for hadrons, and Electro-Magnetic
Calorimeters for measuring energies of electrons and photons.
These detectors are positioned radially 
with respect to the $z$-axis
extending from 2 m to 5 m.
Details of operation and performance of the Central Arm detectors
can be found in
~\cite{ref:Jeff_NIM,ref:EMCal_NIM,ref:Jane_thesis}.
Descriptions of the analysis of the $J/\psi \rightarrow e^{+}e^{-}$
channel using the Central Arms are found
in~\cite{ref:PHENIX_jpsi_ee}.

In addition to these four Arms, there are three kinds 
of counters to trigger p+p interactions which will be
described in the next subsection followed by the subsection dedicated to
description of the Muon Arms.

\subsection{Interaction trigger counters}

Three kinds of interaction-trigger counters (ITC) have been used 
to trigger p+p inelastic events and find vertices 
during the Run-2 p+p period. They are Beam-Beam Counters (BBC), 
Normalization Trigger Counters (NTC) and Zero-Degree Calorimeters (ZDC).
Table~\ref{tab:ITC_summary} summarizes acceptance and performance
of each ITC.
Since trigger efficiency of the ZDC for p+p inelastic events
is small (0.01), it was not used for
the $J/\psi$ analysis, hence will not be described here.
Detailed information on the ZDC is found in \cite{ref:ZDC_NIM} 


\begin{table}[htbp]
\begin{center}
\begin{tabular}{|c||c|c|c|c|}\hline
  Counters & Acceptance & Sensitive particle & Typical  &
   Typical vertex \\ 
   & & type & efficiency & resolution (cm) \\ \hline

   \raisebox{-2.5mm}[0mm][0mm]{BBC}
   & 3.0$<|\eta|<$3.9 
   &  \raisebox{-2.5mm}[0mm][0mm]{charged particles} 
   &  \raisebox{-2.5mm}[0mm][0mm]{0.5} 
   &  \raisebox{-2.5mm}[0mm][0mm]{2} \\
   & $\beta>0.7$ & & & \\ \hline
  NTC & 1.1$<|\eta|<$2.8 & charged particles & 0.6 & 10 \\ \hline
  ZDC & $|\eta|>$6.2 & neutral particles & 0.01 & 10 \\ \hline

\end{tabular}
\end{center}
\caption[Acceptances and performances of the interaction-trigger counters]
{Acceptances and performances of the interaction-trigger
   counters. Efficiencies and resolutions are for p+p inelastic events.}
\label{tab:ITC_summary}
\end{table}

\subsubsection{The Beam-Beam Counters}

Two \b{B}eam-\b{B}eam \b{C}ounters (BBC) have been used for the
primary trigger and vertex counter for the p+p interactions.
They are positioned 1.4 meters away from the interaction
point along the beam axis on each side
and cover from 2.4 to 5.7 degrees (3.0 $<|\eta|<$ 3.9) with
a full azimuth.
They determine the vertex position of an event from
the time difference between two counters.
Attached figure~4 shows a photograph 
of a BBC.
Each BBC consists of 64 hexagonal 
$\check{\mbox{C}}$erenkov radiators each of which
is mounted on a 1'' photomultiplier tube (PMT).
They are sensitive to charged particles whose $\beta \equiv v/c$ is greater 
than 0.7.


With a beam test,
intrinsic timing resolution of 50~psec 
was obtained for one module \cite{ref:BBC_NIM}.
In Au+Au collisions, better vertex resolution is expected
because hit multiplicity is higher. Actually 
0.5-cm vertex resolution has been obtained in Run-2 Au+Au collisions.
In p+p collisions,
about 2-cm vertex resolution is expected
which is still 
good enough not to worsen invariant mass resolution for $J/\psi$ particles
measured in the Muon Arm.
Trigger efficiency
for p+p inelastic events is expected to be
about 0.5, which will be discussed in section~\ref{chap:analysis}.

\subsubsection*{Beam-Beam Local-Level-1}

Hit time information of all PMTs is sent to
the Beam-Beam Local-Level-1 (BBLL1) board 
where a trigger decision is made.
Online vertex position is obtained 
from hit time information without
pulse-height corrections (threwing corrections).
During the p+p run, a trigger is fired when there is
at least one hit on both sides of the BBC counters 
and online $z$-vertex position $|z_{vtx}| <$ 75 cm
which is sufficiently large compared to offline vertex cuts
(30 to 40~cm).
Online vertices (without slewing corrections)
and offline vertices (with slewing corrections) agree
within the accuracy of the vertex determination (2~cm).

\subsubsection{The Normalization Trigger Counters}

\b{N}ormalization \b{T}rigger \b{C}ounters (NTC) have been introduced
to increase trigger efficiency for p+p inelastic 
(including diffractive) events 
which is about 0.5 with the BBC only.

Each NTC is located on top of each side of the nosecone 
which is 40-cm away from 
the interaction point.
Pseudo-rapidity coverage is 1.1 $< |\eta| <$ 2.8.
It consists of four fan-shaped scintillators
(called quadrants) each of which is mounted on a PMT to collect
scintillation light emitted when a charged particle traverses.
Single-particle detection efficiency obtained with
a beam test is about 90\%~\cite{ref:NTC}.

A simple NIM-logic makes
NTC triggers using hit time information of all quadrants,
which are sent to the Global-Level-1 board 
(see section \ref{section:trigger}).
A trigger is fired when at least one quadrant on both
sides has a hit.
Trigger efficiency for
p+p inelastic events is estimated with a simulation to be 60\% 
by the NTC itself and go up
to 74\% when combined with the BBC
where statistical errors of the simulation is 1\%.
Although NTC-triggered events
without a BBC-trigger have not been used to increase
the number of $J/\psi$'s because of its poor vertex-resolution (10 cm),
they are
used to confirm trigger efficiency of the BBC which will be
discussed in section~\ref{chap:analysis}.

\subsection{The Muon Arms}


The PHENIX Muon Arms were designed to detect muon pairs 
from decays of 
vector mesons and $Z^{0}$ bosons
produced in the forward rapidity region
($-2.2<y<-1.2$ for the South Muon Arm and
$1.2<y<2.4$ for the North Muon Arm)
as well as single muons from (semi-)leptonic decays of
open heavy flavors and $W$ bosons.
Attached figure~5 shows 
a schematic view of a Muon Arm.
Since only the South Arm was operational in Run-2,
following descriptions specifically focus on it.

In this forward region, 
hadron background is relatively larger
than that in the central region,
since hadrons can be produced in soft
processes with wider rapidity distribution
than hard processes, while signal muons
such as from heavy flavors are produced
in hard processes.
Good hadron rejection while keeping good momentum resolution
is achieved with the following components of the Muon Arms
and PHENIX.

\begin{enumerate}

\item Pre-rejection of hadrons with a five-interaction-length
absorber of the nosecone and Central Magnet.
Interaction length was determined
not to degrade momentum resolution for low momentum (2 to 5~GeV/$c$)
muons and to keep good mass resolution for $J/\psi$'s.
Hadron rejection factor of about 100 is achieved here. 

\item Measurement of particle momentum with a magnetic spectrometer 
 (Muon Tracking Chamber inside the Muon Magnet)

\item Further rejection of hadrons with 
an array of coarse-segmented
tracking chambers and absorbers (Muon Identifier).
Another factor of about 30 is achieved for the hadron rejection.

\end{enumerate}

\begin{figure}[htbp]
\begin{center}
\includegraphics[width=4.5in]{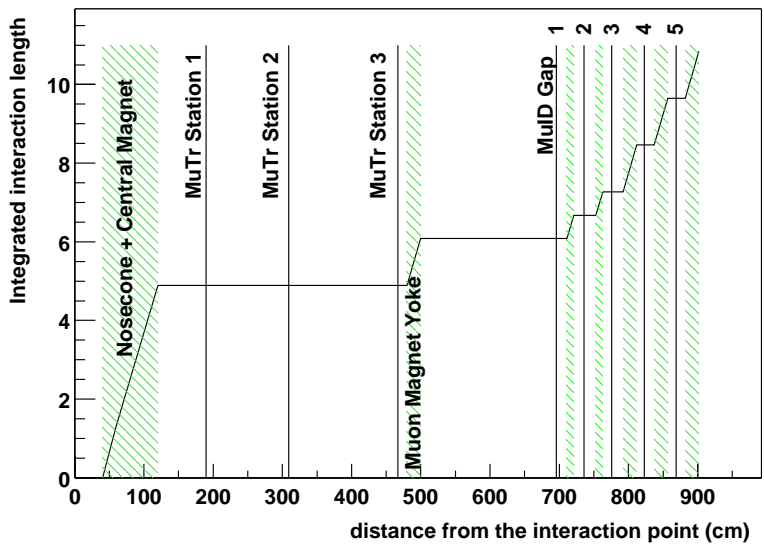}
\end{center}
\caption[Nuclear interaction length of the absorbers
in the South Muon Arm]
{Integrated 
nuclear interaction-length of the absorbers in the South Muon Arm as a function
of the distance 
from the interaction point in the $z$ direction. 
Vertical lines indicate rough positions of the chambers.
Hatched areas represent absorber materials.
}
\label{fig:muon_int_length}
\end{figure}

Figure~\ref{fig:muon_int_length} shows integrated 
nuclear interaction-length ($\lambda_{int}$) 
in the South Muon Arm as a function
of the distance from the interaction point in the $z$ direction.
At the last MuID gap (gap 5), $\lambda_{int}$ becomes 9.65.
The minimum $p_{z}$ ($z$-component of momentum)
for a muon to reach gap 5 is 2.5 GeV/$c$.

\subsubsection{The Muon Tracker}

The PHENIX \b{Mu}on \b{Tr}acker (MuTr)
comprises three stations of tracking chambers inside the Muon Magnet
as shown in Fig.~\ref{fig:muon_north}.
Its design was
driven by requirements from both heavy-ion
physics and spin physics. 
The separation of each charmonium or bottomonium
state from the others, $J/\psi$ from $\psi$' for example, is
essential to find a QGP signal, since the degrees of suppression
for each state are expected to 
vary because of different binding radii.
For spin physics, charges of high-$p_{T}$ 
($p_{T} > $ 20 GeV/$c$)
muons from $W$ and $Z$
boson decays are needed to be identified \cite{ref:RHIC-spin}.
To satisfy the requirements above,
100-$\mu$m resolution is needed for chamber resolution.
In addition, multiple cathode-strip
orientations and read-out planes are required for each station
to reconstruct tracks efficiently
even in the most central Au+Au events.
A MuTr electronics design was also
driven by the requirement of 100-$\mu$m resolution measurements.



\begin{figure}[htbp]
\begin{center}
\includegraphics[width=2.5in,angle=-90]{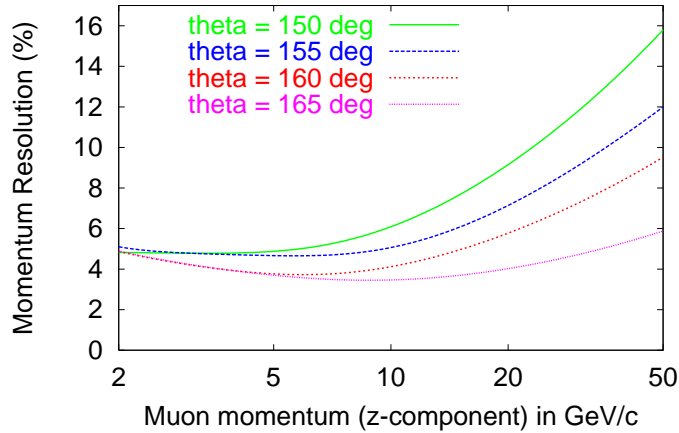}
\end{center}
\caption[Muon momentum resolution]{Momentum resolution for muons 
with different polar angles in the South Arm acceptance
as a function of $p_{z}$ obtained with a simulation.
Typical statistical errors of the simulation is 5\%.
}
\label{fig:mom_reso}
\end{figure}

With 100-$\mu$m position resolution,
momentum resolution $\Delta p/p$ = 3 to 5\% 
is achieved for 2 to 10 GeV/$c$ muons
as shown in Fig.~\ref{fig:mom_reso}.
For low momentum (2 to 5 GeV/$c$) muons, multiple
scattering is the dominant factor to smear muon momenta,
whereas
position resolution of chambers becomes dominant
for high momentum (above 10 GeV/$c$) muons. 
Polar angle dependence of momentum resolution
for high momentum muons
is due to the difference in magnitudes of the magnetic
field inside the South Muon Magnet as shown in Table~\ref{tab:muon_mag_field}.

\begin{table}[htbp]
\begin{center}
\begin{tabular}{c|cccc} \hline
   Polar angle in degrees & 165 & 160 & 155 & 150 \\ \hline
   Integrated magnetic field in T$\cdot$m & 0.774  & 0.494  & 0.344  & 0.255 \\ \hline
\end{tabular}
\end{center}
\caption[Integrated magnetic field inside the South Muon Magnets]
{Integrated magnetic field inside the South Muon Magnets. 
}
\label{tab:muon_mag_field}
\end{table}

\subsubsection*{Mechanical design}

All chamber planes inside the Muon Magnet are perpendicular to the $z$-axis.
Their $z$-positions for each
station are shown in Fig.~\ref{fig:muon_int_length}.
The magnetic field is
in the radial direction so that charged particles
from the interaction vertex 
bend primarily in the azimuthal direction,
which is perpendicular to the direction of cathode strips
to determine the track positions.
Typical bend from the straight line at station 2 
is 1 cm for medium-momentum (5 to 10~GeV/$c$) muons.

In one MuTr station there are three (or two)
gaps each of which consists of two cathode-strip planes
and one anode-wire plane in between
with a 3.2-mm anode-cathode spacing.
Attached figure~6 shows a cross section of a MuTr station.
Station 3 has only two gaps because it can exploit
additional two-dimensional position information of MuID roads.
All gaps are divided into octants electrically
as shown in attached figure~7, 
together with octant numbers.
For station 2 and 3, octants are also the unit of
mechanical assemblies while quadrants are for station 1.
Attached figure~8 shows a photograph of
a station-2 octant.
An octant is further divided into two half-octants
in the middle, in each of which 
directions of anode wires and cathode strips are fixed.




A cathode plane consists of
5-mm width strips with alternate readout
to avoid cross-talks between them.
An anode plane is an alternating structure of 20-$\mu$m 
gold-plated tungsten sense 
wires and 75-$\mu$m gold-plated Cu-Be field wires with a sense wire spacing of 
10 mm.  
Anode wires run in the azimuthal directions
while cathode strips run in the radial directions.
The direction of cathode strips in
one plane of each gap is perpendicular
to that of the anode-wires (called a non-stereo angle plane), with
a goal of 100~$\mu$m position resolution.
The direction of cathode strips in the other plane of the gap
(a stereo angle plane) is tilted by 3.25 to 11.25 degrees depending on gap
and station, which are summarized in Table~\ref{tab:cath_angle}
and illustrated in attached figure~9 for station-1
planes.
Resolutions of stereo-angle planes
are worse (300 $\mu$m) but they are
needed to determine 3-D positions of hits and
reject ghost tracks.

\begin{table}
\begin{center}
\begin{tabular}{|c|c|c|} \hline
Station & Gap & angle (degree) \\ \hline
  & 1 & $-11.25$ \\ 
1 & 2 & $+6$ \\ 
  & 3 & $+11.25$ \\ \hline
  & 1 & $+7.5$ \\ 
2 & 2 & $+3.75$ \\
  & 3 & $+11.25$ \\ \hline
 \raisebox{-2.5mm}[0mm][0mm]{3} & 1 & $-11.25$ \\ 
  & 2 & $-11.25$ \\ \hline
\end{tabular}
\caption[Angles of cathode strips in the stereo-angle planes]
{Relative angles of the cathode strips in stereo-angle planes 
with respect to the non-stereo angle (radial direction).
Positive signs represent the positive $\phi$ direction (counter-clockwise)
for all half-octants in station 1 and half-octant 0 
in station 2 and 3, while the negative $\phi$ direction for 
the others.

}

\label{tab:cath_angle}
\end{center}
\end{table}


Specific technologies were used for each station  
to produce a cathode pattern to an accuracy of better than 25 $\mu$m;
photolithography for station~1, 
electro-mechanical etching for station~2
and mechanical routing for station~3 \cite{ref:muon_NIM}.  
A unique wire laying apparatus was designed and implemented for 
each station.  

In order to reduce the multiple scattering in the 
spectrometer which degrades momentum resolution,
thickness at the station 2 detector was required 
to be less than 10$^{-3}$ of a radiation length.  
To meet this requirement, the 
station-2 octant cathodes were made of etched 25-$\mu$m copper coated mylar 
foils.  The thickness of the copper coat is 600 \AA.  
As a result, the total thickness of $8.5\times 10^{-4}$ radiation lengths
is achieved which satisfies the requirement.


The chamber gas mixture 
was 50\% Ar + 30\% CO$_{2}$ + 20\% CF$_{4}$ with a gas recirculation system included in normal operation.  
The nominal high-voltage potential applied to anode wires was
1850~V with a gain of approximately $2 \times 10^{4}$.

To maintain good momentum resolution, an optical alignment system has been 
installed to calibrate initial placement of the chambers, and to monitor 
displacement of the chambers during their operation
to $\pm 25$ $\mu$m. 
There are 
seven optical beams surrounding each octant chamber, consisting of 
an optical-fiber light source at station 1, 
a convex lens at station 2 and a CCD camera 
at station 3 which are shown in attached figure~10.  


\subsubsection*{Electronics design}
\label{sec:mutr_fee}

Attached figure~11 shows
a schematic diagram for the 
MuTr \b{F}ront \b{E}nd \b{E}lectronics (FEE).
Raw chamber signals are continuously amplified by
CPAs (\b{C}harge \b{P}re-\b{A}mps) and stored in 
AMUs (\b{A}nalog \b{M}emory \b{U}nits) with 64-event buffers
with the 10-MHz beam clock.
Upon receipt of a level-1 trigger bit from
a \b{G}ranule \b{T}iming \b{M}odule (GTM),
stored samples of all channels are 
digitized by 11-bit ADCs (\b{A}nalog to \b{D}igital \b{C}onverters)
and the results are
sent to a \b{D}ata \b{C}ollection \b{M}odule (DCM)
through a \b{F}ront \b{E}nd \b{M}odule (FEM).
GTMs and DCMs are described in section~\ref{sec:daq}.

\begin{figure}[htbp]
\begin{center}
\includegraphics[width=3.5in,angle=0]{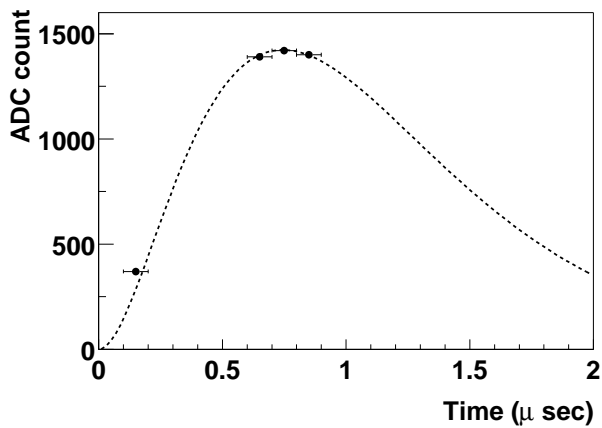}
\end{center}
\caption[Typical ADC counts for each sampling]
{Typical ADC counts for each sampling together with 
a typical signal pulse (dotted line).
The horizontal axis is for the relative sampling time.
One ADC count corresponds to about 0.5 fC ($10^{-15}$ Coulomb).
}
\label{fig:strip_adc}
\end{figure}

Four ADC samples are used to determine the amount
of the charge deposited on a strip to
reject noise hits as much as possible.
Each sample is measured for the duration of 100 nsec.
The second sampling starts 400 nsec after the first
sampling ends. Second to fourth samplings are consecutive.
Timing has been set so that the third sample
comes to the peak of the pulse as shown in 
Fig.~\ref{fig:strip_adc}.
Relative charges
(or ADC counts) of these samples have been monitored
online to guarantee peak positions
not to move around from the third sampling.
The amount of the peak charge of a strip
is obtained offline as the average of the second to fourth samples
with a pedestal subtracted.
These four samples are converted within 40 $\mu$sec 
per event.

Strip by strip calibration is crucial for good position resolution,
since a position of a muon track is determined by 
fitting charges on typically 2 or 3 consecutive strips induced by the track,
which will be described in section~\ref{sec:track}.
A calibration system has been implemented to
inject pulses into all of the chambers.
Four wires in each chamber gap, which span
the entire width of the cathode planes,
are sent a square pulse from a digital to analog 
converter (DAC), thus inducing 
a charge on all cathode strips in
a given gap simultaneously.
Several different pulse amplitudes
are sent to the chambers and many events
are collected at each amplitude so that relative gains of
the cathode strips can be determined over the entire
range of the electronics.
Pedestals are monitored by collecting calibration data with the
DAC amplitude set to zero.

To meet the design requirement of 100-$\mu$m resolution,
rms noise at the input to the preamps is
required to be 0.5 fC for a typical pulse of 80 fC,
which is achieved in a test bench measurement described next.

\subsubsection*{Integrated performance}
\label{sec:mutr_performance}

Integrated performance of chambers and electronics has
been studied in a cosmic-ray test in a test stand
with one station-2 chamber and its 
full complement of electronics, and in readout of the entire South MuTr
system prior to Run-2.  The cosmic-ray test
data showed that the system was capable of meeting the noise 
specifications and that 100-$\mu$m resolution could be achieved.  
The noise specifications have been met on 
the full South MuTr system and the system has 
been shown to be robust over several months of data taking.  

\begin{figure}[htbp]
 \begin{center}
  \includegraphics[width=3.5in]{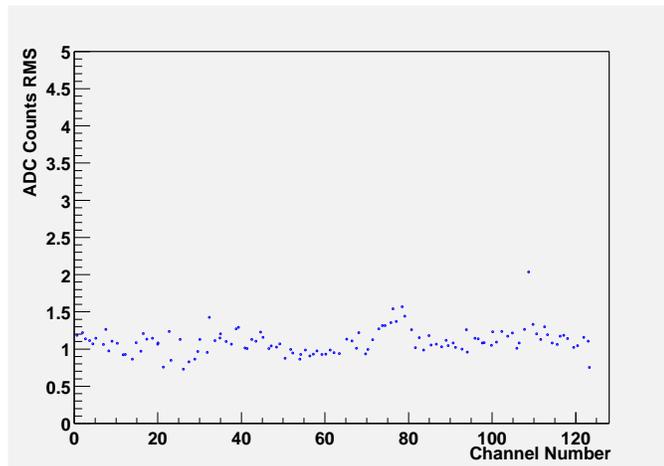}
  \caption[Noise level of the MuTr]
{Measurement of rms noises for 128 typical 
channels in units of ADC counts (1 ADC count 
is roughly 0.5 fC) obtained in the cosmic-ray tests for a 
station-2 octant.}
  \label{fig:mutr_noise}
 \end{center}
\end{figure}

The cosmic-ray test was performed with one station-2 chamber, 960
channels of production front-end electronics, the same high-voltage and 
low-voltage distribution system that is used in the final system, and 
with a copy of the PHENIX data acquisition system.  The noise 
specifications of 0.5~fC (1 ADC count) were met, as can be seen in
Fig.~\ref{fig:mutr_noise}, 
where the rms values of the pedestals on all readout channels are shown.
Noise environment in situ has turned out to be similar to this test 
except for chambers in station 2, which has caused degradation of the mass
resolution for $J/\psi$ in Run-2.

Two scintillators, one on either side of the station-2 chamber, were
used to provide a trigger for cosmic rays going through the chamber.
The data collected from this trigger were searched for clusters in each 
cathode readout plane, the clusters were fit to extract the centroid
strip positions, and 5 out of 6 readout planes were fit to a 
straight line and projected to the sixth, central non-stereo readout 
plane.  A cut was placed on the straight line fit to only the select tracks 
which were approximately perpendicular to the face of the chamber and
the difference between the projected straight-line fit and the measured 
position on the sixth plane was plotted.  The result is shown in
Attached figure~12, where a resolution of approximately 
100~$\mu$m was achieved when the projection error,
position resolution for a track obtained
with the other five planes, 
is removed from the residual.

\begin{figure}[htbp]
 \begin{center}
  \includegraphics[width=3.5in]{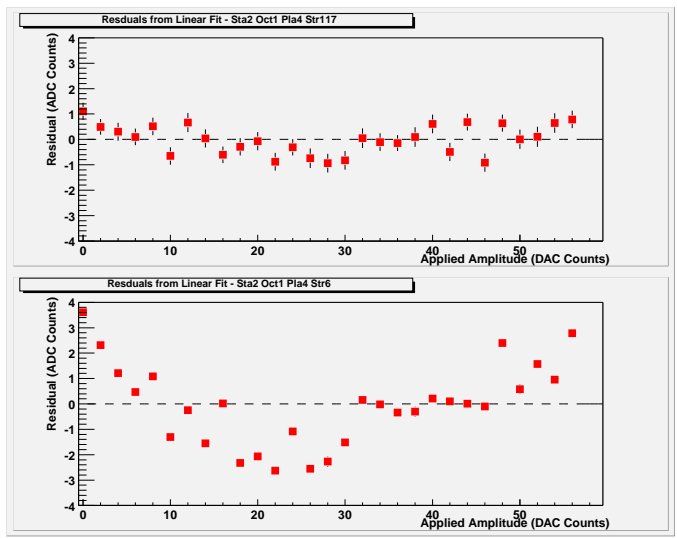}
  \caption[\,\,\,MuTr gain linearity]
{Residuals (in ADC counts) from a straight-line 
fit to the charge measured on a given strip of the ADC vs the DAC pulse 
amplitude applied to the calibration wires.  Two different strips from a 
given cathode plane are shown.  Approximately one half of the ADC range 
(of 2047 channels) is shown and is linear to a few ADC counts and constant 
over time.}
  \label{fig:mutr_gain}
 \end{center}
\end{figure}

Each individual channel's gain, pedestal and variation (or
noise) in the pedestal were measured.  The dynamic range in the charge 
measurement of the system was verified and long runs demonstrated the 
stability of the optical links from
FEMs
to a DCM.
Figure~\ref{fig:mutr_gain} shows the 
residuals from a gain measurement during the
commissioning period. The residuals are shown to be consistent with
a linear gain to a few ADC counts over the operable ADC range.

\subsubsection{The Muon Identifier}

The \b{Mu}on \b{ID}entifier (MuID) consists of five layers of chambers 
interleaved with steel absorbers.
Each chamber plane is called 
gap 1 through gap 5 starting from the nearest gap 
to the interaction point.
The MuID is used for separating muons from charged hadrons and other background
as well as providing triggers for
single muons and dimuons (muon pairs).
It gives another hadron
rejection factor of about 30 in addition
to the Central Magnet and nosecone (about 100), thus
reducing the mis-identification rate for the punch-through hadrons
to $3\times10^{-4}$. This is much smaller
compared to the irreducible hadron weak-decay 
background ($\pi \rightarrow \mu \nu$ and $K \rightarrow \mu \nu$)
in flight
before the nosecone for near-threshold momentum (3 GeV/$c$)
muons ($3\times10^{-3}$).
Minimum $p_{z}$ for a muon produced at the interaction vertex
to reach the MuID is 1.8 GeV/$c$ and 2.5 GeV/$c$ 
to penetrate through it.

Segmentation of the absorber into
multiple layers improves the measurement
of the trajectory in the MuID gaps (chamber layers)
for low momentum muons,
which is desirable to increase the acceptance for
the $\phi$ meson detection.
The segmentation chosen is 20~cm,
10~cm, 10~cm, 20~cm and 20~cm starting from
the South Magnet backplate
as shown in Fig.~\ref{fig:muon_int_length}.
In each gap between these absorbers, chamber
panels are installed.

\subsubsection*{Mechanical Design}
\label{sec:muid_mech}

\begin{figure}[htbp]
\begin{center}
\includegraphics[width=3.5in,angle=0]{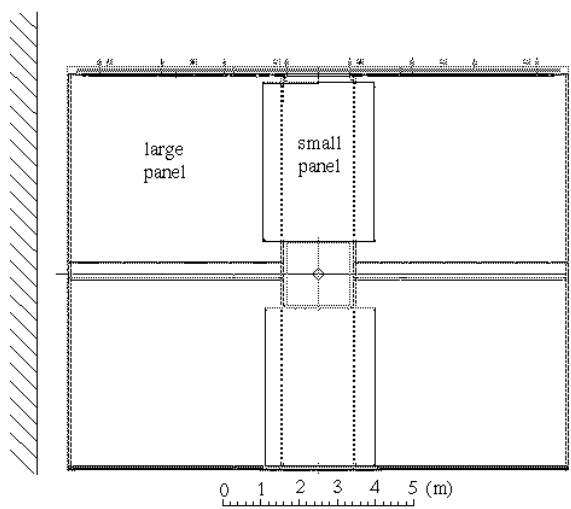}
\end{center}
\caption[\,\,\,Configuration of MuID panels in one gap]{
Configuration of MuID panels in one gap. The beam axis runs
in the middle perpendicularly to the paper.
}
\label{fig:muid_gap}
\end{figure}

\begin{figure}[htbp]
 \begin{center}
  \includegraphics[width=3.5in]{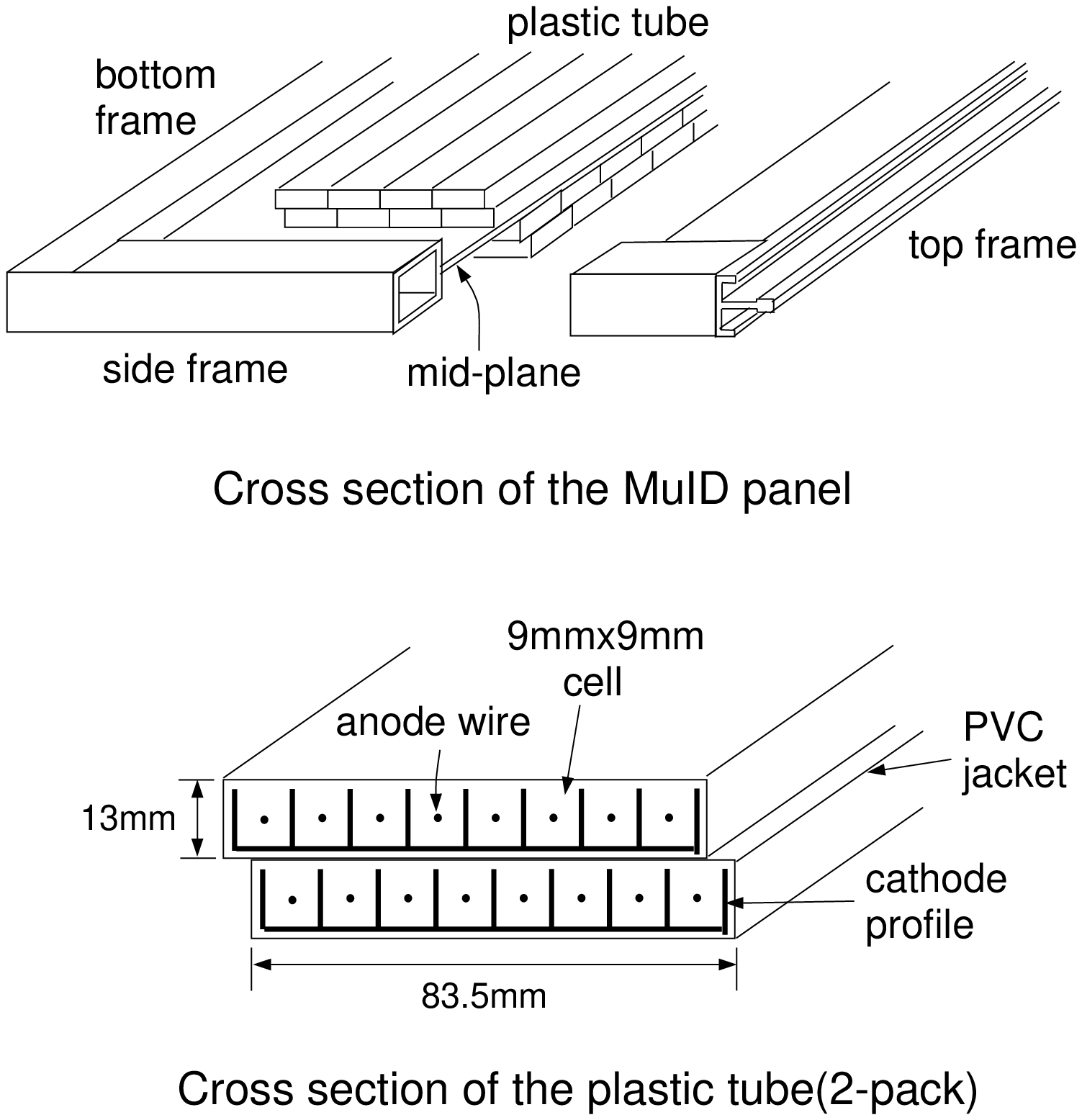}
  \caption[\,\,\,Cross sections of a MuID panel and an Iarocci tube]
  {Cross section of a MuID panel (upper) and an Iarocci tube (lower).
  In one panel-orientation, there are two layers with independent
  gas and high-voltage chains.
  Two adjacent tubes in those layers make one signal channel to be read out.
}
  \label{fig:muid_tube}
 \end{center}
\end{figure}

One MuID plane, or a gap, consists of six panel structures
(called MuID panels) as shown in Fig.~\ref{fig:muid_gap},
into which chambers
are assembled in both horizontal and vertical orientations.
The upper figure of Fig.~\ref{fig:muid_tube} shows
the cross section of a MuID panel.

Iarocci-type plastic tubes have been chosen as MuID chambers for longevity, 
robustness and low cost to cover a large area (13 $\times$ 10 m$^{2}$ 
for each Arm).
One tube has eight cells with a 9 $\times$ 9-mm$^{2}$ cross section
each of which has a gold-coated CuBe anode-wire
with a 100-$\mu$m diameter at its center. 
The lower figure of Fig.~\ref{fig:muid_tube} shows
the cross section of an Iarocci tube
(in this figure, two tubes, or a two-pack is shown which
will be explained later).
The cathode wall
is made of polyvinyl chloride (PVC) coated with carbon.
Length of a tube varies from 2.5 m
to 5.6 m depending on its position to be installed.

Tubes are operated in the proportional mode 
with 4300 to 4500 V potentials and 
isobutane-CO$_{2}$ mixed gas, where
a gain of approximately $2 \times 10^{4}$ is achieved.
As test-bench results, 
92 $\pm$ 1 \% efficiency, which was also
measured in the beam test to be described later,
and 80 nsec (nano-seconds) drift time
width were obtained.  
To achieve a better efficiency and faster drift time,
two tubes with a half cell (5-mm) shift
consist one channel (called a 2-pack) 
as shown in Fig.~\ref{fig:muid_tube}.
For a 2-pack, 97 $\pm$ 1 \% efficiency and 60 nsec (nano-seconds) drift time
have been obtained~\cite{ref:my_master_thesis}.
Drift time is faster enough than the 
bunch-crossing interval (106 nsec), so that a level-1
trigger can uniquely determine the bunch crossing in which
an event has occurred.
Gas mixture ratio is adjustable between 0 to 25\% for isobutane.
In Run-2 it has been set to 7\% to meet the non-flammable
requirement of PHENIX.

Iarocci tubes are assembled in a panel structure
called a MuID panel which consists of Al frames,
cover plates and a mid-plate 
(shown in the upper figure of Fig.~\ref{fig:muid_tube}).
Iarocci tubes are glued 
to both sides of a 3-mm width Al mid-plate
with double-sided tapes at 8.4-cm intervals
for both horizontal and vertical orientations. 
Each orientation has two chamber layers shifted by a half
cell (5 mm) to make 2-packs.
Those layers have independent chains for
gas and high-voltage supplies
to minimize the number of dead channels.

There are two kinds of the panel size which are
5.6 $\times$ 5.2 m$^{2}$ (large panel) and 4.4 (or 4.2) $\times$ 2.9 m$^{2}$
(small panel).
In one gap, four large panels 
and two small panels are installed as shown in
Fig.~\ref{fig:muid_gap}.
Adjacent panels overlap with each other 
so that there is no inactive area
between them.


\subsubsection*{Electronics Design}

A passive OR of signals of 16 wires in a 2-pack is read out 
and amplified by a factor of 150
with an in-panel amplifier on a
high-voltage distribution board which also
provides high voltage to each tube.
Signals are then sent to
the MuID Front-End Electronics (FEE) where
they are again amplified by a factor of 3,
discriminated and stored in the data buffer.
Discriminator threshold values have been set to 90~mV to
minimize the number of noise hits while keeping good efficiency
(typical pulse height of signals is 500~mV to 1~V).
Upon receipt of a level-1 trigger,
all digitized bits are sent to a DCM.
There is additional output of the MuID FEE 
called ``pseudo-trigger output''
which is grand logical OR of a certain fraction of channels.
They are used for the NIM-logic level-1 trigger 
which is described in section~\ref{subsec:muid_trigger}.

\subsubsection*{Integrated performance }
\label{sec:muid_performance}

To confirm hadron-rejection performance and
muon detection efficiency of the entire MuID system experimentally,
a beam test was performed at KEK-PS~\footnote{
Proton Synchrotron at High Energy Accelerator Research Organization, Japan} 
using the same type of Iarocci tubes and steel absorbers
in the same configuration as in PHENIX.

Attached figure~13 shows the experimental setup
for the beam test.
Pion and muon beams were produced by 
bombering an inner target with 12-GeV proton beams, then
1-4 GeV/$c$ momentum was
selected with the magnet system and delivered to the experimental area.
Four scintillation counters (ST1 to ST4) defined the beam. 
Three gaseous $\check{\mbox{C}}$erenkov counters (GC1 to GC3) 
identified pions and muons.
GC1 and GC2 were pressured to distinguish between pions and muons
and GC3 to distinguish between muons and electrons.
Beam qualities obtained were
better than 99\% for muons
and 
better than 99.9 \% for pions 
excluding their weak decays ($\pi^{\pm} \rightarrow \mu^{\pm}\nu$) after GC2.
Five iron slabs were used with a width of
10~cm, 10~cm, 10~cm, 20~cm and 20-cm 
respectively staring from the upper stream.
An additional 10-cm (20-cm)
plate was added in front of the first layer
to simulate the backplate of 
the South (North) Muon Magnet.
The numbers of Iarocci tubes used were 
3, 3, 5, 7 and 9 starting from the first gap 
for each orientation.
They were optimized to 3$\sigma$ dispersion of 
2-GeV/$c$ momentum muons due to multiple scattering.

As a result, muon detection efficiency of
86 $\pm$ 2 \% has been obtained for 1.8-2.5 GeV/$c$ muons
with a small momentum dependence.
This is slightly lower than the test bench
result described before (92 $\pm$ 1 \%), which is explained by
additional inactive volumes between tubes.
Attached figure~14 shows the results of 
pion mis-identification rate 
as a function of pion momentum,
which is consistent with a GEANT~\cite{ref:GEANT} simulation
including weak decays into muons.
For the South Arm, mis-identification rate for
4-GeV/$c$ momentum pions (about 5 GeV/$c$ at the interaction region)
has been determined to be 0.04 excluding decays.
Multiplying the rejection factor of the nosecone and 
Central Magnet ($e^{-5} \sim 7 \times 10^{-3}$),  
the net mis-identification rate of 3 $\times 10^{-4}$ 
is obtained, which satisfies the design value.



\subsubsection{The MuID NIM-logic trigger for LVL-1}
\label{subsec:muid_trigger}

A coarse-segmented trigger system called
the MuID NIM-logic trigger\footnote{
So named because the trigger circuit is constructed using
NIM- and CAMAC-standard electronics.
}
was used to
trigger muons in the South Muon Arm during the Run-2 p+p period.
Attached figure~15 shows a schematic view of the system.


As shown in the figure, all the MuID planes are divided into four 
regions (called quadrants) by horizontal
and vertical lines through the middle.
A trigger decision for a quadrant, that is, whether a muon has passed or not,
is made by taking
the coincidence of fired planes of the quadrant.
Since the number of input channels of the trigger circuit is limited,
only four gaps out of five were used. 
Deeper gaps (gap 3,4 and 5) were included since
they are more important for muon identification than shallower gaps.
Gap 1 was also included since
(1) better tracking performance due to longer tracking length and
(2) better rejection of cosmic rays and
very low angle ($\theta<$ 9 degrees) particles.
The chance for a muon to cross more than one quadrant is small.
Each quadrant is further divided into
two sectors for both horizontal and vertical tube orientations.
Figure~\ref{fig:blt_segment} shows the segmentation for
each gap.
An output of each sector, grand OR of all channels inside,
corresponds to a pseudo-trigger output of a ROC.
in total, 16 pseudo-trigger output signals
are sent to the trigger algorithm.

\begin{figure}[htbp]
\begin{center}

  \begin{picture}(104,80)
    \allinethickness{0.3mm}
    \put(0,0){\line(0,1){80}}
    \put(52,0){\line(0,1){80}}
    \put(104,0){\line(0,1){80}}
    \put(0,0){\line(1,0){104}}
    \put(0,40){\line(1,0){104}}
    \put(0,80){\line(1,0){104}}

    \allinethickness{0.1mm}
    \put(26,0){\line(0,1){80}}
    \put(79,0){\line(0,1){80}}
    \put(0,20){\line(1,0){104}}
    \put(0,60){\line(1,0){104}}

    \multiputlist(13, 10)(26,0){D, B, B, D}
    \multiputlist(13, 30)(26,0){C, A, A, C}
    \multiputlist(13, 50)(26,0){C, A, A, C}
    \multiputlist(13, 70)(26,0){D, B, B, D}

  \end{picture}
  \caption[\,\,\,Segmentation of a MuID plane for the NIM-logic trigger]
{Segmentation of a MuID plane for the NIM-logic trigger. 
Bold lines divide quadrants and thin lines divide segments
in each quadrant. 
Capital letters represent positions of each segment and used in the text.}
  \label{fig:blt_segment}

\end{center}
\end{figure}

Trigger decisions are made by
LeCroy 2372 Memory Lookup Units (MLUs) for each quadrant.
Two MLUs are prepared for each quadrant
for both ``deep'' and ``shallow'' triggers.
A deep trigger requires hits in all four gaps used
and a shallow trigger requires hits up to gap~3.
Hit patterns are required to point
to the event vertex. 
Hit patterns such as
A-A-B-B and A-C-D-D are accepted but
B-B-A-A and D-A-A-C are not, for example,
where four letters represent hit segments in
a quadrant shown in Fig.~\ref{fig:blt_segment}
for each gap starting from the first gap.
An example of an accepted-pattern
is shown in attached figure~16.
To minimize the loss of trigger efficiency
due to finite chamber efficiencies,
the algorithm allows some gaps to miss hits.
For a deep trigger, 6 out of 8 gaps (including
both orientations) are required to have a hit
and 3 out of 4 gaps for a shallow trigger.
Extra hits are allowed,
thus no efficiency loss is expected due to background hits.


Total 8 quadrant trigger signals are sent to
another MLU, called the ``Decision MLU'',
which counts the number of triggered quadrants
and issues five trigger signals:

\begin{tabular}{lrp{7.5cm}} \\
(1) single shallow (1S) & -- & one shallow quadrant is fired, \\
(2) double shallow (2S) & -- & two or more shallow quadrants, \\
(3) single deep (1D) & -- & one deep quadrant, \\
(4) deep-shallow (1D1S) & -- & one deep quadrant and one or more shallow 
quadrants, and \\
(5) double deep (2D) & -- & two or more deep quadrants, \\
\end{tabular}

\vspace{5mm}
\noindent
which are sent to the Global Level-1.
A shallow quadrant-trigger is fired always when the 
deep quadrant-trigger is fired
for that quadrant. Therefore, the following relations hold true:
2D $\subseteq$
1D1S $\subseteq$
1D $\subseteq$
1S
where A $\subseteq$ B denotes that the trigger B is always fired 
when the trigger A is fired.
The reasons to have both shallow and deep 
triggers for each quadrant are
(i) to achieve higher $J/\psi \rightarrow \mu^{+}\mu^{-}$
efficiency obtained with the 1D1S 
trigger than the 2D trigger by 30\% 
and 
(ii) to study hadron punch-through background
with the 1S and 2S triggers (not described in this paper).

\subsection{The PHENIX DAQ system}
\label{sec:daq}

\begin{figure}[htbp]
\begin{center}
\includegraphics[width=5.0in]{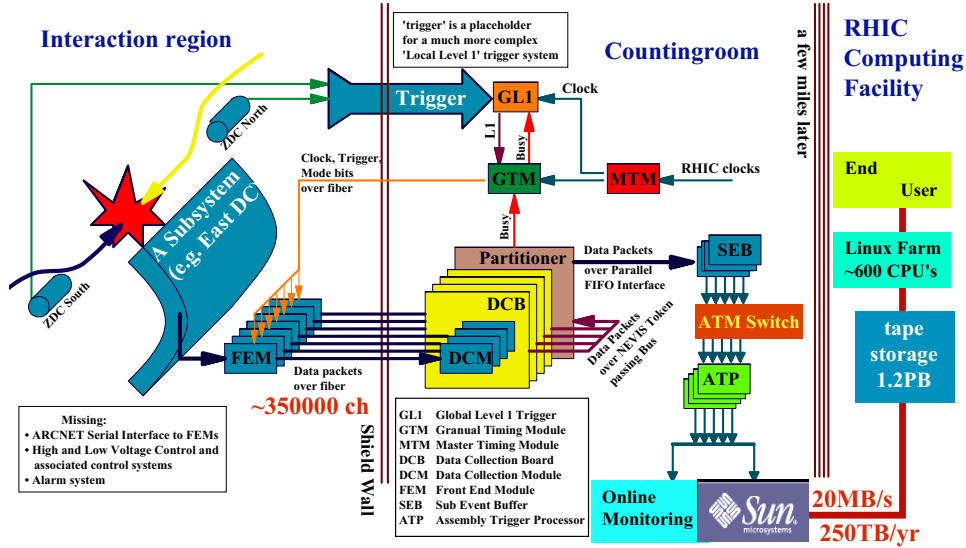}
\end{center}
\caption[The PHENIX data acquisition system]{
The PHENIX data acquisition system, triggered by an event
in the Zero-Degree Calorimeters (Left in diagram)}
\label{fig:phenix_daq}
\end{figure}

Figure~\ref{fig:phenix_daq} shows a schematic view of
the PHENIX Data Acquisition (DAQ) System.
All the subsystems are equipped with the timing modules
called Granule Timing Modules (GTMs).
The \b{M}aster \b{T}iming \b{}Module (MTM) 
delivers the 9.4-MHz RHIC clock to GTMs.
When a trigger is issued by the Global Level-1, 
the clock and trigger bit are transported over
optical fibers to the electronics of each detector,
the Front End Module (FEM). The detector records the data in raw 
digitized format and transports the data packets over optical fibers
to the Data Collection Module (DCM). The data are 
recorded in a buffer disk
in the PHENIX control room where
data quality is monitored online. 
At the maximum 60 mega-bytes per second has been achieved as a rate
of recording data in the disk which corresponds to 
about 1200 events per second
for p+p data whose size is about 50 kilo-bytes per event.
The data are then transferred and stored in a huge storage tape
with 1.2$\times 10^{15}$ bytes capacity 
(sharing with other RHIC experiments) for offline
computing such as calibration and track reconstruction.
In Run-2 PHENIX has recorded about 2$\times 10^{13}$ bytes 
for the p+p data.

\subsection{Global Level-1 triggers \label{section:trigger}}

The level-1 triggers of PHENIX
are issued by the \b{G}lobal \b{L}evel-\b{1} (GL1) module, 
for which local level-1 and NIM-logic triggers 
from various subsystems are input.
The maximum DAQ rate (about 1 kHz) and the number of triggers (10 or more)
limit the rates for each GL1 trigger to about 100 Hz or less.
Level-2 triggers were not used during the p+p Run.

\subsubsection{Minimum bias trigger}
\label{sec:trigger_mb}

Logical OR of the BBC (BBLL1) and NTC triggers,
called the ``\b{M}inimum \b{B}ias (MB)'' trigger was used
to trigger p+p inelastic events with
about 70\% efficiency.
MB-triggered events are useful for studying detector efficiency and background
as well as extracting physics.
At the highest luminosity of RHIC achieved in Run-2
(10$^{30}$~cm$^{-2}$~sec$^{-1}$), MB trigger rate was
typically 
10$^{30}$~cm$^{-2}$~sec$^{-1}$ $\times$ 42~mb $\times$ 0.7 = 30~kHz.
We have applied a prescale factor
to this trigger depending on its rate
to keep the DAQ rate below an affordable level.
Here a prescaled factor $F$ is an integer number and defined as
\[ F = \frac{\mbox{Number of triggers when the DAQ is alive}}
{\mbox{Number of events recorded}} - 1. \] 
During the run, $F$ = 19 to 79 were applied to the MB trigger.

\subsubsection{Muon triggers}
\label{sec:muon_trigger}

Two kinds of muon-related triggers were prepared for physics triggers:

\vspace{1mm}
(A) MB $\otimes$ 1D for the single muon trigger and

\vspace{1mm}
(B) MB $\otimes$ 1D1S for the dimuon trigger,
\vspace{1mm}

\noindent
where $\otimes$ stands for logical AND, ``1D1S'' and ``1D'' have been
already explained in section~\ref{subsec:muid_trigger}.

The rate of the single muon trigger was usually less than 100 Hz and
consistent with that of the irreducible hadron-decay 
background before the nosecone (1/1500 of the minimum bias rate)
within a factor of two.
However, we sometimes observed abnormally high
($\sim$1~kHz) rates.
This was found to be due to blown-up beams
which hit the beam pipe and produced high-energy
secondary particles (including muons) sailing through the MuID.
(This background can be rejected with a vertex cut for offline analyses.) 
We used this trigger
with $F$ = 4 when the trigger rate was too high.
Rates of the dimuon trigger were typically about 1/10 that of
the single muon trigger and never beyond 100 Hz. Therefore no prescale
factor was applied to the dimuon trigger that has been used for
the $J/\psi$ analysis described in the next section.

\begin{figure}[htbp]
\begin{center}
\includegraphics[width=4.5in]{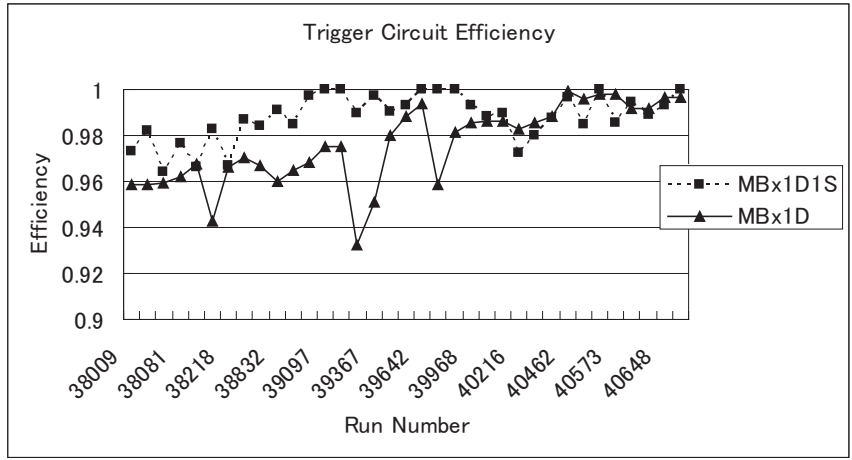}
\caption[Trigger circuit efficiencies]
{Trigger circuit efficiency (defined in the text) 
as a function of the run number
for the single muon (MB $\otimes$ 1D) and 
dimuon (MB $\otimes$ 1D1S) triggers.}
\label{fig:TC_eff_1}
\end{center}
\end{figure}

Figure~\ref{fig:TC_eff_1} shows Trigger Circuit (TC) 
efficiencies ($\varepsilon_{TC}$) for 34 runs randomly
selected from the runs used for the $J/\psi$ analysis.
The efficiencies $\varepsilon_{TC}$ reflect the inefficiency
of the hardware trigger circuit. It is defined as

\[ \varepsilon_{TC} = 
\frac{\mbox{number of events with both hardware and software triggers}}
{\mbox{number of events with software triggers}}
\]

\begin{figure}[htbp]
\begin{center}
\includegraphics[width=4.5in]{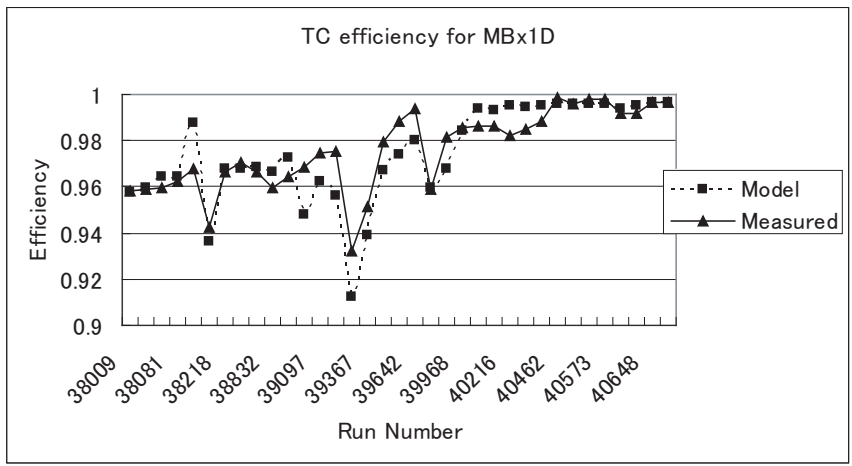}
\caption[Trigger circuit Efficiencies compared with a model calculation]
{Trigger circuit efficiency (defined in the text) 
as a function of the run number
for the single muon trigger,
with that determined by a simple
model calculation.}
\label{fig:TC_eff_model}
\end{center}
\end{figure}

\vspace{2mm}
\noindent
using MB-triggered events, 
where a hardware trigger equals the NIM-logic trigger output 
(GL1 input) and a software trigger is a result of the software 
which emulates the trigger algorithm.
On average, an efficiency of 96.8 \% 
has been obtained for the single muon trigger and 
98.7 \% for the dimuon trigger with small
satistical errors ($<$ 0.1\%).
The inefficiency for the dimuon trigger, 1.3\%, 
is much smaller than other errors 
on the cross sections for $J/\psi$ production,
which will be described in the next section. 
Inefficiencies are ascribed to hardware dead time 
which depends on the trigger rate. 
A simple model calculation, in which 
all efficiency loss is assumed to be due to the dead time,
reproduces $\varepsilon_{TC}$ for the single muon trigger
well, as shown in Fig.~\ref{fig:TC_eff_model}.
This consistency ensures that
both hardware and software triggers
worked as expected.

\subsection{Online Monitoring}

Conditions of the detectors in PHENIX
have been continuously monitored online 
throughout the run
to keep the quality of data.
Monitoring of
high-voltage (HV) status and data qualities 
are described in the following.
In addition, status of such as low-voltage, electronics,
and gas flows was monitored, which was rather stable.

\subsubsection*{High Voltage}

Status of all HV chains was monitored attentively,
which were the most delicate component of a detector.
Both the MuTr and MuID suffered from trip HV chains. 
They were sometimes due to high current caused by
unstable beam condition, but usually due to
spark current caused by humidity or other reasons.
Flowing gas (air or nitrogen)
into the secondary volumes of the detectors
(just outside of the ionization gas volume)
significantly reduced the number of trip chains
for both the detectors.
Trip chains were automatically recovered
by a script monitoring and recording 
the status of all chains in every 10 (MuTr) or 60 (MuID) seconds.
Because of shorter recovery time 
(1 minute) compared to the typical run scale (1 hour)
and small trip frequency ($\ll$ 1 trip per channel per run)
for most of the channels, its effect on
the variation of $J/\psi$ detection efficiency 
is minimal, which is confirmed in 
the next section. 

\subsubsection*{Data quality}

\begin{figure}[htbp]
\begin{center}
\includegraphics[width=4.0in,angle=0]{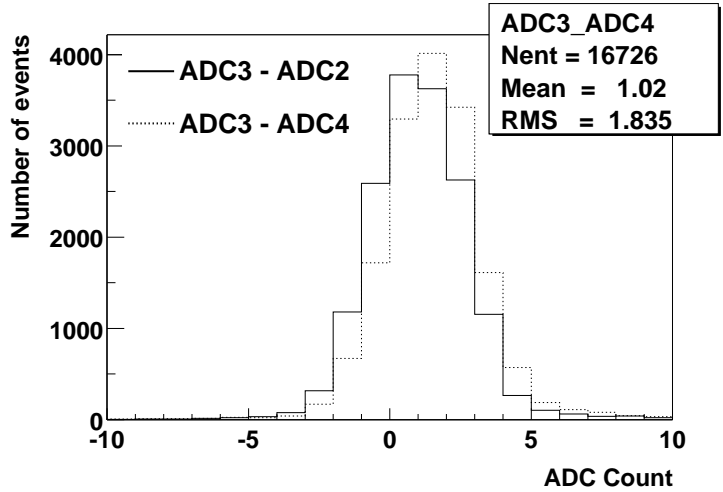}
\end{center}
\caption[Relative ADC count distribution]
{Distributions of relative ADC counts,
(ADC3 $-$ ADC2) and (ADC3 $-$ ADC4),
which were monitored online to confirm their average values
were positive.
}
\label{fig:adc_rel}
\end{figure}

For both the MuTr and MuID, hit occupancies in
each chamber plane were inspected
for dead electronics, dead HV chains, and hot channels.
For the MuTr, distributions of relative ADC counts of samples
were monitored as shown in Fig.~\ref{fig:adc_rel},
to ensure the proper timing of the read-out
(peaks in the third sample).
For the MuID, TC efficiency, described in section~\ref{section:trigger},
was checked to make sure the NIM-logic trigger was working properly.
No significant deviation of these quantities 
from criteria was observed throughout the p+p run.

\subsection{Run and data summary}
\label{sec:run}

Attached figure~17 shows the integrated
luminosity RHIC has delivered to PHENIX in the Run-2
p+p period as a function of day starting from December 21, 2001 
to January 23, 2002.
It has continuously grown since the beginning
of January and finally reached 700~nb$^{-1}$.
Due to finite live-time of the PHENIX DAQ system
and the vertex cut of the BBLL1 algorithm (75~cm), 
about a 150-nb$^{-1}$ luminosity has been recorded with PHENIX.
Table~\ref{tab:number_of_events} summarizes the numbers of events
with the triggers related to muon analyses.


\begin{table}[htbp]
\begin{center}
\begin{tabular}{c|c} \hline
Trigger type & Number of events \\ \hline \hline
Minimum bias & 196 $\times 10^{6}$ \\ \hline
Single muon & 34 $\times 10^{6}$ \\ \hline
Dimuon & 4.8 $\times 10^{6}$ \\ \hline
\end{tabular}
\end{center}
\caption[Numbers of events with the triggers related to muon analyses]{
Numbers of events with the triggers related to muon analyses
obtained in Run-2 p+p collisions.}
\label{tab:number_of_events}
\end{table}

For the $J/\psi$ analysis, the following run 
and event selections have been applied:
(i) runs after RHIC has achieved stable beam 
condition and luminosity (January 8, or the 19th day 
in Fig.~\ref{fig:int_lumi}),
(ii) runs useful for physics analysis, excluding 
dedicated runs for beam and detector response studies as well as
runs with HV or LV off, and
(iii) events with a useful $z$-vertex, which is $|z_{vtx}| <$ 38~cm.

Finally, a 81-nb$^{-1}$ luminosity has been used for the $J/\psi$ analysis.
The detail procedure to obtain it 
as well as $J/\psi$ cross sections will be described in the next section.

\section{Data analysis}
\label{chap:analysis}

In this section, 
the branching fraction for the decay $J/\psi \rightarrow
\mu^{+}\mu^{-}$ ($B_{\mu\mu}$)
times rapidity-differential cross section for
inclusive $J/\psi$ production in the South Muon Arm acceptance
$B_{\mu\mu}d\sigma_{J/\psi}/dy|_{y=1.7}$
is obtained, which is decomposed into
\begin{equation}
\label{eq:xs1}
  B_{\mu\mu}\frac{d\sigma_{J/\psi}}{dy}\bigg|_{y=1.7} =
  \frac{N_{J/\psi}}{\varepsilon_{tot}^{J/\psi} \cdot {\cal L} \cdot \Delta y }.
\end{equation}
$J/\psi$ particles are identified in an
invariant mass spectrum of $\mu^{+}\mu^{-}$ pairs,
where the number of $J/\psi$'s, $N_{J/\psi}$, is counted
with a reasonable background subtraction.
The detection efficiency for the $J/\psi \rightarrow \mu^{+}\mu^{-}$
events, $\varepsilon_{tot}^{J/\psi}$, is determined with a simulation, where
detector response is well tuned to the real data. 
An integrated luminosity, ${\cal L}$, is estimated 
also with a simulation, where consistency with the real data is confirmed.
The rapidity width $\Delta y$ is 1.0 for this case.
Details of analysis procedures to obtain $N_{J/\psi}$,
$\varepsilon_{tot}^{J/\psi}$
and  ${\cal L}$ are described in the following.

The rapidity range is further divided into
two in each of which differential cross section is obtained
($1.2<y<1.7$ and $1.7<y<2.2$) since distribution 
is expected to change significantly in this region.
Similarly double-differential cross section in the invariant form,
$B_{\mu\mu} d^{2}\sigma_{J/\psi} / 2 \pi p_{T} dp_{T} dy |_{y=1.7}$,
and the average value of the transverse momenta of $J/\psi$'s,
$\langle p_{T} \rangle$,
are obtained.

\subsection{Data sample}

The number of events with the triggers related to muon analyses
have been shown in Table~\ref{tab:number_of_events}
in section~\ref{sec:run}.
Dimuon triggered events, which contain most
of the $J/\psi$ yield ($>$ 90\%), have been
used to obtain $N_{J/\psi}$.
Minimum-bias triggered events and single-muon triggered events 
have been used to evaluate detector performance
such as chamber efficiencies.


\subsection{Muon track reconstruction}

Muon tracks are reconstructed using
offline software 
from raw data which contain information
on MuID channel hits, MuTr cathode-strip charges
and BBC $z$-vertices.
This section is devoted to describe
the muon reconstruction algorithm,
which starts from finding a track seed
(a road) in the MuID then grows by attaching clusters
(consecutive cathode strips) at each MuTr station.

\subsubsection{MuID road finding}
\label{sec:road}

First, one-dimensional roads are searched
using either horizontal or vertical tubes.
A road seed is constructed from a hit tube in the seed gap 
(or the first gap in the ``search order'') and
event vertex position measured with the BBC
or other interaction trigger counters.
It is then extrapolated to the next gap in the search order,
where additional hits are searched which are
consistent with the road trajectory.
The hit closest to the projection of the road to the gap
within the search window is generally accepted
and attached to the road. The road is fitted including
the new hit and extrapolated to the next gap.
The size of the search window is
set to 15~cm (about two-tube widths) 
to allow a deviation from a straight-line
trajectory due to multiple scattering in the steel absorbers.
Two search orders,
{2$\rightarrow$1$\rightarrow$3$\rightarrow$4$\rightarrow$5}
and
{3$\rightarrow$2$\rightarrow$1$\rightarrow$4$\rightarrow$5}
where each number $n$ stands for the $n$-th gap of the MuID,
are used to enable roads to be reconstructed even 
the seed gap has inefficient tubes.
Generally shallower gaps come
earlier because of smaller multiple 
scattering, except for the first gap
because of its higher hit occupancy
from hadron and soft-electron background
than the second or third gap.
The algorithm is allowed to skip gaps without hits,
in order to keep higher efficiency despite
low-efficiency tubes in one or two gaps.
Finally both horizontal and vertical roads are
combined and final two-dimensional roads are reconstructed.
Typical road-multiplicity is small ($<$ 0.01)
for minimum-bias events.

Road finding efficiency is expected to be over 99\% for
muons with a momentum $p>$ 3~GeV/$c$ with good chamber 
efficiencies (97\%) and low hit occupancies
as in p+p collisions. 
However, it was actually lower than this since chamber efficiencies 
measured during Run-2 were lower ($\sim$90\%),
which will be described in the later section.

\subsubsection{MuTr cluster finding / fitting}
\label{sec:track}

When a charged particle passes through a MuTr gap
which is composed of two cathode planes and one
anode-wire plane,
charges are usually induced in 
two or three consecutive cathode-strips
in each cathode plane.
The hit position of the particle in each cathode plane 
is reconstructed with the following method.

\begin{figure}[htbp]
\begin{center}
\includegraphics[width=3.0in,angle=0]{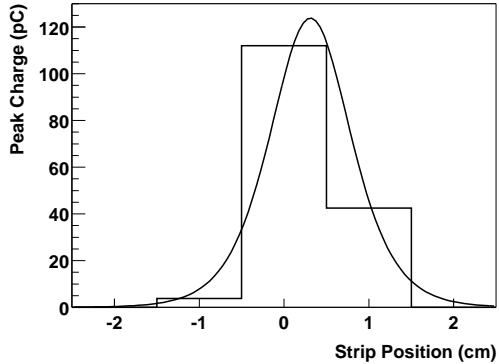}
\end{center}
\caption[An example of peak charge distribution of a cluster]
{An example of peak charge values of sequential cathode-strips 
(a cluster) induced by a charged track, together with a fit of
the Mathieson function. 
In this example, the cluster has three hit strips and its
position is determined to be 3.1-mm off 
the position of the central strip.}
\label{fig:strip_hit}
\end{figure}

The amount of the peak charge in each cathode strip
is determined by four ADC samples
as described in section~\ref{sec:mutr_fee}.
A sequence of consecutive hit strips, called a cluster, is searched and
fitted with an empirical formula (Mathieson function~\cite{ref:Mathieson})
to find the one-dimensional position at which a particle
would have passed in each cathode plane. Figure~\ref{fig:strip_hit} shows
an example of peak charge distribution for a cluster.
The cluster position is obtained as the peak position of 
the fit function.

If relative gain fluctuation and noise level
are 1\% of a typical signal pulse or less,
position resolution of 100~$\mu$m is obtained,
which is confirmed with the cosmic ray test 
described in section~\ref{sec:mutr_performance}.
With this resolution, about 110~MeV/$c^{2}$ is expected as
the mass resolution for a $J/\psi$.
However the actual noise level has turned out to be
worse during the run at station~2, which was typically 3\%.
Degradation in $J/\psi$ mass resolution due to higher noise level
is expected to be about 30\% (140~MeV/$c^{2}$),
which is compared with the real data in section~\ref{sec:dimuon}.

\subsubsection{Muon track finding / fitting}

Starting from a road found in the MuID,
a track grows 
by attaching clusters in the MuTr stations 
from backward to forward, that is,
station~3 to station~1.

Each road
is extended to each cathode plane in station~3, where
clusters are found consistent with the road
inside its search window.
Table~\ref{tab:mask_area} shows widths of the search windows for both the
$\theta$ and $\phi$ directions
at each station in the South MuTr.
The search window in station~3 is large enough
compared to position resolution of roads 
(10 to 20~cm depending on quality cuts on them).
It gets smaller from station~3 to station~1
because track information such as position and momentum
is getting more accurate
in the process of track finding.
By fitting clusters, a local vector called a stub is found. 
A stub requires at least two cathode planes with a cluster in each station.
A crudely estimated momentum is assigned to the track 
from the last penetration gap of the MuID road.

\begin{table}[htbp]
\begin{center}
\begin{tabular}{|c|c|c|} \hline
station & $\theta$ direction (cm) & $\phi$ direction (cm) \\ \hline
1 & 20 & 10 \\
2 & 30 & 25 \\
3 & 50 & 40 \\ \hline
\end{tabular}
\end{center}
\caption[Widths of search windows for each MuTr station]
{Widths of the search windows for the $\theta$ and
$\phi$ directions at each MuTr station.}
\label{tab:mask_area}
\end{table}

The track is then extrapolated to station~2, using
an effective bend-plane and a momentum kick,
which are determined by the approximate momentum
and the magnetic field 
inside the Muon Magnet.
Clusters are searched again in station~2 within the search window.
If a stub is found in station~2, 
the stubs in station~2 and 3 are fitted
altogether to assign more accurate
momentum to the track.
It is again extrapolated to 
station~1 using better momentum information,
where clusters are searched and attached to the track
once more.
The minimum number of hit planes required for a track
is 12 out of all 16 planes.

Finally all the cluster positions and 
the event vertex position are fitted with 
the Kalman-Filter algorithm~\cite{ref:Kalman_Filter},
which is a recursive technique to obtain the 
solution to a least-squares fit,
to determine the momentum vector of the track at the event vertex 
taking energy loss in the absorber (the Central Magnet and copper nosecone)
into account. 

\begin{figure}[htbp]
\begin{center}
\includegraphics[width=4.0in,angle=0]{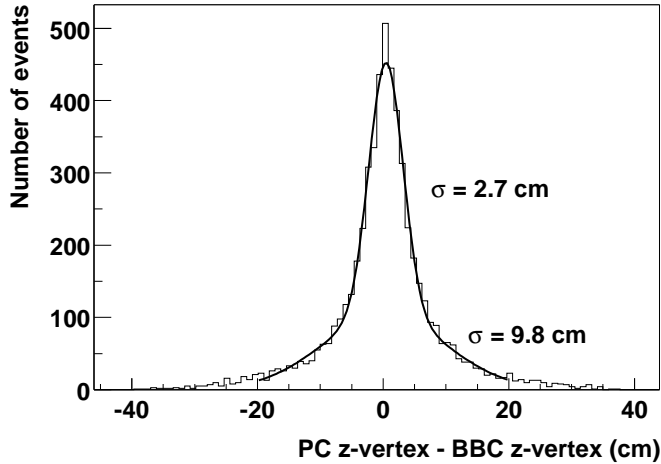}
\end{center}
\caption[BBC $z$-vertex resolution]
{Distribution of the differences between $z$-vertices
found with the BBC and the Pad Chambers (PC) in the
Central Arms.
The distribution is fitted with a double Gaussian function.
The larger Gaussian has a 2.7~cm width, while 
the smaller one has a 9.8~cm width which is supposedly
due to background contribution.}
\label{fig:bbc_resolution}
\end{figure}

Position resolution of the BBC, expected to be 
about 2~cm with a simulation, is confirmed with the real data.
Figure~\ref{fig:bbc_resolution} shows
distribution of the differences between $z$-vertices
found with the BBC and the Pad Chambers (PC) in the
Central Arms~\cite{ref:Jeff_NIM}.
The distribution is fitted with a double Gaussian function.
The larger Gaussian has a 2.7~cm width, while
the smaller one has a 9.8~cm width which is supposedly
due to background contribution.
Resolution of the PC is expected to be also about 
2~cm with a simulation, which is consistent with
the real data with background subtracted.

%
%
%
%
%
%

\subsection{Single muons}

In this section, some properties of reconstructed muons
are shown to demonstrate their qualities and South Muon Arm performance.


  

\subsubsection{Event vertex distribution}

Low-momentum ($p <$ 5~GeV/$c$) single muons,
the majority of the inclusive muon yield,
are expected to be dominated by
charged hadrons decaying weakly into muons ($\pi^{\pm} \rightarrow 
\mu^{\pm} \nu$
and $K^{\pm} \rightarrow \mu^{\pm} \nu$)
in flight before the nosecone.
Contribution of punch-through hadrons 
is small,
which has been confirmed by the beam test
described in section~\ref{sec:muid_performance}.
The decay probability for charged pions 
$P(\pi \rightarrow \mu)$ is given by
\[ P(\pi \rightarrow \mu) = 1 - \exp ( -L B_{r} /\gamma c\tau ) 
  \simeq L B_{r} /\gamma c \tau \]
where $L$ is the distance between the vertex point and nosecone
including one hadronic interaction length of the nosecone (15~cm), 
$B_{r}$ is the branching fraction for the decay
$\pi^{\pm} \rightarrow \mu^{\pm} \nu$,
$\gamma$ 
is the ratio of the particle energy to mass and $c\tau$ is the 
decay length of the particle.
Typical value of $LB_{r}/\gamma c \tau$ is 3$\times 10^{-3}$
for $p \sim 3$~GeV/$c$ muons.
The same kind of formula holds also for charged kaons.
The net contribution of charged kaons to single muon spectra
is expected to be about the same as that of charged pions.

\begin{figure}[htbp]
\begin{center}
\includegraphics[width=4.in,angle=0]{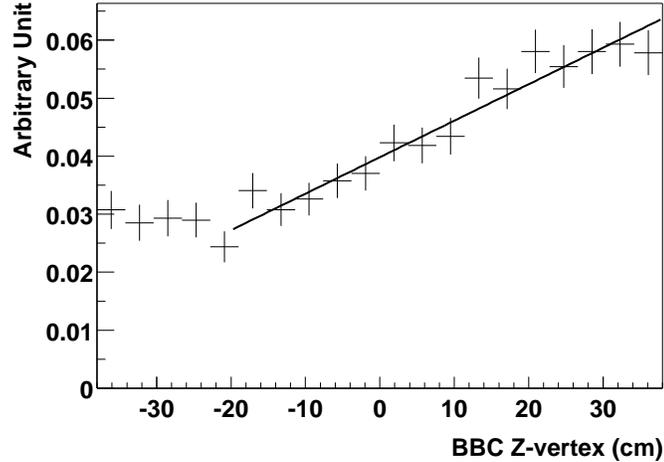}
\end{center}
\caption[BBC $z$-vertex distribution for single-muon events]{
  BBC $z$-vertex distribution for single-muon events divided 
  by that for minimum bias events.
  Error bars include statistical errors only.  
  The solid line shows a linear fit to the data points 
  assuming constant distribution for non-decay components.
}
\label{fig:singlemu_zvertex}
\end{figure}

Figure~\ref{fig:singlemu_zvertex} shows BBC $z$-vertex ($z_{vtx}$) distribution
for single-muon events divided by that for minimum-bias events
in the range $|z_{vtx}| < 38$~cm which is a cut 
used for the $J/\psi$ analysis.
Error bars indicate statistical errors of the particular run.
The solid line shows a straight line fit to the data
assuming flat distribution for non-decay components,
which reproduces the data very well in the range
$-20 < z_{vtx} <$ 38~cm.
The deviation from the fit in the range
$-38 < z_{vtx} < -20$~cm is due to 
the increase in background induced by very low angle particles.
From the fit, the fraction of non-decay
components is determined to be about 20\%, which
supposedly includes background such as punch-through
hadrons and ghost tracks as well as physics signals
such as charm and bottom mesons
decaying semi-leptonically.

The dominance of decay muons to the measured
single-muon yield is consistent with an expectation,
which confirms that the South Muon Arm
has been worked as expected and there is no significant 
contribution of ghost tracks nor punch-through
hadron background to reconstructed muons. 

\subsubsection{Track-road matching}

\begin{figure}[htbp]
\begin{center}
\includegraphics[width=4.in,angle=0]{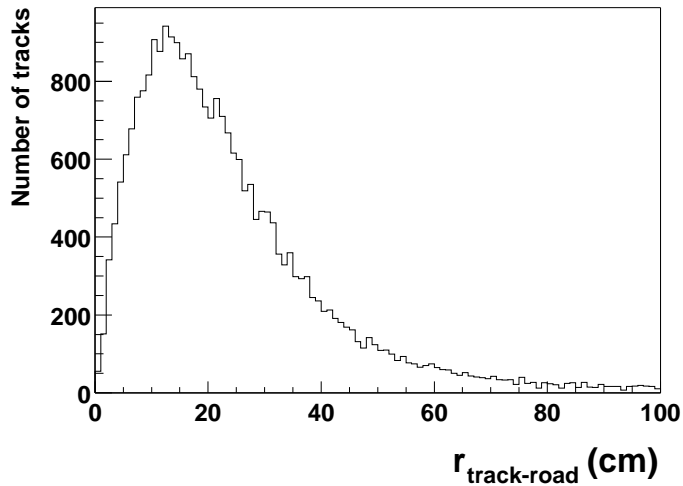}
\end{center}
\caption[Track-road residual distribution]{
  The distribution of the distance between a track and a road 
  ($r_{track-road}$) at station~3.
  The peak position, around 12~cm, roughly represents 
  the rms in the two-dimensional space.
}
\label{fig:match_position}
\end{figure}

Figure~\ref{fig:match_position} shows distribution of the distance
between intersections to a station-3 plane
of a track and a corresponding road,
$r_{track-road}$.
The peak is around 10~cm which is consistent with
the expectation from the position resolution 
of a road (8.4~cm) and multiple scattering
in the absorbers,
thus demonstrating that the two detectors match
well as expected.

\subsection{Dimuon mass and $N_{J/\psi}$}
\label{sec:dimuon}

\begin{figure}[htbp]
\begin{center}
\includegraphics[width=3.5in]{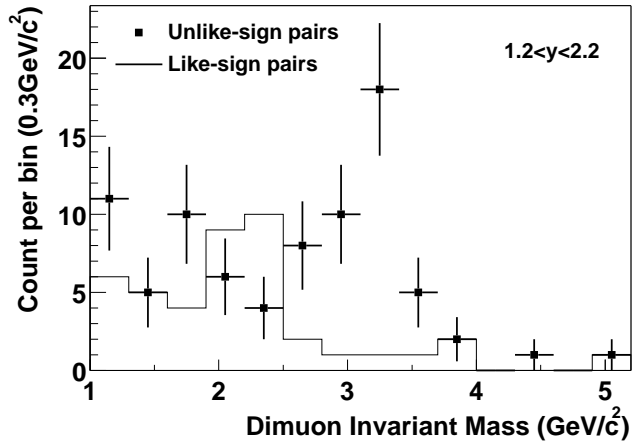}
\caption[Invariant mass spectra for unlike-sign and like-sign muon pairs]
{
Invariant mass spectra for unlike-sign (points with statistical
errors) and like-sign (line) muon pairs. 
Cuts are described in the text.}
\label{fig:jpsi_mass_all}
\end{center}
\end{figure}

Figure~\ref{fig:jpsi_mass_all} shows invariant mass $M_{inv}$
spectra for both unlike-sign and like-sign muon pairs 
with the following cuts:

\begin{itemize}
\item BBC $z$-vertex position $|z_{vtx} \mbox{(BBC)}|<$ 38~cm,
which is between the north and south copper nosecones 
($|z|<$ 40~cm) minus resolution of the BBC (2~cm), and
\item track $z$-vertex consistency $|z_{vtx} \mbox{(Tr1)} 
- z_{vtx} \mbox{(Tr2)}|<$ 30~cm where $z_{vtx} \mbox{(Tr1)}$
and $z_{vtx} \mbox{(Tr2)}$
represent the $z$-vertex position of 
each track obtained as the closest approach 
to the $z$-axis.
This cut corresponds to about 2$\sigma$ of the
position smearing of a track due to multiple scattering 
in the Central Magnet steel (about 10~cm for each track).

\end{itemize}

\noindent
A significant enhancement is found 
in the $J/\psi$ mass region
for unlike-sign pairs
while no such a peak is found for like-sign pairs.

\begin{figure}[htbp]
\begin{center}

  \begin{minipage}{6cm}
  \begin{center}
  \includegraphics[width=6cm]{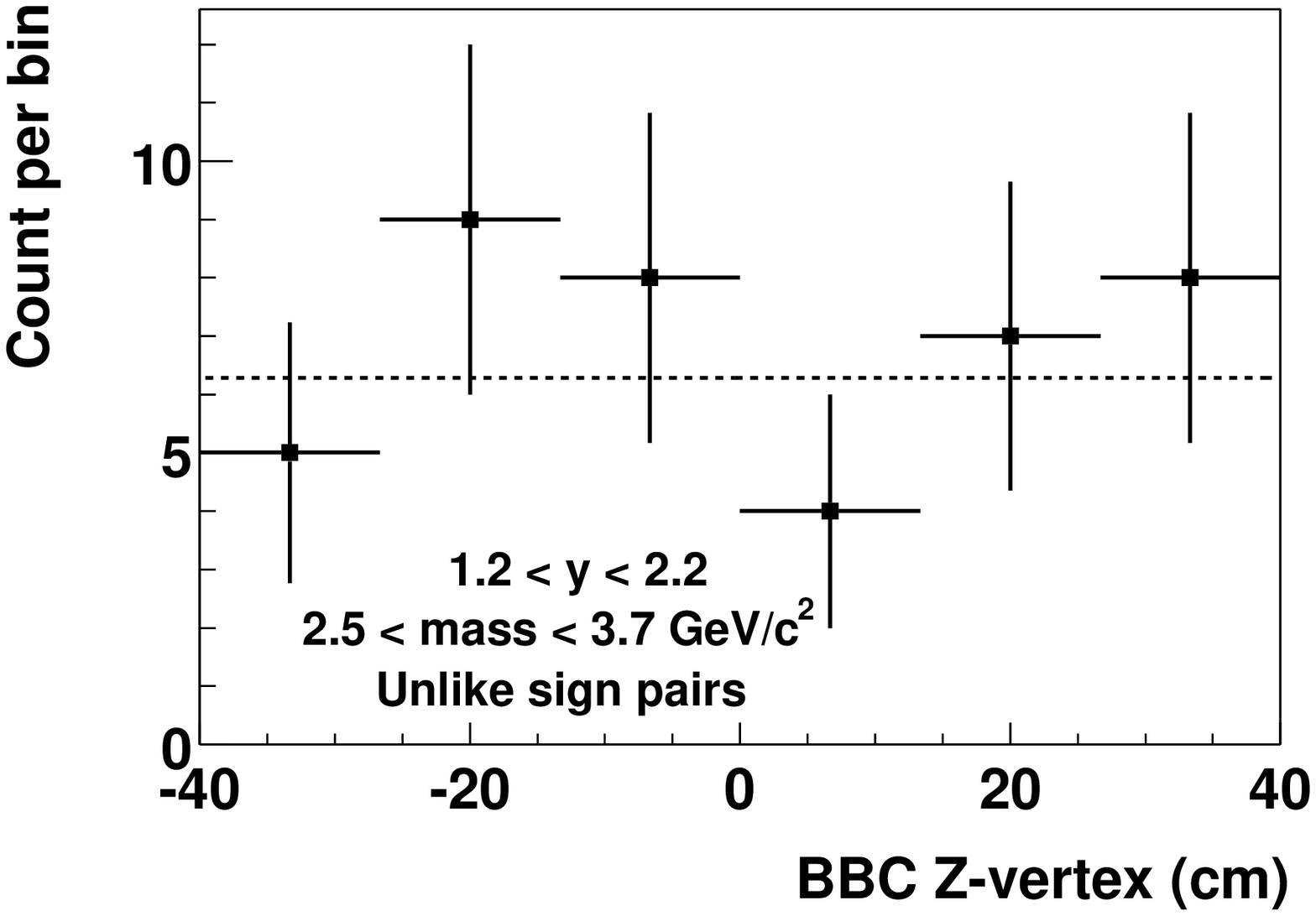}
  \caption{BBC $z$-vertex distribution for unlike-sign muon pairs with
  2.5 $<$ $M_{inv}$ $<$ 3.7 GeV/$c^{2}$.
  The line shows a straight line fit to the data.
  }
  \label{fig:bbc_zvertex_jpsi}
  \end{center}
  \end{minipage}
\hspace{1mm}
  \begin{minipage}{6cm}
  \begin{center}
  \includegraphics[width=6cm]{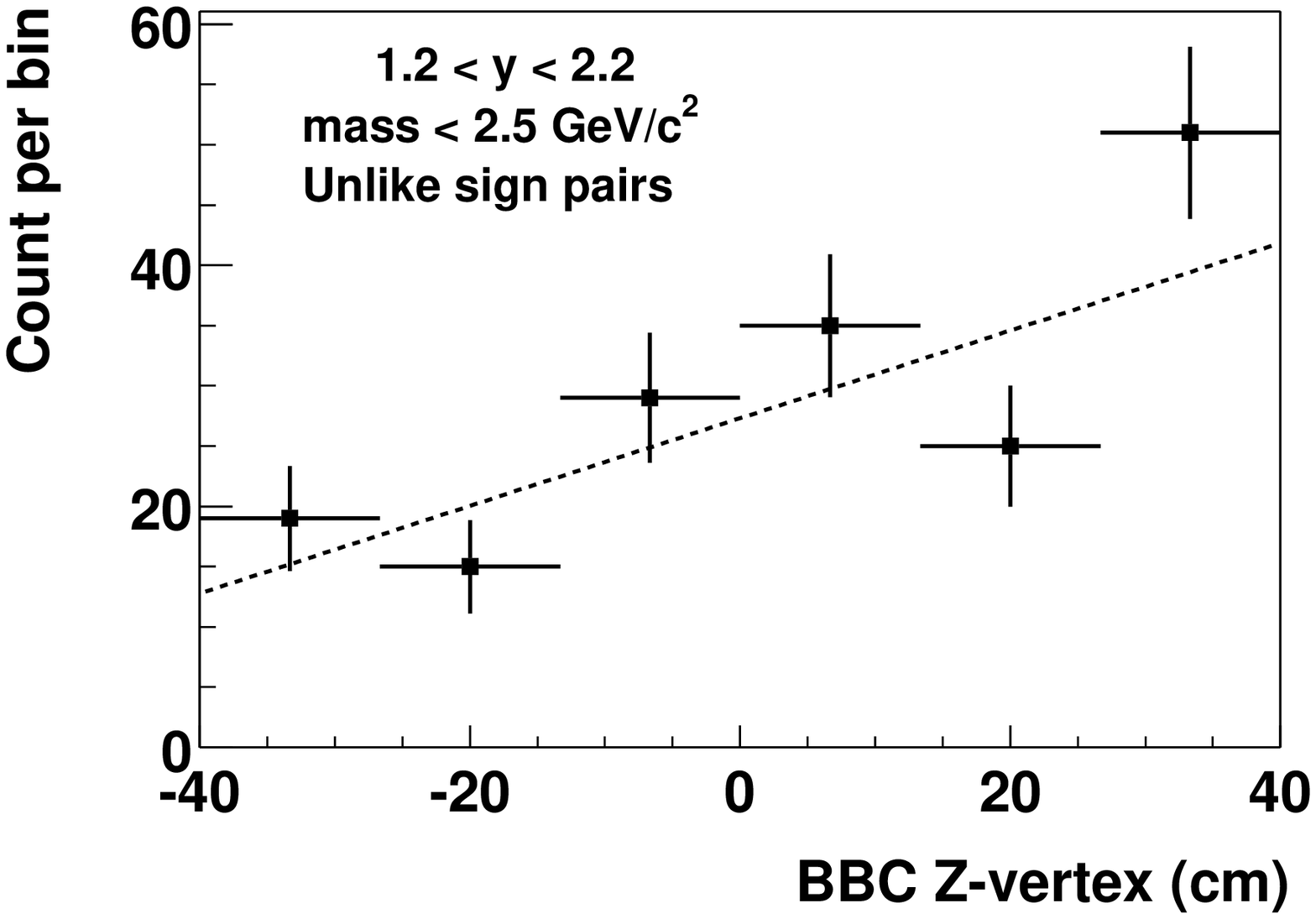}
  \caption{BBC $z$-vertex distribution for unlike-sign muon pairs with
  $M_{inv}$ $<$ 2.5 GeV/$c^{2}$.
  The line shows a straight line fit to the data.
  }
  \label{fig:bbc_zvertex_bg}
  \end{center}
  \end{minipage}
\end{center}
\end{figure}

Figure~\ref{fig:bbc_zvertex_jpsi}
and \ref{fig:bbc_zvertex_bg} show BBC
$z$-vertex distributions for unlike-sign muon pairs
in the $J/\psi$ mass region (2.5~GeV/$c^{2} < M_{inv} <$ 3.7~GeV/$c^{2}$)
and the background region ($M_{inv} <$ 2.5~GeV/$c^{2}$) respectively.
Larger contribution from 
the far-side of the South Arm (positive $z$) 
in the background region 
confirms
that the dominant fraction of background
comes from hadron decays as shown 
also for inclusive single muons (Fig.~\ref{fig:singlemu_zvertex}).
On the other hand, the flat distribution for
the $J/\psi$ candidates is consistent with
the much shorter decay-length of a $J/\psi$ ($c\tau = 2.3 \times
10^{-12}$~m).

From a simulation study using 
the PYTHIA event generator~\cite{ref:pythia}
and GEANT~\cite{ref:GEANT}, 
it is found that the combinatoric
hadron-decay background does not produce a large ($\sim$30\%)
difference between the numbers of unlike-sign
and like-sign muon pairs in the $J/\psi$ mass region
including  
dead electronics and chamber efficiencies during the run.
The consistency of PYTHIA simulation with experimental data
will be examined in section~\ref{sec:bbc_eff}.
The number of $\mu^{+}\mu^{-}$ pairs from 
the Drell-Yan process is estimated to be small ($\sim$0.03 events)
according to a PYTHIA simulation.
Thus we have simply assumed that the number of background for
unlike-sign pairs is the same as the number of like-sign pairs then obtain
\[ N_{J/\psi} = 41\,(\mbox{unlike-sign pairs}) - 5\,(\mbox{like-sign
  pairs}) = 36 \pm 6.8 (\mbox{stat.}).\]

\begin{figure}[htbp]
\begin{center}
\includegraphics[width=3.5in]{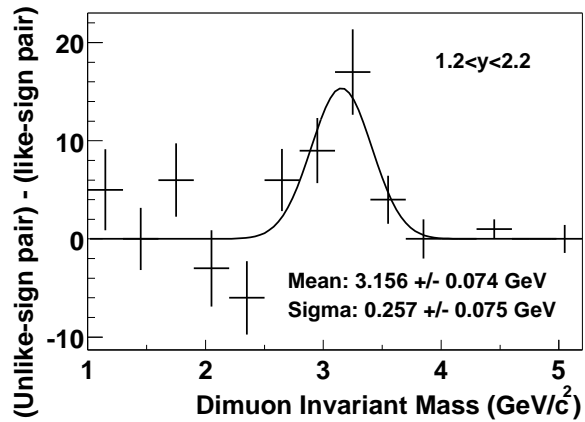}
\caption[The difference between the invariant mass spectrum of
unlike-sign muon pairs and like-sign pairs
with a Gaussian fit]
{The difference between the invariant mass spectrum of
unlike-sign muon pairs and that of like-sign pairs,
which is fit to a Gaussian distribution.
Cuts are described in the text.}
\label{fig:jpsi_mass_fit}
\end{center}
\end{figure}

\begin{figure}[htbp]
\begin{center}
\includegraphics[width=3.5in]{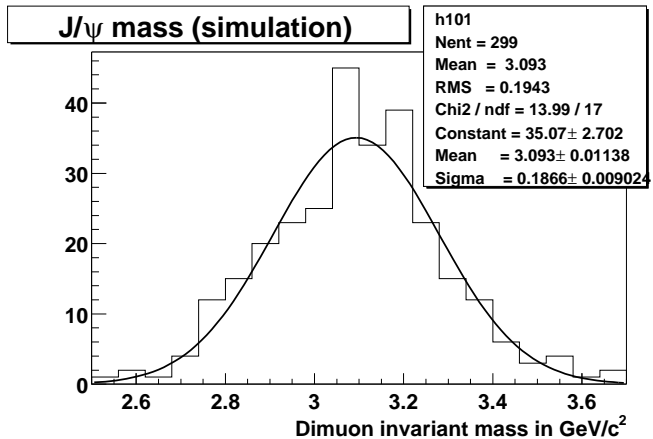}
\caption[
Invariant mass spectrum for unlike-sign muon pairs
from $J/\psi \rightarrow \mu^{+} \mu^{-}$ events obtained with a simulation
]
{
Invariant mass spectrum for unlike-sign muon pairs
from $J/\psi \rightarrow \mu^{+} \mu^{-}$ events obtained with a simulation.
MuTr cathode planes are mis-aligned intentionally by about 300~$\mu$m.
The curve shows a fit with the Gaussian function.
}
\label{fig:jpsi_mass_sim}
\end{center}
\end{figure}

Figure~\ref{fig:jpsi_mass_fit} shows 
the difference
between the invariant mass spectrum of
unlike-sign muon pairs and that of like-sign pairs,
which is fit to a Gaussian distribution.
The center position, 
3156 $\pm$ 74 MeV/$c^2$, is consistent with the $J/\psi$ mass
(3097 MeV/$c^2$~\cite{ref:RPP}).
The width, 257 $\pm$ 75 MeV/$c^{2}$, 
is slightly worse than the expectation with the realistic chamber noise
(3\% at station~2), which is roughly 140~MeV/$c^2$.
This discrepancy is supposedly due to 
the alignment of MuTr chamber positions
done only to 
300~$\mu$m
due to the limited number of straight tracks
with the magnetic field off used for the alignment. 
Mis-alignment effectively worsens the chamber 
resolution since it can differ octant by octant
and has different effects on each charge.
The net effect on the $J/\psi$ mass resolution 
is studied with a simulation and about 190~MeV/$c^{2}$ is obtained
as shown in Fig.~\ref{fig:jpsi_mass_sim},
which is consistent with the real data.

We estimate a systematic error
of 10\% on the background-subtracted yield 
adjusting the mass window by
300~MeV/$c^{2}$ that is roughly the resolution for a $J/\psi$. 
The Gaussian fit gives $N_{J/\psi}$ =
33.0 which is consistent with the simple counting method 
within the systematic error.

\begin{figure}[htbp]
\begin{center}
  \begin{minipage}{6cm}
  \begin{center}
  \includegraphics[width=6cm]{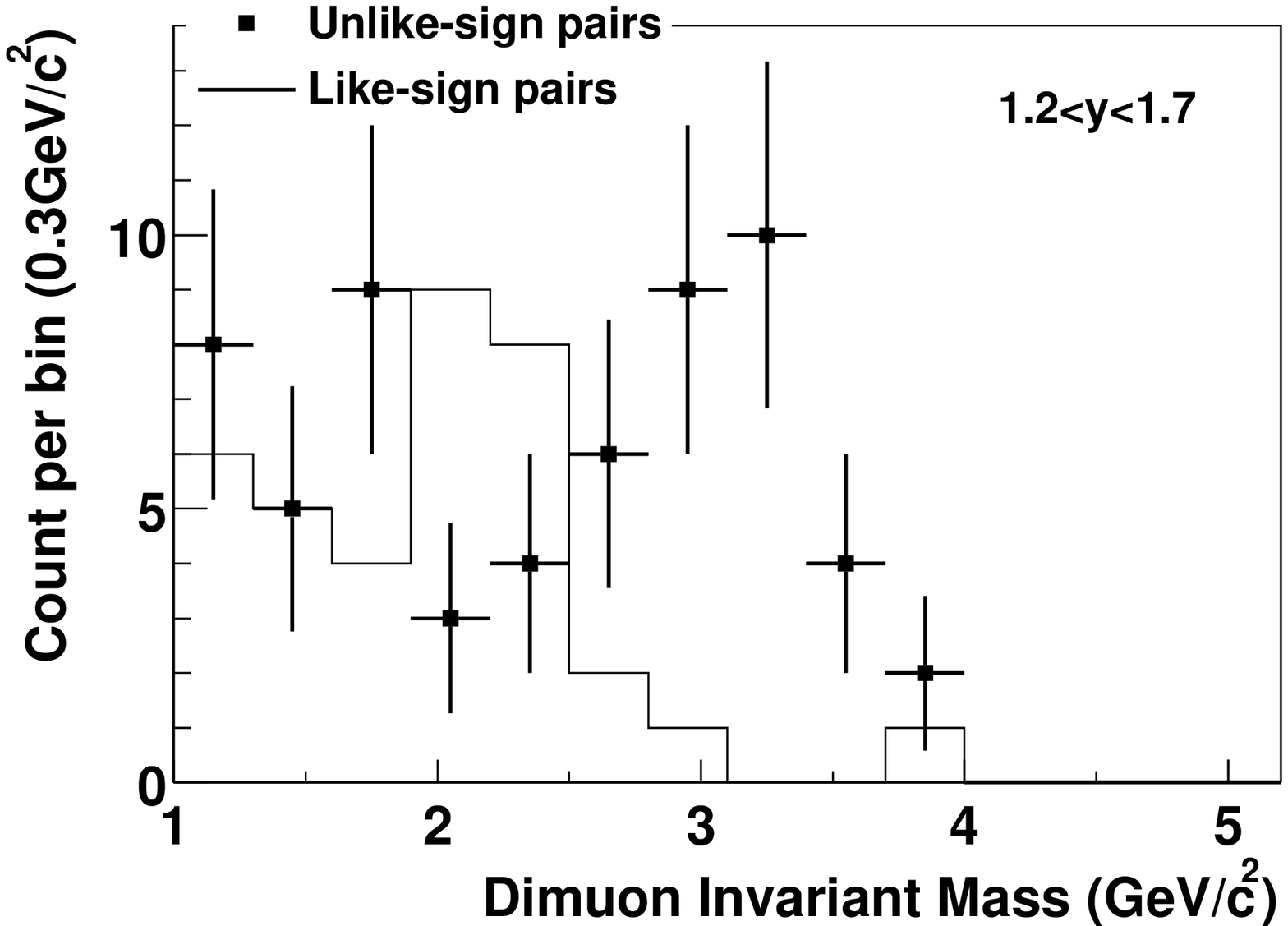}
  \caption[Invariant mass spectra for unlike-sign and like-sign muon
  pairs with 1.2 $<$ y $<$ 1.7]{Invariant mass spectra for unlike-sign
  and like-sign muon pairs in 1.2 $<$ y $<$ 1.7.}
  \label{fig:jpsi_mass_y1.2-1.7}
  \end{center}
  \end{minipage}
\hspace{1mm}
  \begin{minipage}{6cm}
  \begin{center}
  \includegraphics[width=6cm]{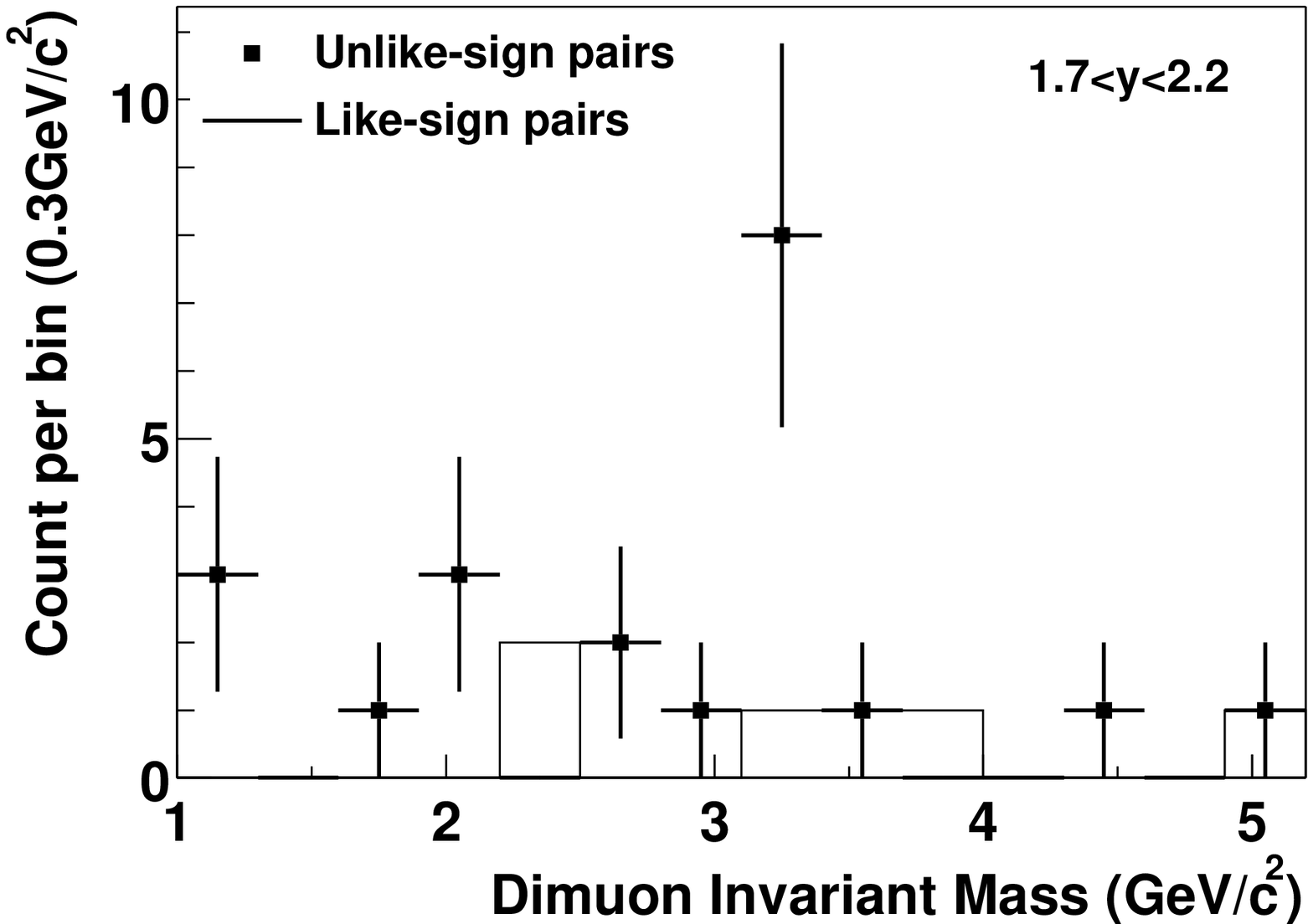}
  \caption[Invariant mass spectra
  for unlike-sign and like-sign muon pairs with
  1.7 $<$ y $<$ 2.2]
  {Invariant mass spectra for unlike-sign and like-sign muon pairs
  in 1.7 $<$ y $<$ 2.2.}
  \label{fig:jpsi_mass_y1.7-2.2}
  \end{center}
  \end{minipage}
 \end{center}
\end{figure}

\begin{figure}[htbp]
\begin{center}
  \begin{minipage}{6cm}
  \begin{center}
  \includegraphics[width=6cm]{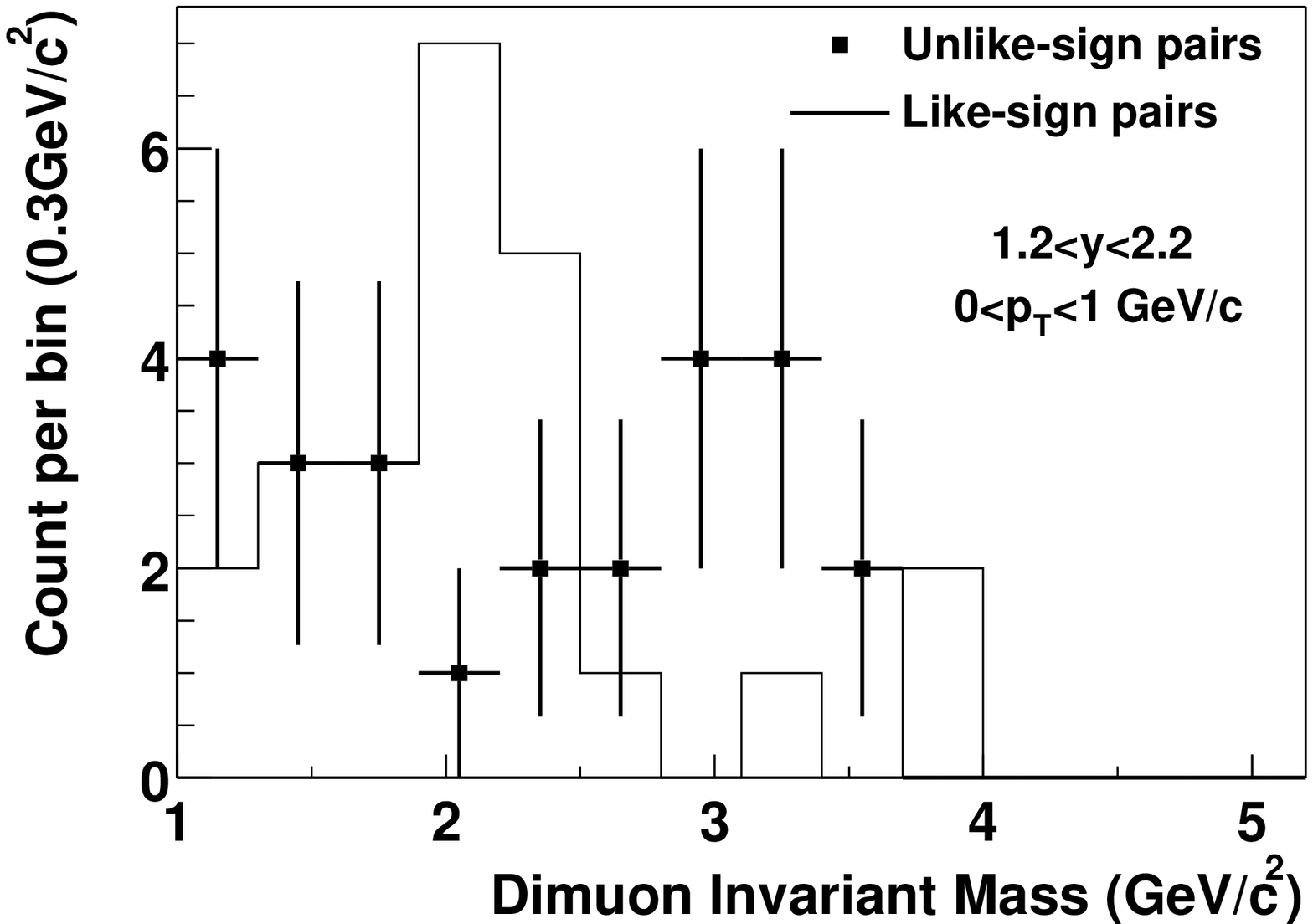}
  \caption[Invariant mass spectra
  for unlike-sign and like-sign muon pairs with
  0 $< p_{T} <$ 1 GeV/$c$]
  {Invariant mass spectra for unlike-sign and like-sign muon pairs
  with 0 $< p_{T} <$ 1 GeV/$c$.}
  \label{fig:jpsi_mass_pt0-1}
  \end{center}
  \end{minipage}
\hspace{1mm}
  \begin{minipage}{6cm}
  \begin{center}
  \includegraphics[width=6cm]{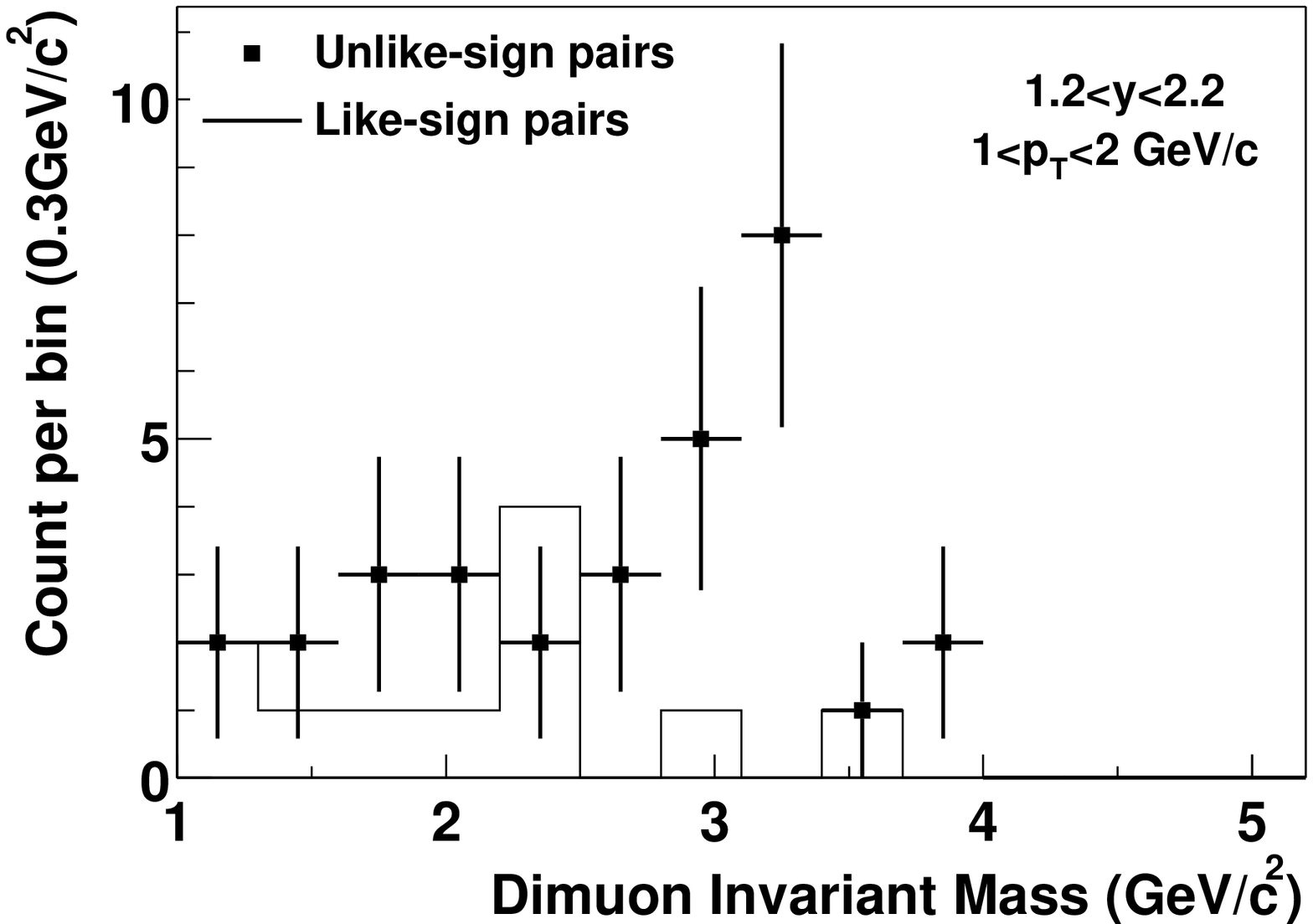}
  \caption[Invariant mass spectra
   for unlike-sign and like-sign muon pairs with
  1 $< p_{T} <$ 2 GeV/$c$]
  {Invariant mass spectra for unlike-sign and like-sign muon pairs
  with 1 $< p_{T} <$ 2 GeV/$c$.}
  \label{fig:jpsi_mass_pt1-2}
  \end{center}
  \end{minipage}
\end{center}
\end{figure}

\begin{figure}[htbp]
\begin{center}
  \begin{minipage}{6cm}
  \begin{center}
  \includegraphics[width=6cm]{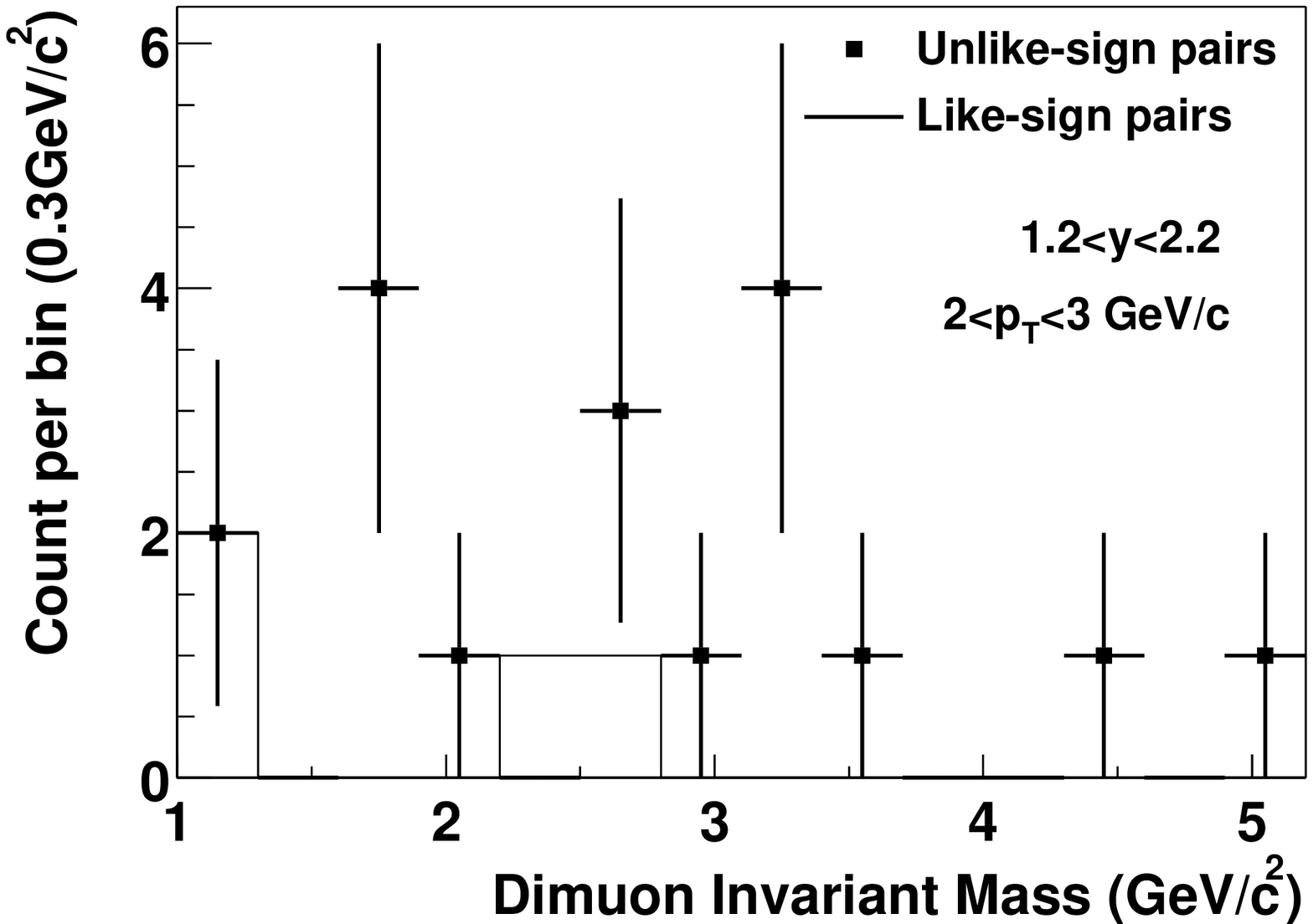}
  \caption[\,\,\,Invariant mass spectra
  for unlike-sign and like-sign muon pairs with
  2 $< p_{T} <$ 3 GeV/$c$]
  {Invariant mass spectra for unlike-sign and like-sign muon pairs
  with 2 $< p_{T} <$ 3 GeV/$c$.}
  \label{fig:jpsi_mass_pt2-3}
  \end{center}
  \end{minipage}
\hspace{1mm}
  \begin{minipage}{6cm}
  \begin{center}
  \includegraphics[width=6cm]{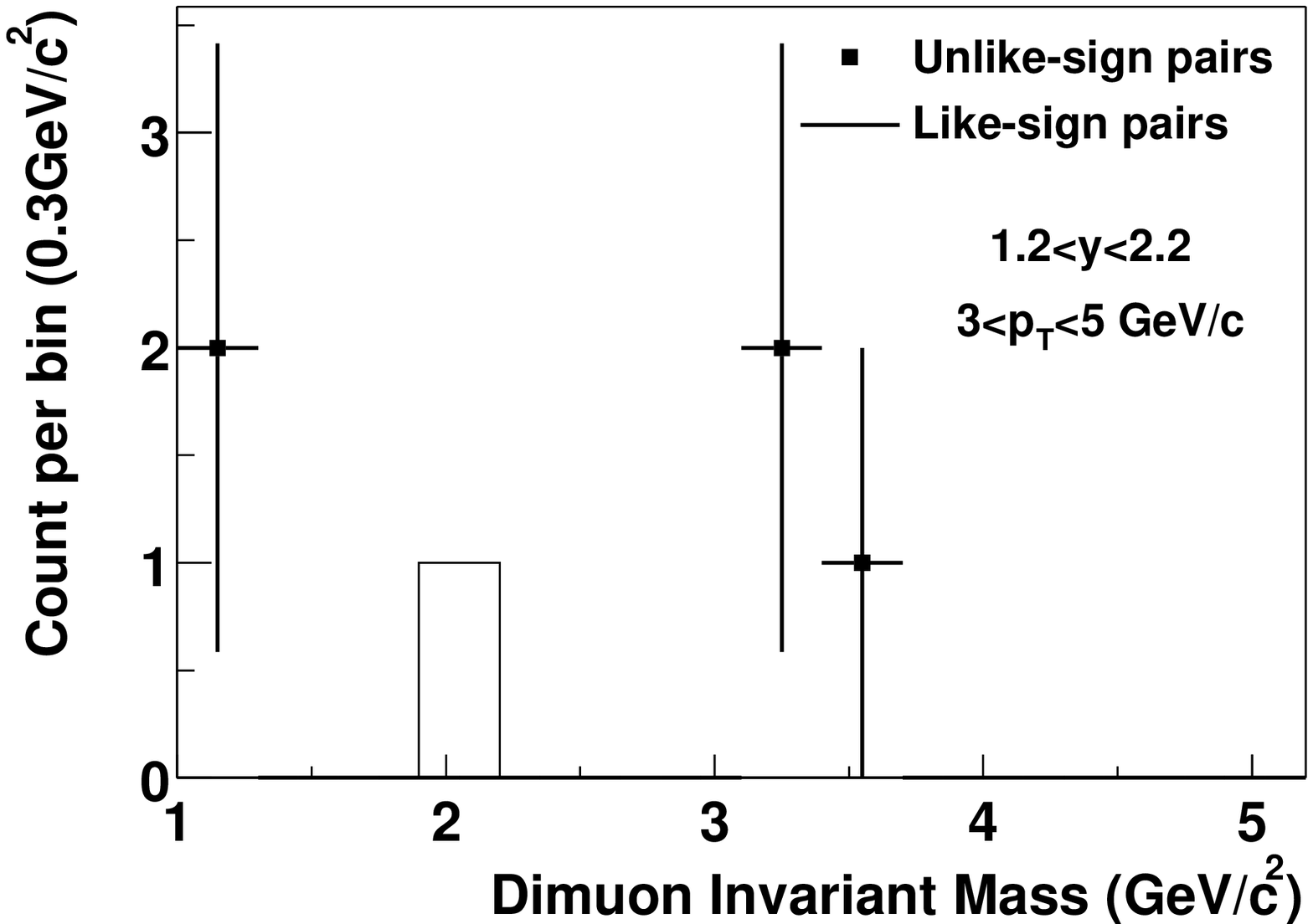}
  \caption[\,\,\,Invariant mass spectra
  for unlike-sign and like-sign muon pairs with
  3 $< p_{T} <$ 5 GeV/$c$]
  {Invariant mass spectra for unlike-sign and like-sign muon pairs
  with 3 $< p_{T} <$ 5 GeV/$c$.}
  \label{fig:jpsi_mass_pt3-5}
  \end{center}
  \end{minipage}
\end{center}
\end{figure}

To obtain $p_{T}$ and rapidity differential cross sections,
invariant mass distributions for
two rapidity bins 
(1.2 $< y <$ 1.7 and 
1.7 $< y <$ 2.2)
and four $p_{T}$ bins 
(0 $< p_{T} <$ 1, 
 1 $< p_{T} <$ 2,
 2 $< p_{T} <$ 3 and
 3 $< p_{T} <$ 5
 in unit of GeV/$c$)
are shown in 
Figs.~\ref{fig:jpsi_mass_y1.2-1.7} through \ref{fig:jpsi_mass_y1.7-2.2}
and 
Figs.~\ref{fig:jpsi_mass_pt0-1} through \ref{fig:jpsi_mass_pt3-5}
respectively.
The numbers of $J/\psi$'s in each bin
are obtained in the same method described above and 
summarized in Tables~\ref{tab:result_summary_rap}
and \ref{tab:result_summary_pt}
in section~\ref{section:result_summary}.
Fluctuations of the background yields in each range
seem to be of statistical nature
and no effect is expected on $N_{J/\psi}$.
For the highest $p_{T}$ bin ($3 < p_{T} <5$ GeV/$c$)
with $N_{J/\psi}$ = 3,
background contamination 
due to such as mis-alignment of MuTr cathode strips
is estimated to be small ($\ll$1 count)
from the counts in the higher mass and higher $p_{T}$
regions as summarized in Table~\ref{tab:high_pt_counts}.

\begin{table}[htbp]
\begin{center} 
\begin{tabular}{|c|c|c|} \hline
       & 2.5$<M_{J/\psi}<$3.7~GeV/$c^{2}$ &
$M_{J/\psi}>$3.7~GeV/$c^{2}$ \\ \hline
3$<p_{T}<$ 5~GeV/$c$ & \b{3}/0 & 0/1 \\ \hline
$p_{T}>$ 5~GeV/$c$ & 0/0 & 2/2 \\ \hline

\end{tabular}
\end{center}
\caption[Dimuon counts in high-$p_{T}$ and high mass regions]
{Dimuon counts in high-$p_{T}$ and high mass regions.
Numbers of both unlike-sign pairs (left-hand side) and like-sign pairs
(right-hand side) are shown in each column.
The underline represents that the number
is for the signal.
}
\label{tab:high_pt_counts}
\end{table}

\subsection{$J/\psi$ detection efficiency}

Total detection efficiency for $J/\psi \rightarrow \mu^{+}\mu^{-}$
events in p+p collisions,
$\varepsilon_{tot}^{J/\psi}$, is decomposed into four factors:
\begin{equation}
\label{eq:eff_decompose}
  \varepsilon_{tot}^{J/\psi} = \eta_{acc}\cdot\varepsilon_{MuID}^{J/\psi}\cdot\varepsilon_{MuTr}^{J/\psi}\cdot
  \varepsilon_{BBC}^{J/\psi}
\end{equation}
where 


\begin{tabular}{lp{11.cm}} \\
$\eta_{acc}$: & South Muon Arm acceptance times reconstruction 
  efficiency for muon pairs from $J/\psi$'s produced in 1.2 $< y <$ 2.2
with 100\% chamber efficiencies, \\
$\varepsilon_{MuID}^{J/\psi}$:&  Efficiency correction due to real
  chamber efficiencies of the MuID, \\
$\varepsilon_{MuTr}^{J/\psi}$: & 
   Efficiency correction due to real chamber efficiencies of the MuTr, and \\
$\varepsilon_{BBC}^{J/\psi}$:  &
Efficiency of BBC for p+p $\rightarrow J/\psi X$ (1.2 $<y^{J/\psi}<$ 2.2)
events. \\
\end{tabular}

\vspace{5mm}

\noindent
Analysis procedures and results of each factor
will be described in the following.

\subsubsection{Detector acceptance}

The geometrical acceptance of the South Muon Arm
is represented as $\eta_{acc}$
including reconstruction
efficiency for muon pairs from $J/\psi$'s 
produced in 1.2 $< y <$ 2.2 with 100\% chamber efficiencies.
In p+p events, multiplicity is sufficiently small to
achieve high ($\sim$90\%) reconstruction efficiency when hits are found
in chambers.
Therefore, $\eta_{acc}$ is close to the geometrical acceptance itself.
Using simulation, $\eta_{acc}$ is calculated as
\[ \eta_{acc} = \frac{\mbox{the number of reconstructed $J/\psi$ events with
1.2 $< y <$ 2.2}}{\mbox{the number of simulated $J/\psi$ events with
1.2 $< y <$ 2.2}}
\]
where PYTHIA has been used to produce $J/\psi$ events,
based on the color-singlet model.
Rapidity and $p_{T}$ distributions (at lower $p_{T}$) of $J/\psi$
do not depend on the production model
unlike a total cross section,
and are consistent with
the real data as discussed in 
section~\ref{section:result_summary}.
The same reconstruction software, parameters and
cuts are used as the real data.
The result of the average value of $\eta_{acc}$
is 0.11 requiring a dimuon trigger.
Rapidity and $p_{T}$ dependence
will be described in section~\ref{sec:total_eff}
including finite chamber and BBC efficiencies ($\varepsilon_{tot}^{J/\psi}$).

\begin{figure}[htbp]
\begin{center}
\includegraphics[width=3.5in]{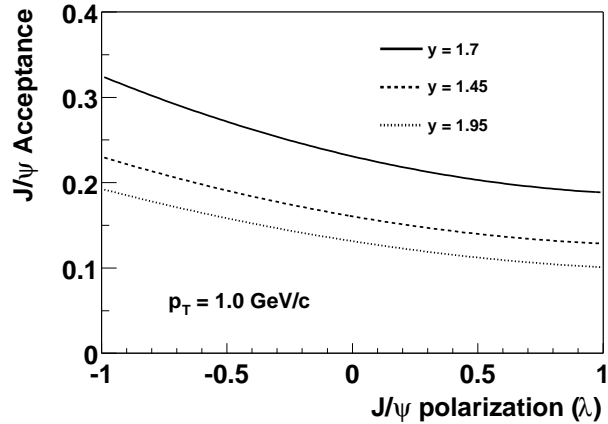}
\caption[$J/\psi$ polarization dependence of the $J/\psi$ acceptance]
{$J/\psi$ polarization dependence of the $J/\psi$ acceptance
at $p_{T}$ = 1 GeV/$c$ and some $y$ (rapidity) points.}
\label{fig:polarization}
\end{center}
\end{figure}

Polarization of $J/\psi$
or spin-alignment (denoted as $\lambda$), 
which is sensitive to its production mechanism, 
is unknown at $\sqrt{s}$ = 200 GeV
and not possible to be determined with the limited number 
of $J/\psi$'s obtained in Run-2.
The South Muon Arm
acceptance has significant $\lambda$ dependence as shown in
Fig.~\ref{fig:polarization}, 
giving a systematic uncertainty of $\eta_{acc}$. 
This dependence is primarily a result
of the $p_{z} < 2$\,GeV/$c$ lower-momentum cutoff which
cuts off the backward-going daughter muons in the $J/\psi$ rest-frame
with respect to the momentum direction of the $J/\psi$ in the laboratory
frame. 
Results of both lower-energy experiments and Tevatron
indicate that $|\lambda|$ is no larger than 0.3 
especially for low-$p_{T}$ ($\sim$ 1 GeV/$c$) 
and low-$x_{F}$ ($\sim$ 0.1) $J/\psi$'s 
\cite{ref:cdf_jpsi_pol, ref:jpsi_pol_fixed_target}
which dominate our yield. 
On the assumption of $|\lambda|<$ 0.3,
we have assigned a 10\% systematic error of $\eta_{acc}$.

\subsubsection{MuID efficiencies}

MuID efficiencies for single-muons and dimuons
are estimated using MuID-panel 
efficiencies obtained from a sample of 37 runs
randomly picked up to cover the entire run-period.

Efficiency of the specific MuID plane $d$ for a muon,
$\varepsilon_{d}$, is given by
\[ \varepsilon_{d} = 
 \frac{\mbox{the number of reconstructed roads with a hit in $d$}}
{\mbox{the number of all reconstructed roads}} \]
for each orientation.
Only minimum-bias triggered events have been used 
for this efficiency study since
MuID-triggered (single-muon and dimuon trigger)
events are possibly biased by the trigger algorithms.
Since we allow ``skipped gaps'' 
in the road finding algorithm described in section~\ref{sec:road},
the plane $d$ is not necessarily excluded from the search order.
This is confirmed by comparing results with
two search orders with and without the plane~$d$.
Since efficiencies 
for planes in the seed gaps\footnote{The first and second
search-order gaps in this case.
}
are not possible to be extracted, 
two sets of search orders are used 
to obtain panel efficiencies for all the gaps:

\begin{itemize}
\item search order set $A$:
{2$\rightarrow$1$\rightarrow$3$\rightarrow$4$\rightarrow$5}
and 
{1$\rightarrow$2$\rightarrow$3$\rightarrow$4$\rightarrow$5}
for gap 3,4,5 and
\vspace{-2mm}
\item search order set $B$: 
{4$\rightarrow$5$\rightarrow$3$\rightarrow$2$\rightarrow$1}
and 
{5$\rightarrow$4$\rightarrow$3$\rightarrow$2$\rightarrow$1}
for gap 1,2,3
\end{itemize}
where two search orders in each set are tried in
the algorithm.
Panel efficiencies in gap~3 obtained
with $A$ and $B$ are compared and found to be consistent
within their statistical uncertainties (2\%), which indicates 
panel efficiencies are not sensitive to the search order.

Tighter cuts in the road finding algorithm,
such as the $r_{z=0}$ 
(the distance between the origin and the intersection of a road 
to the $z$ = 0 plane)
cut, are used
than those for the usual muon reconstruction
to reject ghost roads as much as possible and select only good quality roads.
Furthermore, additional cuts are applied to sample roads:
(i) each road should penetrate through all the five gaps,
(ii) only one road is found in each event,
(iii) each has a hit in the other orientation in the panel of interest and
(iv) each has a hit of either orientation in both adjacent planes.
Cut dependence on panel efficiencies is studied
for example, change
(iv) $\rightarrow$ (iv)':
has a hit of both orientations in both adjacent planes.
The result is consistent with the original one 
to the statistical accuracy of the result (2\%). 

\begin{figure}[htbp]
\begin{center}
\includegraphics[width=3.5in]{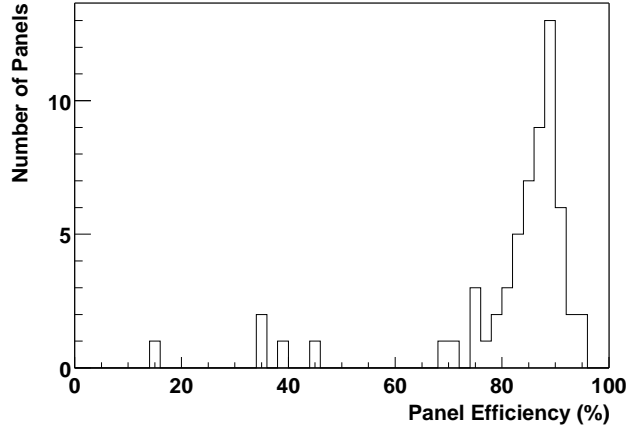}
\caption[MuID panel efficiencies]
{Statistics of MuID panel efficiencies for a single muon.}
\label{fig:panel_eff}
\end{center}
\end{figure}

Figure~\ref{fig:panel_eff} shows distribution of panel efficiencies
for all the 60 panels in the South MuID including both 
orientations.
Most of them gather around 90\% but some are smaller
because HV (high-voltage) potentials applied to
both layers of each panel-orientation
(see Fig.~\ref{fig:muid_tube} in section \ref{sec:muid_mech})
were not high enough ($<$ 4300~V) 
in order to avoid sparks or some problems.
In addition, about one third of all HV chains
had one dead or lower-potential layer,
for which about 90\% efficiency is expected.
However, other two thirds had good potentials on both layers,
where about 98\% efficiency is expected \cite{ref:my_master_thesis}.
The panel efficiencies obtained from the real data
are somewhat smaller than the naive expectation.
A possible explanation is we have applied 
potentials of 4350~V nominally during the run which
may not be high enough since it is 
about the edge of the plateau of the excitation curve
(typically, 4300~V).
For example, an effective potential may drop 
by 100~V with a 100~nA current flow on an anode wire, where
a typical measured-current on one HV chain (about 20 tubes)
was 10 to 100~nA.

Trigger and road-finding efficiencies
have been estimated with simulation including 
the trigger emulator and the MuID chamber 
efficiencies obtained above for
both single muon and dimuon ($J/\psi$) events.
Definition of MuID efficiency for single muons
$\varepsilon_{MuID}^{\mu}$ is
given by 
\[ \varepsilon_{MuID}^{\mu} = 
\frac{\mbox{$N_{\mu}$ with the real MuID efficiencies}}{
\mbox{$N_{\mu}$ with 100\% MuID efficiencies}} \]
where $N_{\mu}$ represents the number of reconstructed muon
tracks through a full simulation.
The similar definition is
applicable for $J/\psi$'s, which is equivalent to $\varepsilon_{MuID}^{J/\psi}$
requiring a dimuon trigger in both the numerator and denominator.
For a single muon with $p_{z}$ = 5~GeV/$c$,
80\% efficiency is obtained
and 62\% for a $J/\psi$ requiring a single or dimuon trigger respectively.
These are consistent with a very crude prediction that 
$\varepsilon_{MuID}^{J/\psi}$ $\sim$ $(\varepsilon_{MuID}^{\mu})^{2}$
since the average momentum from $J/\psi$'s ($p_{z}$ = 5~GeV/$c$)
is used for the single muon simulation.

Without trigger requirements
higher efficiency is expected, since
the offline road-finder is more flexible than the online trigger algorithm,
that is, less susceptible to chamber efficiencies.
For single muons, 86\% efficiency is obtained using minimum-bias
triggered events.
The ratio ($\varepsilon_{MuID}^{\mu}$ requiring a trigger) to
($\varepsilon_{MuID}^{\mu}$ not requiring a trigger) is
obtained, which is 0.93 and consistent with that of the real data (0.92)
with about 2\% statistical errors for each.
This consistency verifies the chamber efficiencies obtained with
the method described above.

\begin{figure}[htbp]
\begin{center}
\includegraphics[width=3.5in]{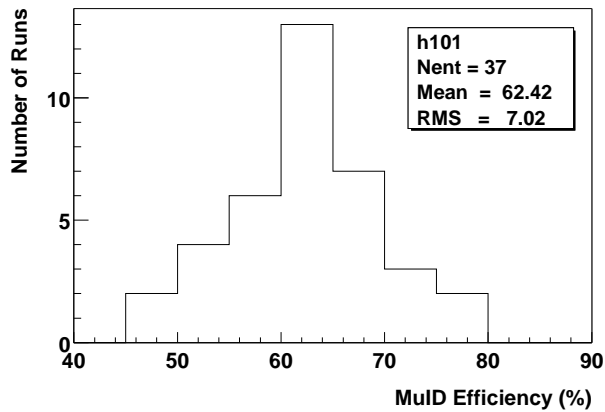}
\caption[Run by run fluctuation of MuID efficiencies for $J/\psi$'s]
{Run by run fluctuation of MuID efficiencies for $J/\psi$'s 
($\varepsilon_{MuID}^{J/\psi}$).
Dimuon triggers are required.}
\label{fig:muid_eff_rundep}
\end{center}
\end{figure}

Run dependence of $\varepsilon_{MuID}^{J/\psi}$
is also studied with a simulation and found that 
its variation is 11\%
which is shown in Fig.~\ref{fig:muid_eff_rundep}.
This fluctuation is roughly consistent with statistical
uncertainties of panel efficiencies which are typically 3\%
for each panel and the minimum requirement for the number
of hit planes which is 9 for a dimuon trigger
(3\% $\times \sqrt{9}$ = 9\%).

\subsubsection*{MuID Hit Occupancy}

\begin{figure}[htbp]
\begin{center}
\includegraphics[width=3.5in,angle=0]{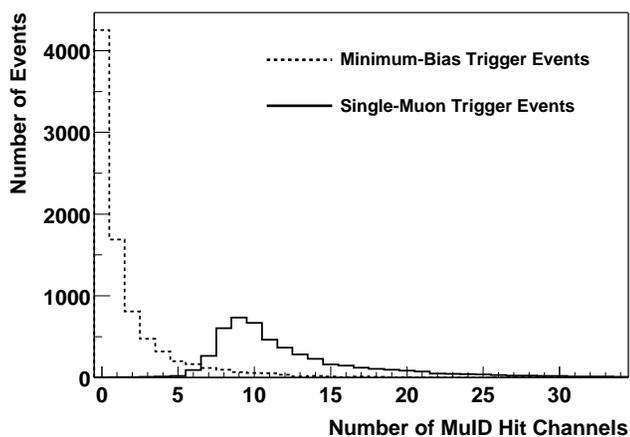}
\end{center}
\caption[MuID hit multiplicity]{
MuID hit multiplicity in minimum-bias triggered events
(dotted line) and single-muon triggered events (solid line).
}
\label{fig:muid_occ}
\end{figure}

In high particle-multiplicity environment such as 
in Au+Au collisions, road finding efficiency may drop
because of difficulty in hit-road associations.
However in p+p collisions, much smaller hit occupancy
ensures high road-finding efficiency. 
Figure~\ref{fig:muid_occ} shows distribution of the 
number of MuID hit channels per event.
Most of the minimum-bias triggered events have zero or small
hit multiplicities. On the contrary, the hit multiplicity
distribution
for the single-muon triggered events has a peak around 9 or 10
which is the expected number of hits when a single muon penetrates
through the entire MuID (5 gaps $\times$ 2 orientations).
This indicates that most of MuID hits are induced by
signal muons and not by background hits such as
from soft electrons.

\begin{figure}[htbp]
\begin{center}
\includegraphics[width=3.5in,angle=0]{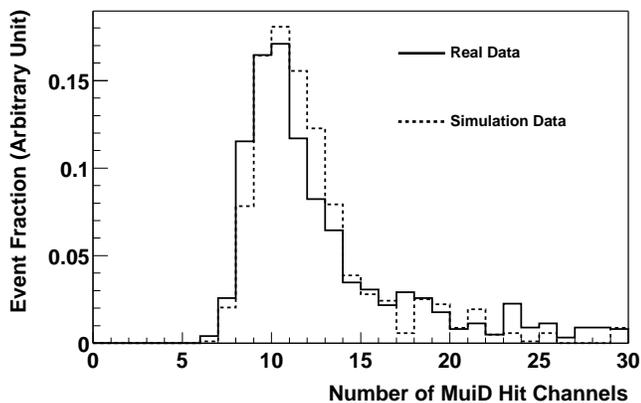}
\end{center}
\caption[MuID hit multiplicity for single-road events]{
MuID hit multiplicity for single-road events.
The solid line shows real data and the dotted line shows simulation data.
}
\label{fig:nhits_muid_singlemu}
\end{figure}

\begin{figure}[htbp]
\begin{center}
\includegraphics[width=3.5in,angle=0]{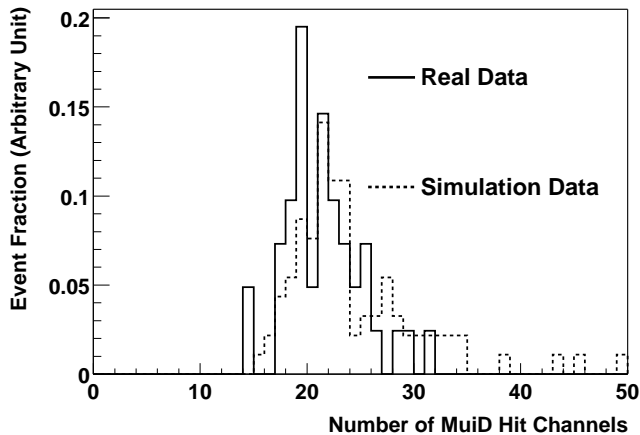}
\end{center}
\caption[MuID hit multiplicity for dimuon events]{
MuID hit multiplicity for $J/\psi$ (candidate) events.
The solid line shows real data and the dotted line shows simulation data.
}
\label{fig:nhits_muid_jpsi}
\end{figure}

Figures~\ref{fig:nhits_muid_singlemu} and \ref{fig:nhits_muid_jpsi}
also show MuID hit multiplicities in single-road events and 
$J/\psi$ candidate events 
(unlike-sign dimuons in the $J/\psi$ mass region)
respectively, both of which are compared with simulations.
For both cases, hit multiplicities agree with simulations
and again confirm 
hit occupancies in the real data are small as expected.
Provided valid particle multiplicity 
of the event generator (PYTHIA)
which is confirmed in the next section, this consistency
assures response and performance of the detectors 
are as expected;
thus, no efficiency degradation is suspected.

\subsubsection{MuTr efficiencies}

Efficiencies of each MuTr plane for a muon
are known to be high (99\%)
inside active detector volumes
according to a cosmic ray test.
Inactive volumes between octants, 
which are their frames, are  
included in the detector acceptance (about 70\% for a single track).
Some dead Front-End Modules (FEMs)
and high-voltage supply chains during the run
caused inefficiencies primarily. 

\begin{figure}[htbp]
\begin{center}
\includegraphics[width=3.5in]{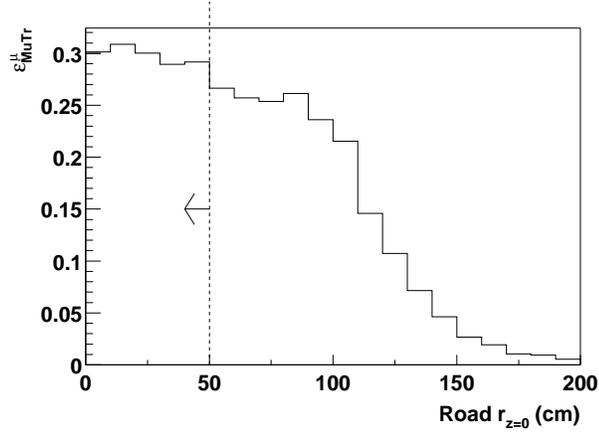}
\caption[MuTr efficiency $r_{z=0}$ dependence]
{MuTr efficiency 
$\varepsilon_{MuTr}^{\mu}$ as a function of $r_{z=0}$ of roads.
See the text for the definitions of $\varepsilon_{MuTr}^{\mu}$
and $r_{z=0}$.
The dotted line and the arrow represent the quality cut to select
good roads used for the MuTr efficiency calculation.
}
\label{fig:mutr_eff_refposition}
\end{center}
\end{figure}

\begin{figure}[htbp]
\begin{center}
\includegraphics[width=3.5in]{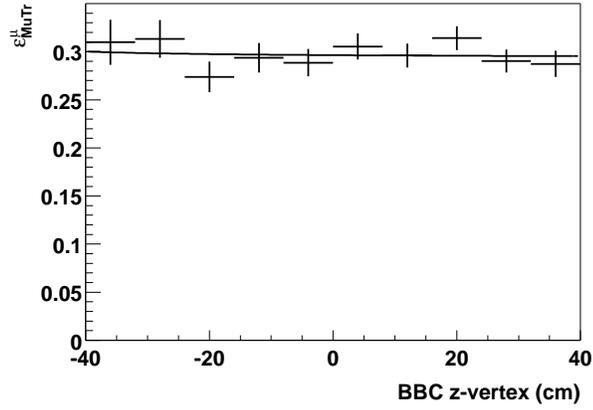}
\caption[MuTr efficiency $z$-vertex dependence]
{MuTr efficiency $\varepsilon_{MuTr}^{\mu}$ as a function of 
the BBC $z$-vertex.
See the text for the definition of $\varepsilon_{MuTr}^{\mu}$.
The error bars are statistical errors for one particular run.
The line shows a fit to the data with a function assuming 
a constant fraction of ghost roads, which is determined to be
less than 5\%.
}
\label{fig:mutr_eff_zvertex}
\end{center}
\end{figure}

A sample run was selected to model
MuTr dead FEMs, dead electronics channels and dead high-voltage chains. 
To examine consistency of the model with the real data, MuTr efficiencies
for a single muon $\varepsilon_{MuTr}^{\mu}$ defined as 
\[ \varepsilon_{MuTr}^{\mu} = \frac{\mbox{the number of roads with a track}}
{\mbox{the number of roads}} \]
are obtained using both the real data and simulation to be compared.
Tighter cuts are applied to the sample roads to reject as many
ghosts as possible.
Figure~\ref{fig:mutr_eff_refposition} shows $\varepsilon_{MuTr}^{\mu}$
as a function of $r_{z=0}$,
which is the distance between the origin and 
the intersection of a road to the $z=0$ plane.
The decrease in efficiency at large $r_{z=0}$
is due to contamination of background roads presumably from
beam scraping that has caused high trigger rates as described 
in section~\ref{sec:muon_trigger}.
The cut $r_{z=0} < 50$~cm is applied to eliminate them.
Other road quality parameters 
are checked in the same way and confirmed to
be valid.
About 80\% of the road sample is dominated by hadron-decays 
as demonstrated by the BBC $z$-vertex distribution as already shown in
Fig.~\ref{fig:singlemu_zvertex}. A fraction of ghost roads
is further constrained by $z$-vertex dependence of
$\varepsilon_{MuTr}^{\mu}$, because efficiency 
is sensitive to the fraction of non-ghost roads with $z$-vertex dependence,
which is dominanted by decay muons. 
Actually,
there is no significant dependence on BBC $z$-vertex
as shown in Fig.~\ref{fig:mutr_eff_zvertex}.
The fraction of the ghost roads to the road sample is 
constrained by fitting it assuming a flat $z$-vertex distribution
for them and estimated to be less than 5\%. 


\begin{figure}[htbp]
\begin{center}
\includegraphics[width=3.5in]{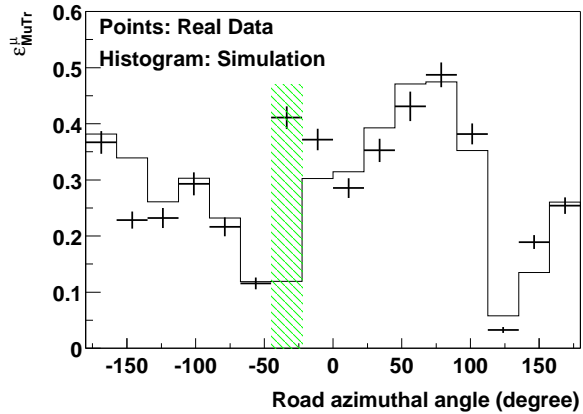}
\caption[MuTr efficiency $\phi$ dependence]
{MuTr efficiency $\varepsilon_{MuTr}^{\mu}$
as a function of the azimuthal angle of roads.
See the text for the definition of $\varepsilon_{MuTr}^{\mu}$.
The points with a statistical error are for the real data 
and the histogram is for the simulation data.
Statistical errors of the simulation data 
are omitted, which are comparable with those of the real data.
}
\label{fig:mutr_eff_phi}
\end{center}
\end{figure}

Figure~\ref{fig:mutr_eff_phi} shows MuTr efficiencies
for both the real and simulation data as a function of
the azimuthal angle of the qualified roads.
There is a reasonably good agreement of the real data and
simulation with exception of the region in
$\phi = -45^{\circ}$ to $-22.5^{\circ}$ (octant 8, half-octant 1 or 8-1).
This discrepancy is turned out to be due to 
the improper treatment of the simulation data 
in 8-1, which has reduced its acceptance significantly (25\%).
Since an opening angle between two muons
from a $J/\psi$ decay is large (30 $\sim$ 60 degrees),
the probability for either muon to fall into
the half-octant 8-1 is roughly
(1/16) $\times$ 2 = 1/8 where 16 is the number of all
half-octants.
In that case, efficiency is
25\% of the ordinary case.
Therefore, the overall efficiency for a $J/\psi$,
$\varepsilon_{MuTr}^{J/\psi}$ 
is estimated to be
\[ \left( 1-\frac{1}{8} \right)
  + \frac{1}{8} \times 0.25  
  \sim 0.90 \]
compared to the normal case.
Therefore we assign a systematic error of 10\% to 
$\varepsilon_{MuTr}^{J/\psi}$ of this particular run based on the
disagreement between the simulation and real data. 
The $\phi$-averaged values of $\varepsilon_{MuTr}^{\mu}$ obtained are
(29.6 $\pm$ 0.3) \% for the real data and (28.2 $\pm$ 0.3) \% 
for simulation with statistical errors, 
whose difference is consistent with 
the systematic error of $\varepsilon_{MuTr}^{J/\psi}$ 
obtained above assuming a crude relation
$\varepsilon_{MuTr}^{J/\psi}$ $\sim$
$(\varepsilon_{MuTr}^{\mu})^{2}$.



\begin{figure}[htbp]
\begin{center}
\includegraphics[width=3.5in]{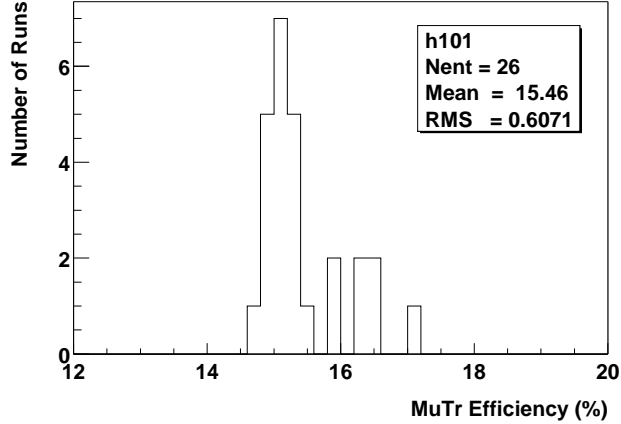}
\caption[\,\,\,MuTr efficiency run statistics]
{Run statistics of MuTr efficiency 
for $J/\psi$'s ($\varepsilon_{MuTr}^{J/\psi}$)
for 26 sample runs.}
\label{fig:mutr_eff_rundep}
\end{center}
\end{figure}

To minimize an uncertainty due to run dependence
of MuTr efficiency, only runs with a high duty fraction 
are selected and used for the analysis.
Here we define the duty fraction $DF$ for each run as
\[ DF = \frac{\mbox{the number of chains with high-voltage on}}
{\mbox{the number of all chains}} \]
each integrated over all measurements 
in every 10 seconds during the run.
We have selected runs with $DF >$ 0.89.
To estimate run dependence of MuTr efficiencies,
26 sample runs with $DF >$ 0.89 
and relatively longer run time
are selected to see their variations.
Figure~\ref{fig:mutr_eff_rundep} shows distribution of
$\varepsilon_{MuTr}^{J/\psi}$ for those runs and variation
is found to be 4\%.
It should be noted that $\varepsilon_{MuTr}^{J/\psi}$ is slightly higher than
$(\varepsilon_{MuTr}^{\mu})^{2} \sim 9\%$ since 
the average momentum of muons from $J/\psi$ decays (about 5~GeV/$c$)
is higher than that of muons used for 
the calculation of $\varepsilon_{MuTr}^{\mu}$ (about 3~GeV/$c$)
with lower efficiency.

\begin{figure}[htbp]
\begin{center}
\includegraphics[width=3.5in]{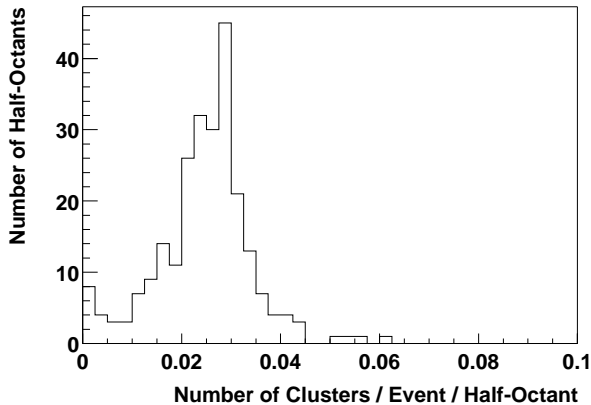}
\caption[\,\,\,Cluster occupancy distribution]
{Distribution of the average numbers of MuTr clusters 
 in each half octant for minimum-bias events.}
\label{fig:clus_occ}
\end{center}
\end{figure}

Cluster hit occupancies in the real data have been examined.
Figure~\ref{fig:clus_occ} shows distribution
of the average number of clusters
in each half-octant for minimum bias events. 
They are typically 0.02 or 0.03 and sufficiently small
not to worsen reconstruction efficiency for muon tracks.

\subsubsection{BBC efficiency for p+p $\rightarrow J/\psi X$ events}

Efficiency of the BBC for p+p $\rightarrow J/\psi X$ events
($\varepsilon_{BBC}^{J/\psi}$),
where a $J/\psi$ decays into a muon pair detected in 
the South Muon Arm,
is estimated to be 0.74 $\pm$ 0.01 (stat.) from simulation studies using
PYTHIA and GEANT.
No significant $p_{T}$ nor rapidity dependence has been found.
This factor is closely related to 
$\varepsilon_{BBC}^{inela}$ which is 
BBC efficiency for p+p inelastic events and
will be described in section~\ref{sec:bbc_eff}.
The systematic error of $\varepsilon_{BBC}^{J/\psi}$
will also be discussed there.

\subsubsection{Total detection efficiency}
\label{sec:total_eff}

Total detection efficiency $\varepsilon_{tot}^{J/\psi}$ is given by
\[ \varepsilon_{tot}^{J/\psi} = 
\frac{\mbox{the number of reconstructed $J/\psi$'s in 1.2 $< y <$ 2.2}} 
{\mbox{the number of simulated $J/\psi$'s in 1.2 $< y <$ 2.2}} \]
with a simulation
including the MuID, MuTr and BBC efficiencies obtained
in the previous subsections.
Run-averaged values are used for the MuID part.
For the MuTr part, 
dead FEMs, high-voltage chains and real ADC gains with the particular
run resulting 15.3\% efficiency (see Fig.~\ref{fig:mutr_eff_rundep})
are used. 
About 88 000 PYTHIA events with unpolarized $J/\psi$'s
are simulated and $\varepsilon_{tot}^{J/\psi}$ is obtained which is 
(1.19 $\pm$ 0.03)\%.
where the error stands for the statistical error 
of the simulation.
PYTHIA's $p_{T}$ and rapidity distributions are
consistent with the real data as shown in the next section.

\begin{figure}[htbp]
\begin{center}
\includegraphics[width=3.5in,angle=0]{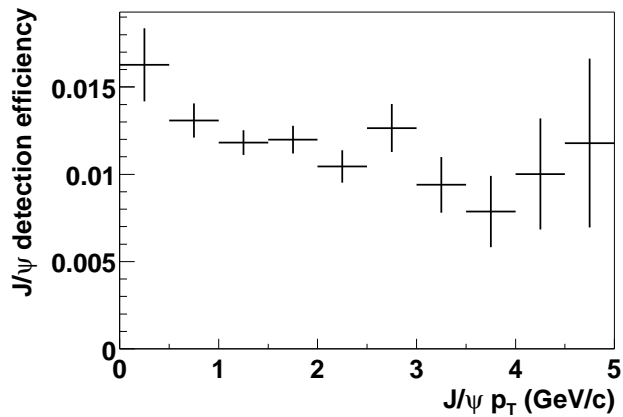}
\end{center}
\caption[\,\,\,$p_{T}$ dependence of $J/\psi$ efficiency]{
Transverse momentum ($p_{T}$) dependence of $J/\psi$ detection efficiency 
($\varepsilon_{tot}^{J/\psi}$).
Error bars are for statistical errors of the simulation.
}
\label{fig:jpsi_eff_pt}
\end{figure}

\begin{figure}[htbp]
\begin{center}
\includegraphics[width=3.5in,angle=0]{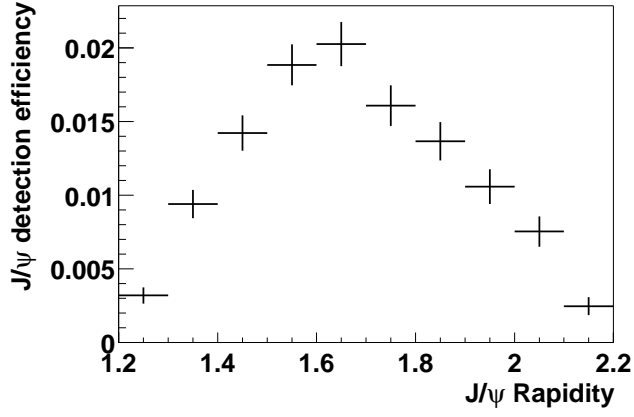}
\end{center}
\caption[\,\,\,Rapidity dependence of $J/\psi$ efficiency]{
Rapidity dependence of $J/\psi$ detection efficiency
($\varepsilon_{tot}^{J/\psi}$).
Error bars are for statistical errors of the simulation.
}
\label{fig:jpsi_eff_rapidity}
\end{figure}

Transverse momentum ($p_{T}$) and rapidity ($y$) dependence are shown 
in Figs.~\ref{fig:jpsi_eff_pt} and \ref{fig:jpsi_eff_rapidity}
respectively.
Error bars represent statistical errors of
the simulation.
Relatively small $p_{T}$ dependence of $\varepsilon_{tot}^{J/\psi}$ 
is found in the range $0 < p_{T} < 5$ GeV/$c$ which is
the statistical limit of our measurement.
For the calculation
of differential cross sections for $J/\psi$ production, 
averaged values
are used in each rapidity or $p_{T}$ 
bin shown in Tables~\ref{tab:result_summary_rap}
and \ref{tab:result_summary_pt} in section~\ref{section:result_summary}. 
A slight decrease in $\varepsilon_{tot}^{J/\psi}$
at higher $p_{T}$ is due to the bias of
the dimuon trigger which requires two different
quadrants to be fired, 
thus losing some $J/\psi$'s with a smaller opening angle
between two muons.
A systematic error of $\varepsilon_{tot}^{J/\psi}$ 
due to the uncertainty of $J/\psi$'s $p_{T}$ distribution
is estimated to be small ($<$ 2\%)
by changing efficiency values
in each $p_{T}$ bin
within their statistical uncertainties.
Rapidity distribution is
sensitive to gluon density in the proton.
A systematic error of $\varepsilon_{tot}^{J/\psi}$
due to the uncertainty of the rapidity 
distribution is studied using PYTHIA with various parton
distribution functions. As a result the variation of 
$\varepsilon_{tot}^{J/\psi}$ is found to be 
negligible ($<$ 2\%).

The systematic errors of each factor of $\varepsilon_{tot}^{J/\psi}$
is summarized in
Table~\ref{tab:systematic_errors} in section~\ref{section:result_summary}.


\subsection{Integrated luminosity}
\label{sec:lumi}

The integrated luminosity used for this analysis is given by

\begin{equation}
\label{eq:lumino_decompose}
   {\cal L} = \frac{N_{MB}^{|z_{vtx}|<38\,{\rm 
  cm}}}{\varepsilon_{BBC}^{inela}\,\sigma_{inela}}
\end{equation}
where

\begin{tabular}{lp{10.cm}} \\
$N_{MB}^{|z_{vtx}|<38\,{\rm cm}}$: & the number of minimum bias triggers
 with an offline BBC $z$-vertex cut, \\
$\varepsilon_{BBC}^{inela}$: & efficiency of 
the BBC for p+p inelastic events with the same vertex cut, and \\
$\sigma_{inela}$: & p+p inelastic cross section. \\
\end{tabular}

\vspace{5mm}

\noindent
Analysis procedures to obtain these factors
will be described in the following.

\subsubsection{Number of minimum bias triggers}
\label{sec:num_triggers}

The number of minimum-bias triggered events integrated over all the 
selected runs is represented as
$N_{MB}^{|z_{vtx}|<38\,{\rm cm}}$, 
with an offline BBC $z$-vertex cut $|z_{vtx}|<38$\,cm.
Since most of minimum bias triggers have been prescaled
(see section~\ref{sec:trigger_mb}),
it is not possible to obtain un-prescaled values of
$N_{MB}^{|z_{vtx}|<38\,{\rm cm}}$\,.
Instead it is calculated as
\begin{equation}
N_{MB}^{run_i, |z_{vtx}|<38\,{\rm cm}} = 
N_{MB,\,{\rm live}}^{run_i} \times
\frac {N_{MB,\,{\rm prescaled}}^{run_i, |z_{vtx}|<38\,{\rm cm}}}
{N_{MB,\,{\rm prescaled}}^{run_i}}
\end{equation}
for each run where un-prescaled (live) counts of all minimum-bias triggers
$N_{MB,{\rm live}}^{run_i}$ are obtained
from the run control log and
the ratios of 
minimum bias events with a BBC $z$-vertex cut
($N_{MB,\,{\rm prescaled}}^{run_i, |z_{vtx}|<38\,{\rm cm}}$)
to all minimum bias events
($N_{MB,\,{\rm prescaled}}^{run_i}$)
for each
run are obtained from the prescaled data.
Intergated over all the runs used for the $J/\psi$ analysis,  
1.72 $\times 10^9$ has been obtained
with a very small statistical error.

\begin{figure}[htbp]
\begin{center}
\includegraphics[width=4.5in,angle=0]{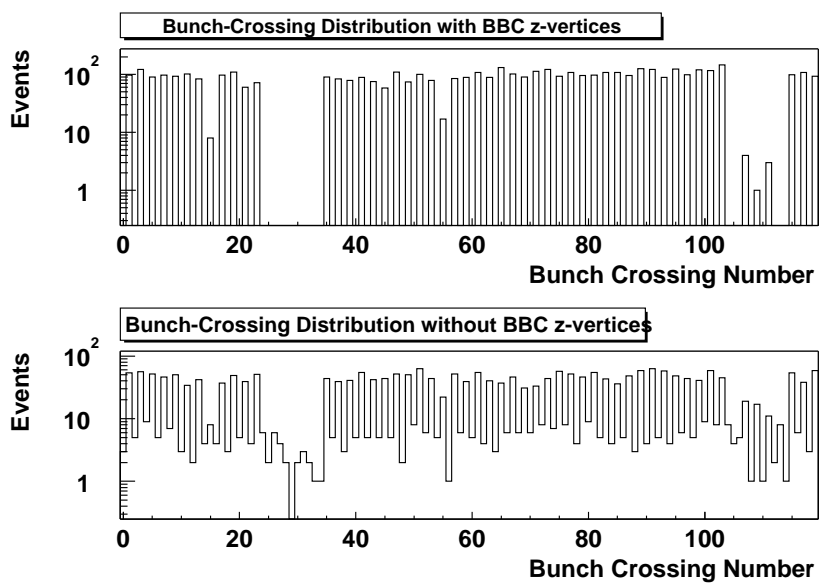}
\end{center}
\caption[Bunch-crossing number distributions with (upper) and without (lower)
BBC z-vertices.]
{Crossing-by-crossing variation for minimum-bias triggered events
with (upper) and without (lower) a BBC $z$-vertex cut
($|z_{vtx}|<$ 75~cm). 
Even numbers are for non-collision crossings and 
odd numbers are for collision crossings.
Events in non-collision crossings are supposedly 
induced by beam-related background.
}
\label{fig:bunch_cross}
\end{figure}

The fraction of the number of triggers from beam-related background
is found to be less than 0.1\% with the beam test, where
the rate of the BBLL1 trigger was reduced from 
10~kHz to 2$\sim$3~Hz when beams were mis-steered.
Crossing-by-crossing variation for minimum-bias triggered events,
shown in Fig.~\ref{fig:bunch_cross},
also give us an estimation of the magnitude of the background.
Since beam-related background events happen randomly,
its fraction is estimated 
by the ratio of the number of events in non-collision crossings
(even numbers in the figure) to
the number of events in collision crossings
(odd numbers in the figure).
If a BBC $z$-vertex ($|z_{vtx}|<$ 75~cm) 
is required (the upper figure), the ratio
is less than 0.1\%.
On the other hand, it goes up to
10\% if no BBC $z$-vertex required (the lower figure), which means
NTC-exclusive triggered events suffer from
beam-related background by 10\%.

\subsubsection{p+p inelastic cross section}

The total p+p cross section $\sigma_{tot}$ 
at $\sqrt{s} = 200$\,GeV has not been measured yet. 
Following parametrization for $\sigma_{tot}$ 
as a function of $\sqrt{s}$ is given in \cite{ref:RPP}
as well as fit parameters,
which reproduce experimental data well.
\begin{equation}
\sigma_{tot} = Z + B \log^{2}(s/s_{0}) + Y_{1}(s)^{-\eta_{1}}
- Y_{2}(s)^{-\eta_{2}} 
\end{equation}
where $Z$, $B$, $Y_{i}$ are in mb and $s$ and $s_{0}$ are in GeV$^{2}$.
The first and second terms, which represent
the pomeron contribution, dominates at high $\sqrt{s}$.
The exponents $\eta_{1}$ and $\eta_{2}$ represent 
lower-lying $C$-even and $C$-odd exchanges, respectively.
From the fitted parameters given there, 
$\sigma_{tot}$ 
is determined to be 51.5 $\pm$ 1.1 mb at $\sqrt{s}$ = 200 GeV.
%
Elastic cross section $\sigma_{ela}$ is also obtained 
(10.2 $\pm$ 0.3 mb), yielding
the inelastic cross section of $\sigma_{inela}
= \sigma_{tot} - \sigma_{ela} = 41.3 \pm 1.2$ mb.

%


Pseudo-rapidity range of protons from elastic
p+p scattering is 5 $< |\eta| <$ 9 which is out of
BBC acceptance, thus being excluded from the 
luminosity calculation.


\subsubsection{BBC efficiencies}
\label{sec:bbc_eff}

Efficiencies of the BBC for 
inclusive p+p inelastic events,
$\varepsilon_{BBC}^{inela}$, and
for p+p $\rightarrow J/\psi X$ events,
$\varepsilon_{BBC}^{J/\psi}$,
are sensitive to charged particle multiplicity
which has not been measured at $\sqrt{s}$ = 200~GeV
before. 
We have used the event generator
PYTHIA to estimate them, whose consistency with the real data
is fully examined.


PYTHIA
and the PHENIX detector simulation using GEANT give
$\varepsilon_{BBC}^{J/\psi}$ = 0.74 $\pm$ 0.01 and
$\varepsilon_{BBC}^{inela}$ = 0.51 $\pm$ 0.01 respectively
with a $|z_{vtx}|<$ 38 cm cut,
where errors are statistical errors of the simulation.
Because of higher multiplicities in
p+p $\rightarrow J/\psi X$ events than in 
inelastic events on average,
$\varepsilon_{BBC}^{J/\psi}$
is larger than $\varepsilon_{BBC}^{inela}$.
No significant $p_{T}$ nor rapidity dependence has been found
for $\varepsilon_{BBC}^{J/\psi}$.

To cross-check these results for their validities,
relative efficiency $R = \varepsilon_{BBC}^{inela} / 
\varepsilon_{MB}^{inela}$ is compared with the real data,
where $\varepsilon_{MB}^{inela}$ is efficiency of
the MB (minimum bias) trigger,
logical OR of BBC and NTC triggers,
for p+p inelastic events.
From simulation, $R$ = 0.51/0.70 = 0.73 is obtained
with a 1\% statistical error.
However, NTC trigger rate has suffered from beam-related background 
which is not included in the simulation.
Its fraction is about 10\% at maximum
as described in section~\ref{sec:num_triggers}.
Therefore $R$ is expected to vary from 0.67 to 0.73.
Figure~\ref{fig:rel_liverate} shows run statistics for the
relative rate of the BBC trigger to the minimum-bias trigger 
which is identical to $R$.
The distribution is
consistent with the expectation from the simulation and background rate.
The lower tail is due to runs with higher background 
(or NTC trigger) rate.
Even at the highest rate, contamination of the beam scraping background
to physics events is expected to be small 
($< 10^{-3}$).

\begin{figure}[htbp]
\begin{center}
\includegraphics[width=3.5in,angle=0]{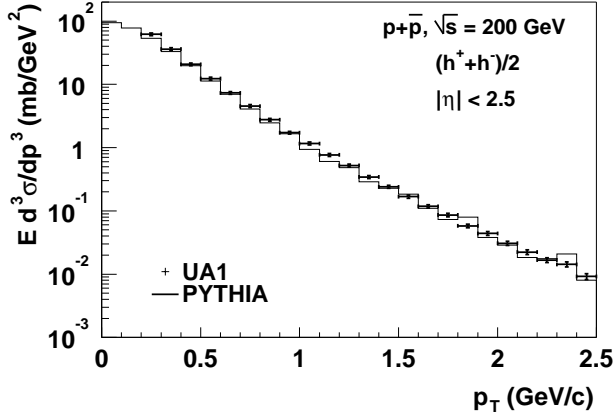}
\end{center}
\caption[
Inclusive cross section for single charged hadrons
]{
Inclusive cross section for single charged hadrons 
in p+$\bar{\mbox{p}}$ collisions at $\sqrt{s}$ = 200 GeV
as a function of $p_{T}$ of hadrons.
The points with error bars are for the UA1 data~\cite{ref:UA1_charged_hadrons}
and the histogram is for a PYTHIA prediction.
}
\label{fig:pythia_ua1_pt}
\end{figure}

\begin{figure}[htbp]
\begin{center}
\includegraphics[width=3.5in,angle=0]{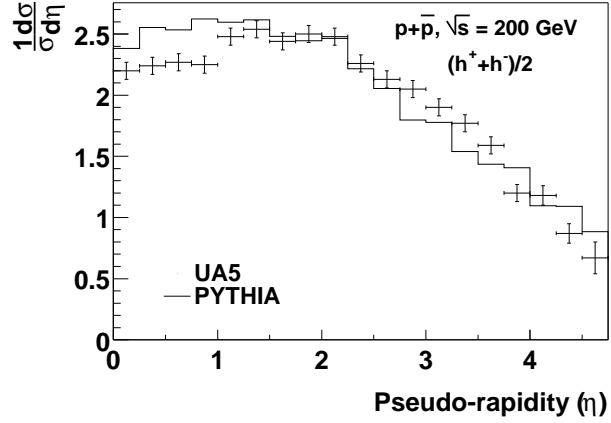}
\end{center}
\caption[
Charged particle pseudo-rapidity distribution
in p+$\bar{\mbox{p}}$ collisions at $\sqrt{s}$ = 200 GeV.
]{
Charged particle pseudo-rapidity distribution 
in p+$\bar{\mbox{p}}$ collisions at $\sqrt{s}$ = 200 GeV.
The points with error bars are UA5 data~\cite{ref:UA5_charged_hadrons}
and the histogram is for PYTHIA.
}
\label{fig:pythia_ua5_rap}
\end{figure}

\begin{figure}[htbp]
\begin{center}
\includegraphics[width=3.5in,angle=0]{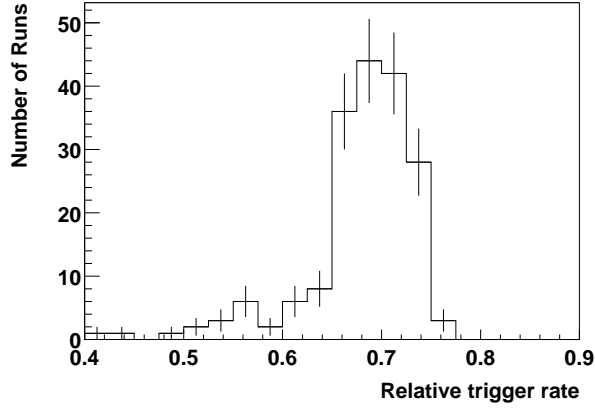}
\end{center}
\caption[
Run statistics for the relative trigger rate
]{
Run statistics for the relative trigger rate, BBC/MB, 
where MB (minimum bias) is the logical OR of the BBC and NTC triggers.
The lower tail is due to the runs with higher NTC trigger rate
from background.
}
\label{fig:rel_liverate}
\end{figure}

BBC efficiencies are sensitive to particle multiplicities
especially in and near the BBC acceptance (3.0 $<|\eta|<$ 3.9)
given by the event generator.
Figure~\ref{fig:pythia_ua1_pt} shows
comparison of PYTHIA with the UA1 
data \cite{ref:UA1_charged_hadrons} for the 
$p_{T}$ spectrum for inclusive charged 
hadrons at central rapidity ($|\eta| < 2.5$) in p+$\bar{\mbox{p}}$ 
collisions at $\sqrt{s}$ = 200 GeV.
They are in good agreement to 20\% precision.
Since gluon-gluon scattering dominates 
the inclusive charged-particle yield
for both p+p and p+$\bar{\mbox{p}}$ collisions,
the consistency in p+$\bar{\mbox{p}}$ collisions
indicate the consistency in p+p collisions.
In Run-2 at RHIC, $p_{T}$ differential 
cross sections for the neutral pion production
have been actually measured in the PHENIX Central Arms ($|\eta|<0.35$)
in p+p collisions at $\sqrt{s}$ = 200 GeV which 
agree with the UA1 results
in the range $p_{T} < $ 4 GeV/$c$ to about 10\% precision
~\cite{ref:PHENIX_pi0_run2}.
Figure~\ref{fig:pythia_ua5_rap} shows
comparison of PYTHIA with the UA5 data \cite{ref:UA5_charged_hadrons}
for the 
charged-particle pseudo-rapidity distributions 
in p+$\bar{\mbox{p}}$ collisions at $\sqrt{s}$ = 200 GeV.
PYTHIA distribution is again consistent with
the real data to 20\% precision.
A systematic error of $\varepsilon_{BBC}^{inela}$ from
uncertainty of particle distribution is estimated by changing 
charged particle multiplicity in PYTHIA by
20\% tuning parameters on the fragmentation function 
\footnote{for example, change PARJ(41) from 0.3 to 1.3 and PARJ(42) 
from 0.58 to 1.58, which represent the parameters $a$ and
$b$ respectively in the Lund fragmentation
function $f(x) \propto z^{-1}(1-z)^{a}\exp(-bm_{\bot}^{2}/z)$}
and changes of $\varepsilon_{BBC}^{inela}$ are
found to be 15\% or less.
PDF (parton distribution function) dependence is also studied and 
is found to be smaller (3\%).
Therefore a 15\% systematic error is assigned to $\varepsilon_{BBC}^{inela}$
from the uncertainty of the initial particle distribution.
The relative trigger rate $R$ is 0.80 when particle multiplicity is increased
by 20\% and 0.66 when decreased, both of which are deviated
from the measurement and indicating the 
particle multiplicity with the original PYTHIA
parameters reproduces the real data well.

In the p+p run, event vertex distribution 
varied from 40 cm to 70 cm in rms.
If $\varepsilon_{BBC}^{inela}$ had a significant dependence on
the $z$-vertex position, it would be affected by the change in the vertex
distribution. 
However, no $z$-vertex dependence within $|z_{vtx}|<$ 38 cm
is actually found over the statistical uncertainty of the simulation
study ($<$ 3\%); hence,
no fluctuation in $\varepsilon_{BBC}^{inela}$ is expected
during the run.

High-voltage potentials applied to each PMT (photomultiplier tube)
of the BBC were stable during the run. 
However there is PMT by PMT
fluctuation in discriminator-threshold values
possibly by 10 to 20\% which is not reflected
in the simulation.
It gives rise to an uncertainty of $\varepsilon_{BBC}^{inela}$. 
However it is found to be small ($<$~5\%)
with a simulation by conservatively doubling and halving 
all the PMT threshold values.

As the ver der Meer scan~\footnote{
a technique to measure the cross section of a beam by steering}
result, 
$\varepsilon_{BBC}^{inela} \times \sigma_{inela}$
has also been determined to be
18.5 $\pm$ 1.9 mb. 
This is consistent with $\varepsilon_{BBC}^{inela}$
obtained with the simulation times $ \sigma_{inela}$
obtained in the previous subsection within their errors.

The same magnitude of uncertainty is 
expected on $\varepsilon_{BBC}^{J/\psi}$.
However, 
some systematic uncertainties of 
$\varepsilon_{BBC}^{inela}$ and $\varepsilon_{BBC}^{J/\psi}$
will be canceled out 
since the total cross section is proportional to their ratio.
For example,
if particle multiplicity is higher
in inelastic events, it
tends to have higher multiplicity
also in $J/\psi \rightarrow \mu^{+}\mu^{-}$ events. 
Therefore a smaller systematic error of  $\varepsilon_{BBC}^{J/\psi}$
is assigned which is 10\%. 



BBC efficiency for another hard scattering events in p+p collisions,
$\pi^{0}$ production events, has been measured 
in the Central Arms ($|\eta|<$ 0.35) in Run-2
\cite{ref:PHENIX_pi0_run2}.
It is relatively $p_{T}$ independent
and consistent with a PYTHIA simulation within a few percent of errors,
as shown in attached figure~18. 
It confirms that PYTHIA reproduces 
particle multiplicities in p+p hard scattering events
for which BBC efficiency is scale
and (supposedly) process independent.

\subsection{Average $p_{T}$}
\label{sec:meanpt}

The average value of the transverse momentum ($p_{T}$) of $J/\psi$'s,
$\langle p_{T} \rangle$ is calculated as
\[ \langle p_{T} \rangle = C_{eff} 
C_{mom-offset} \langle p_{T} \rangle_{uncorr} \]
where $\langle p_{T} \rangle_{uncorr}$ is
an uncorrected value of $\langle p_{T} \rangle$,
$C_{eff}$ is a correction factor due to
$p_{T}$ dependence of efficiency and
$C_{mom-offset}$ is another correction factor 
due to the shift and resolution of momentum.
Following descriptions are
for procedures to obtain each factor.

\subsubsection{Uncorrected $\langle p_{T} \rangle$}

All data points are $p_{T}<$ 5 GeV/$c$, which is 
reasonable with the number of events obtained.
A statistical error of $\langle p_{T} \rangle$, $\Delta \langle p_{T}
\rangle$ is calculated as
$\sqrt{\langle p_{T}^{2}\rangle/N}$, where 
$N$ is the number of events.
Table \ref{tab:average_pt} shows $\langle p_{T} \rangle$
and $\Delta \langle p_{T}\rangle$ for both unlike-sign and like-sign
muon pairs with the $J/\psi$ identification 
cuts described in section~\ref{sec:dimuon}.

\begin{table}[htbp]
\begin{center}
\begin{tabular}{|c|c|c|c|}\hline 
     & count  & $\langle p_{T} \rangle$ & $\Delta \langle p_{T} \rangle$ \\ \hline
 unlike-sign pairs & 41 & 1.59 GeV/$c$ & 0.15 GeV/$c$\\ \hline
 like-sign pairs & 5 & 1.51 GeV/$c$ & 0.31 GeV/$c$ \\ \hline

\end{tabular}
\end{center}
\caption[Raw values for the average $p_{T}$]
{Average $p_{T}$ values and their statistical errors for unlike-sign and
  like-sign muon pairs with no corrections.}
\label{tab:average_pt}
\end{table}

Assuming the same $p_{T}$ distribution for unlike-sign and like-sign muon
pairs from background,
\[ \langle p_{T} \rangle_{uncorr} = \frac{N_{+-}}{N_{+-}-N_{++}} 
\left[\langle p_{T} \rangle_{+-} - \frac{N_{++}}{N_{+-}}\langle p_{T} \rangle_{++}\right] = 
1.60 \pm 0.17 \,\mbox{GeV}/c \]
is obtained where a suffix $+-$ ($++$) represents unlike (like)-sign pairs.


\subsubsection{Efficiency corrections}

The $p_{T}$ dependence of the $J/\psi$ detection efficiency 
has been obtained which is relatively small (shown in
Fig.~\ref{fig:jpsi_eff_pt}).
The correction factor $C_{eff}$ = 1.06 is obtained
with a 3\% statistical error
because of slightly higher efficiency for low $p_{T}$ $J/\psi$'s.

A systematic error of $C_{eff}$ has been estimated to be $\pm$4\%,
by increasing
efficiency for lower $p_{T}$ ($p_{T} <$ 2 GeV/c) $J/\psi$'s while
decreasing for higher $p_{T}$ ($p_{T} >$ 4 GeV/c) by
statistical errors of the simulation and vice versa.

It is confirmed that 
a correction of $\langle p_{T} \rangle$
due to rapidity dependence of efficiency
is small (1\%) since 
the correlation between $p_{T}$ and rapidity is small.


\subsubsection{Momentum offset and resolution correction}

The change in $\langle p_{T}\rangle$ from finite
momentum resolution and momentum shift has been estimated using a simulation,
based on the observed center position of the $J/\psi$ mass,
3156 $\pm$ 74~MeV/$c^{2}$, and its resolution, 257 $\pm$ 75~MeV/$c^{2}$, which are
shown in Fig.~\ref{fig:jpsi_mass_all}.

Assuming the shift of momentum is a constant (momentum-independent) and
the effect of momentum resolution on $\langle
p_{T}\rangle$ is small, momentum shift is naively expected to be
(3156+74-3097)/3097 $\sim$+4\%
at the worst case (using $M_{J/\psi}$ = 3097 MeV/$c^{2}$ \cite{ref:RPP}).
To estimate the entangled effect of both the
momentum shift and resolution on $\langle p_{T}\rangle$,
following two models are assumed for the shift
$\delta p/p$ and resolution $\Delta p/p$:
(I) $\delta p/p = a$, $\Delta p/p = b$  and 
(II) $\delta p/p = a'p$, $\Delta p/p = b'p$,
where $a$, $a'$, $b$ and $b'$ are parameters
which have been determined to reproduce the real data. 
For example, mis-calibrated magnetic field 
can cause (I), whereas mis-alignment of the MuTr chambers causes (II).
The shifts of $\langle p_{T} \rangle$ are estimated
using PYTHIA events whose $p_{T}$ shape is consistent with the real
data as shown in the next section.
The correction factor $C_{mom-offset}$ is calculated as
$C_{mom-offset} = \langle p_{T}\rangle_{generated}/\langle
p_{T}\rangle_{smeared} $
where $\langle p_{T}\rangle_{generated}$ is the average $p_{T}$
of unpolarized $J/\psi$'s generated by PYTHIA and 
$\langle p_{T}\rangle_{smeared}$ is the average $p_{T}$ of
$J/\psi$'s reconstructed from the smeared-momentum muons with the formula
(I) or (II) above.
In the worst case, where parameters are fixed to reproduce
the 3230-MeV/$c^{2}$ center position  
and 250-MeV/$c^{2}$ resolution of $J/\psi$,
$C_{mom-offset}$ = 0.95 is obtained for (I) and 0.94 for (II)
with 2\% statistical errors of the simulation
both of which are consistent with the naive expectation.
Without mass shift, where parameters are tuned to
the 3097-MeV/$c^{2}$ center position and 
250-MeV/$c^{2}$ resolution,
$\langle p_{T} \rangle$ does not alter over the statistical errors of
the simulation (2\%) for both (I) and (II) cases.
Consequently,
the correction factor $C_{mom-offset}$ = 0.97 $\pm$ 0.05 is obtained.


\section{Results and Discussions}
\label{chap:discussions}

\subsection{Differential and total cross sections}
\label{section:result_summary}

Results of 
rapidity-differential and
double-differential cross sections for $J/\psi$ production
measured in the South Muon Arm
are summarized in 
Table~\ref{tab:result_summary_rap} and \ref{tab:result_summary_pt} 
respectively
with each factor for them shown in the equation~(\ref{eq:xs1}).
Systematic errors of each factor are summarized in
Table~\ref{tab:systematic_errors}.

\begin{table}[htbp]
\begin{center}
\begin{tabular}{|c|c|c|c|c|} \hline
  $y$, $dy$, $p_{T}$, $dp_{T}$ &
  $N_{+-} - N_{++}$ &
  $\varepsilon_{tot}^{J/\psi}$ & 
  ${\cal L}$ &
  $Br\,d\sigma/dy$ (nb) \\
  & = $N_{J/\psi}$ & (\%) & (nb$^{-1}$) &
  \\ \hline\hline

  1.70, 1.0, all, --  & $41-5=36$      
               & 1.21 & 81 & 37.4 $\pm$ 7.1 (stat.) $\pm$ 10.1 (syst.)
               \\ \hline
  1.45, 0.5, all, --  & $29-3=26$     
               & 1.31 & 81 & 49.1 $\pm$ 10.8 $\pm$ 13.3 \\ \hline
  1.95, 0.5, all, --  & $12-2=10$       
               & 1.08 & 81 & 22.9 $\pm$ 8.5 $\pm$ 6.2 \\ \hline		     
\end{tabular}
\end{center}
\caption[
Results of rapidity-differential cross-sections and their input values]
{Results of rapidity-differential cross-sections and their input values.}
\label{tab:result_summary_rap}
\end{table}

\begin{table}[htbp]
\begin{center}
\begin{tabular}{|c|c|c|c|c|c|} \hline
  $y$, $dy$, $p_{T}$, $dp_{T}$ &
  $N_{+-} - N_{++}$ &
  $\varepsilon_{tot}^{J/\psi}$ & 
  ${\cal L}$ &
  $Br (1/2\pi p_{T}) d\sigma/dp_{T}dy$ \\

  & = $N_{J/\psi}$ & (\%) & (nb$^{-1}$) &  [nb/(GeV/$c$)$^{2}$]
  \\ \hline\hline

  1.70, 1.0, 0.5, 1.0 & $12-2=10$    
               & 1.40 & 81 & 2.81 $\pm$ 1.04 (stat.) $\pm$ 0.76
               (syst.) \\ \hline  
  1.70, 1.0, 1.5, 1.0 & $17-2=15$     
               & 1.21 & 81 & 1.63 $\pm$ 0.47 $\pm$ 0.44 \\ \hline
  1.70, 1.0, 2.5, 1.0 & $9-1=8$     
               & 1.14 & 81 & 0.55 $\pm$ 0.22 $\pm$ 0.15 \\ \hline
  1.70, 1.0, 4.0, 2.0 & $3-0=3$     
               & 0.92 & 81 & 0.08 $\pm$ 0.05 $\pm$ 0.02 \\ \hline
\end{tabular}
\end{center}
\caption[Results of double-differential cross-sections and their input values]
{Results of double-differential cross-sections and their input values.}
\label{tab:result_summary_pt}
\end{table}

\begin{table}[htbp]
\begin{center}
\begin{tabular}{|c|c|c|} \hline
\multicolumn{2}{|c|}{Variables} & Syst. errors (\%)\\ \hline\hline
\multicolumn{2}{|c|}{$N_{J/\psi}$} & 10 \\ \hline

	      & $A$ ($J/\psi$ pol.) & 10 \\ \cline{2-3}
              \raisebox{-2.5mm}[0mm][0mm]{$\varepsilon_{tot}^{J/\psi}$} 
              & $\varepsilon_{MuID}^{J/\psi}$ & 11 \\ \cline{2-3}
              & $\varepsilon_{MuTr}^{J/\psi}$ & 10 \\ \cline{2-3}
              & $\varepsilon_{BBC}^{J/\psi}$ & 10 \\ \hline
           & $N_{J/\psi}$     & negligible \\ \cline{2-3} 
${\cal L}$ & $\sigma_{inela}$ & 3 \\ \cline{2-3}
           & $\varepsilon_{BBC}^{inela}$ & 15 \\ \hline\hline


\multicolumn{2}{|c|}{Total} & 27 \\ \hline
\end{tabular}
\end{center}
\caption[Summary of the systematic errors of
each factor of the differential cross sections]
 {Summary of the systematic errors of 
each factor for the differential cross sections.
}

\label{tab:systematic_errors}
\end{table}

\subsubsection{Transverse momentum distribution}

Transverse momentum ($p_{T}$) distribution for
$J/\psi$ production cross-sections
at low $p_{T}$ ($p_{T} < M_{J/\psi}$), 
which dominates the sample obtained, is
sensitive to the intrinsic parton transverse-momentum
($k_{T}$) as well as to initial momentum distribution of partons 
(mainly gluons),
but not to the production mechanism much.
With an increased yield especially at higher $p_{T}$
($p_{T} \gg M_{J/\psi}$), 
we can discuss discrimination between the production models
based upon a $p_{T}$ shape.
Here our data is compared with a phenomenological function
which has been used to fit lower-energy results as well
as a PYTHIA prediction based on the color-singlet model.

\begin{figure}[htbp]
\begin{center}
\includegraphics[width=4in]{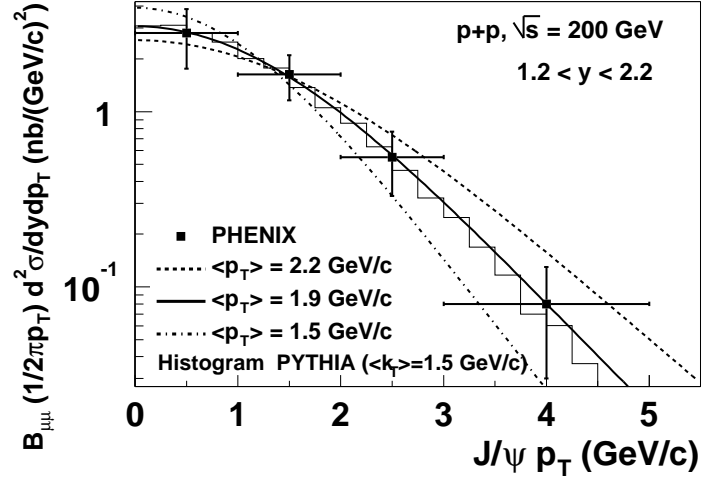}
\caption[Invariant cross sections for the $J/\psi$
production as a function of $p_{T}$]
{Invariant cross sections for the $J/\psi$
production as a function of $p_{T}$.
The error bars include statistical errors only.
The lines shows the function form
$d\sigma/dp_{T} \propto [1+(p_{T}/B)^{2}]^{-6}$
with different values of the average $p_{T}$,
$\langle p_{T}\rangle$ = (105$\pi$/768) $B$ GeV/c.
All the lines are normalized to the data.
The histogram is a PYTHIA prediction
with $\langle k_{T}\rangle$ = 1.5~GeV$/c$. 
}

\label{fig:dsigma_dpt}
\end{center}
\end{figure}

\begin{figure}[htbp]
\begin{center}
\includegraphics[width=4in]{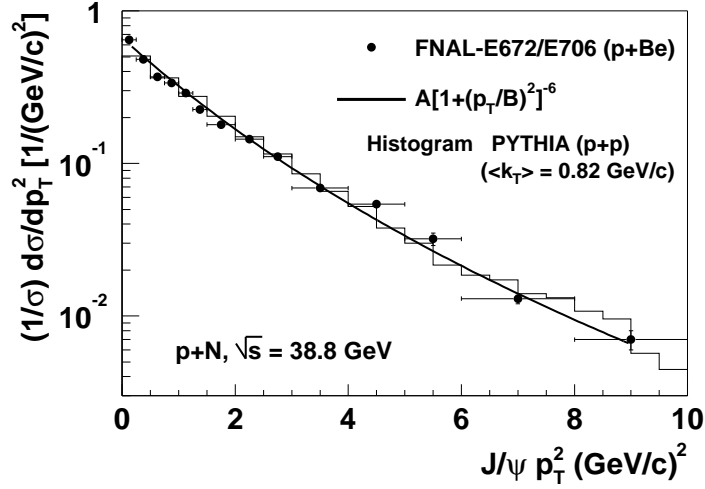}
\caption[
$p_{T}$ differential cross sections for the $J/\psi$
production at $\sqrt{s}$ = 38.8 GeV]
{
Differential cross sections for $J/\psi$
production as a function of $p_{T}^{2}$
at $\sqrt{s}$ = 38.8 GeV.
Data points are taken from the FNAL-E672/706 data
in p+Be collisions \cite{ref:E672-706}.
The line shows the fit of the function
$d\sigma/dp_{T}^{2} \propto [1+(p_{T}/B)^{2}]^{-6}$ to the data.
The histogram shows a PYTHIA prediction 
with $\langle k_{T}\rangle$ = 0.82 GeV/$c$.
}
\label{fig:dsigma_dpt_sqrts40}
\end{center}
\end{figure}

Figure~\ref{fig:dsigma_dpt} shows
invariant (double-differential) cross sections for $J/\psi$ production
as a function of $p_{T}$.
Error bars include statistical errors only.
Lines show the function form
\begin{equation} 
  \frac{d\sigma}{dp_{T}^{2}}
 \propto \left[1+\left(\frac{p_{T}}{B}\right)^{2}\right]^{-6}
\label{eq:pt_spectrum}
\end{equation}
with different values of a parameter $B$,
which is related to the average $p_{T}$ 
($\langle p_{T}\rangle$)
with $\langle p_{T}\rangle = (105\pi/768)B$
by a numerical integration.
All the lines are normalized to the data.
This function form is consistent with our data with $\langle p_{T} \rangle$ 
in the range
1.5 $< \langle p_{T} \rangle <$ 2.2~GeV/$c$.
It also fits the FNAL-E672/E706 data~\cite{ref:E672-706},
shown in Fig.~\ref{fig:dsigma_dpt_sqrts40},
and the FNAL-E789 data~\cite{ref:exp_sqrt_31.5_E789}
at $\sqrt{s}$ = 38.8 GeV well.
The histograms in those figures show the 
PYTHIA distributions which are 
consistent with both the real data and the fits
with appropriate parameters.\footnote{
The low-$p_{T}$ cut off for partonic subprocesses
is deactivated (default = 1~GeV/$c$). 
} 
Contribution of $b$-quark decays is neglected,
since it is estimated to be small (1\%) 
at $\sqrt{s}$ = 200 GeV
from other experiments and simulation,
which will be described in Appendix~\ref{chap:feeddown}.
The transverse momentum $p_{T}$ of a $J/\psi$ is sensitive to 
the rms of $k_{T}$, $\langle k_{T}\rangle$, which is set to
1.5 GeV/$c$ at $\sqrt{s}$ = 200 GeV
and 0.82 GeV/$c$ at $\sqrt{s}$ = 38.8 GeV
to reproduce the real data as shown 
in Figs.~\ref{fig:dsigma_dpt} and \ref{fig:dsigma_dpt_sqrts40}.
The result of $\langle p_{T} \rangle$ 
will be discussed in the next section.


\subsubsection{Rapidity distribution and gluon distribution}

Since the gluon-fusion subprocess
dominates at high energies,
rapidity distribution for $J/\psi$
production cross-sections, $d\sigma/dy$,
is sensitive to gluon
distribution $g(x,Q)$ in the proton, 
where $x$ is a momentum fraction of a gluon (Bjorken-$x$)
and $Q$ is a scale for the partonic interaction,
which is in the order of $M_{c}$ 
(charm quark mass)
for the case of $J/\psi$ production.
Sensitive $x$-region probed by our measurements
is discussed, 
followed by the comparison of our data with some typical
gluon distributions currently available.

\begin{figure}[htbp]
\begin{center}
\includegraphics[width=3.5in]{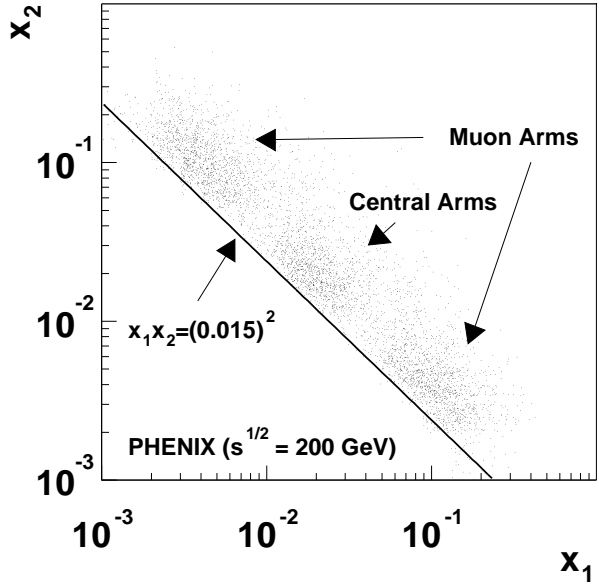}
\caption[The correlation of $x_{1}$ and $x_{2}$ of gluons probed by
the $J/\psi$ measurement]
{The correlation of $x_{1}$ and $x_{2}$ of gluons probed by
$J/\psi$ measurements with the Central Arms and Muon Arms.
The line shows the low-$p_{T}$ limit, where
$x_{1}x_{2} = \tau \sim (0.015)^{2}$.}
\label{fig:x1_x2}
\end{center}
\end{figure}

\begin{figure}[htbp]
\begin{center}
\includegraphics[width=4in]{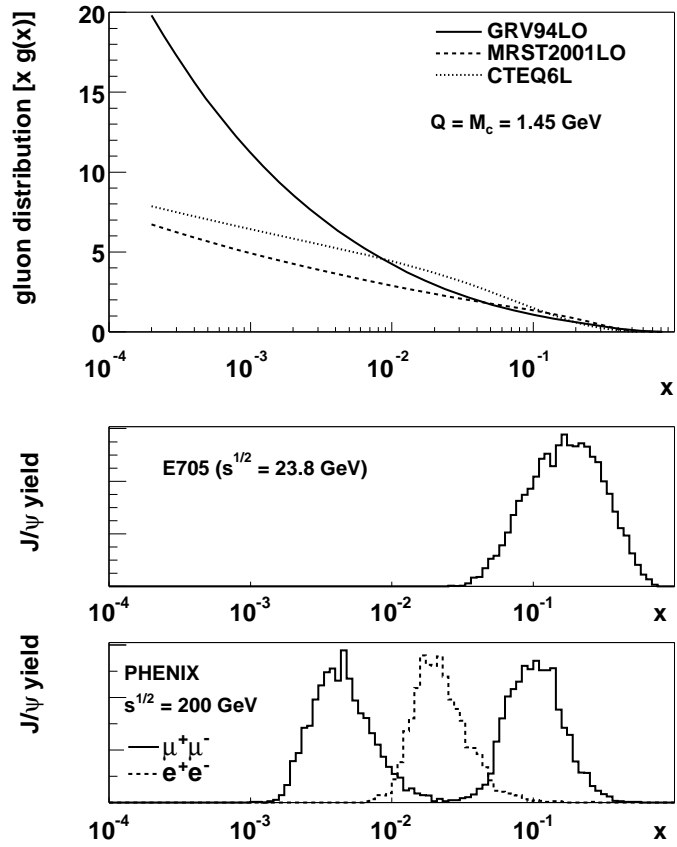}
\caption[The gluon distribution function $x g(x)$]
{The gluon density $x g(x)$
of three typical PDF sets with a scale $Q = M_{c}$ = 1.45~GeV,
together with sensitive $x$ regions
probed by $J/\psi$ measurements with the PHENIX and FNAL-E705 experiments.
}
\label{fig:gluon_dist}
\end{center}
\end{figure}

\begin{figure}[htbp]
\begin{center}
\includegraphics[width=4in]{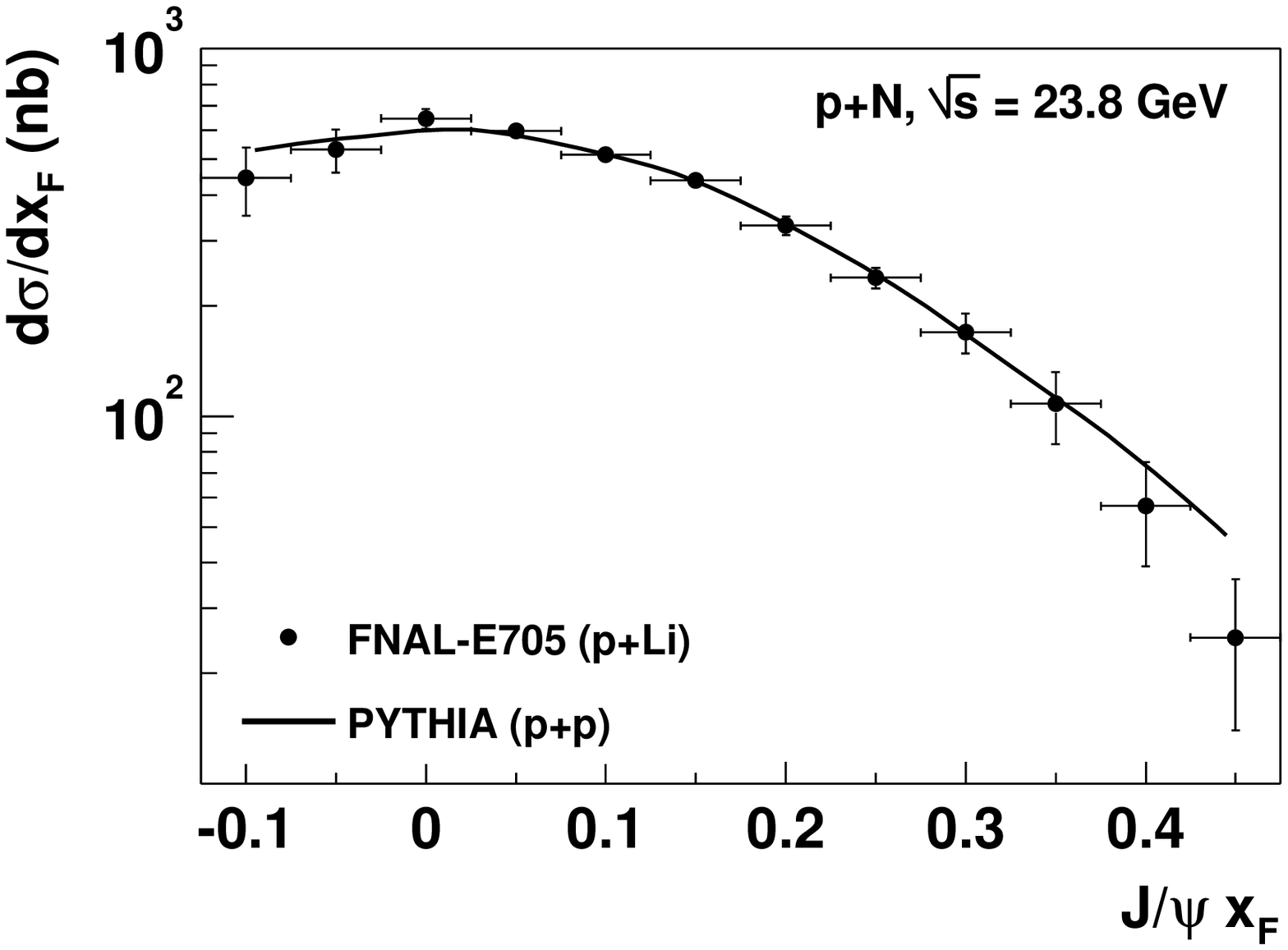}
\caption[$x_{F}$ differential distribution of $J/\psi$ of FNAL-E705
  data]
{
Feynman-$x$
($x_{F}$) differential cross sections for $J/\psi$ production obtained
by the FNAL-E705 experiment in p+Li collisions at $\sqrt{s}$ = 23.8
GeV \cite{ref:exp_sqrt_23.7}.
The line shows a PYTHIA prediction with GRV94LO PDFs.
}
\label{fig:dsigma_dxf_sqrts23.8}
\end{center}
\end{figure}

For low-$p_{T}$ $J/\psi$ production ($p_{T} < M_{J/\psi}$),
$d\sigma/dy$ is roughly proportional to 
the product of gluon distribution function
$g(x_{1})g(x_{2})$ where $x_{1(2)}$ is Bjorken-$x$ of each gluon. 
They are related with $y$ (rapidity) of a $J/\psi$ by
\begin{equation*}
x_{1} = \sqrt{\tau}\exp(y)\,\mbox{ and }\, x_{2} = \sqrt{\tau}\exp(-y)
\end{equation*}
where $\tau \equiv (M_{J/\psi}/\sqrt{s})^{2}$. 
The relation of $x_{1}$ and $x_{2}$ for $J/\psi$ production events, 
\begin{equation}
x_{1}x_{2} = \tau \propto \frac{1}{s}
\label{eq:x1x2}
\end{equation}
is shown in Fig.~\ref{fig:x1_x2} using PYTHIA
and the GRV94LO PDFs with a scale $Q = M_{c}$ = 1.45~GeV.
The clear correlation $x_{1}x_{2} = \tau \sim
(0.015)^{2}$ is seen with a band which represents finite
$p_{T}$ of $J/\psi$'s.
The sensitive regions of $x$ are
$0.01 < x < 0.05$ for $J/\psi$'s measured in the Central Arms and
$2 \times 10^{-3} < x < 0.01$ 
and $0.05 < x < 0.2$ measured in the Muon Arms,
which are also shown in Fig.~\ref{fig:gluon_dist}
together with some gluon distributions of some typical
PDF sets with a scale $Q = M_{c}$ = 1.45~GeV.
Distributions of Feynman-$x$ ($x_{F} = x_{1} - x_{2}$)
of lower-energy experiments,
obtained as $x_{F} = 2p_{z}/\sqrt{s}$
where $p_{z}$ is the $z$-component of the momentum of a $J/\psi$,
are well reproduced by these gluon distributions.
For example, the $x_{F}$
distribution of the FNAL-E705 experiment
at $\sqrt{s}$ = 23.8 GeV~\cite{ref:exp_sqrt_23.7}
is well reproduced by the GRV94LO PDFs as shown in
Fig.~\ref{fig:dsigma_dxf_sqrts23.8}.
This agreement confirms that $J/\psi$ production
is dominated by the fusion of gluons, whose
distribution is reasonable in the corresponding $x$-range,
which is around $x \sim 0.1$ covered also by the PHENIX Muon Arms
as shown in Fig.~\ref{fig:gluon_dist}.
Another $J/\psi$ event generator, the RV generator~\cite{ref:RV},
gives consistent $x_{F}$ distribution 
with PYTHIA, which confirms
that $x_{F}$ distribution (or rapidity distribution at
low $p_{T}$) depends on gluon
density but not on the event generator
with a specific production mechanism.


\begin{figure}[htbp]
\begin{center}
\includegraphics[width=4in]{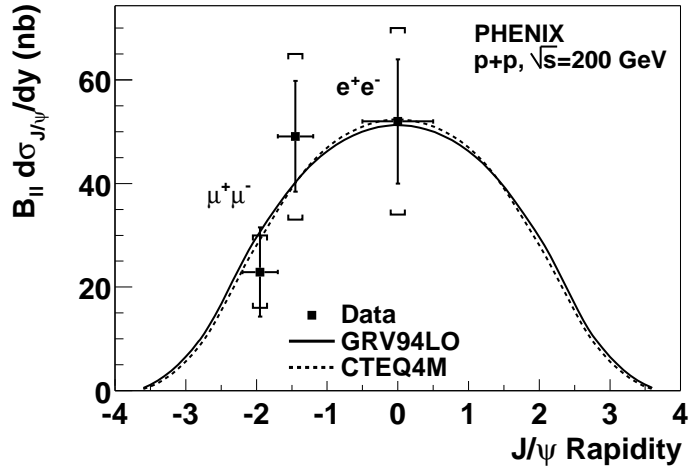}
\caption[Rapidity differential cross section]
{Rapidity differential cross sections for $J/\psi$ production. 
Both statistical and systematic errors are shown,
which are in bars and in brackets respectively. 
Two lines show PYTHIA predictions
with different parton distribution functions
(GRV94LO and CTEQ4M) fitted to the data.
}
\label{fig:dsigma_dy}
\end{center}
\end{figure}

Using these PDFs, our rapidity distribution is
well reproduced as shown in 
Fig.~\ref{fig:dsigma_dy}
including the result of the $J/\psi \rightarrow e^{+}e^{-}$
channel measured in the Central Arms 
($d\sigma/dy\,|_{\,y=0} = 52 \pm 13 \pm 18$ nb)
\cite{ref:PHENIX_jpsi_ee}.
Two curves in the figure show PYTHIA predictions
with the GRV94LO and CTEQ4M PDF sets fitted to the data,
both of which are consistent with the results.
Their difference is small compared to the errors
of our measurement. Other PDF sets are also tried and
fluctuation is found to be small (3\%)
in the determination of $\sigma_{J/\psi}$ described next.
This consistency confirms reasonable gluon density
in the $x$-region $2 \times 10^{-3} < x < $ 0.2. 

\subsubsection{Total cross section and its comparison with lower-energy results}

Assuming a PYTHIA rapidity distribution
which reproduces our results well
and the common
branching fraction $B_{ll}$ for the electron decay channel ($B_{ee}$) and
the muon decay channel ($B_{\mu\mu}$),
\[ B_{ll} \sigma_{J/\psi} = 226 \pm 36\,(\mbox{stat.}) \pm 79\,(\mbox{syst.})
  \mbox{ nb} \]
is obtained. Systematic error is obtained as 
the maximum plausible systematic spreads including 
both the electron and muon measurements.
Using the average value of $B_{ee}$ and $B_{\mu\mu}$~\footnote{
$B_{ee}$ = 0.0593 $\pm$ 0.0010 and $B_{\mu\mu}$ = 
0.0588 $\pm$ 0.0010~\cite{ref:RPP}} 
for $B_{ll}$, total production cross-section
\[ \sigma_{J/\psi} = 3.8 \pm 0.6\,(\mbox{stat.}) \pm 1.3\,(\mbox{syst.})
  \mbox{ } \mu\mbox{b} \]
is extracted. 



\begin{figure}[htbp]
\begin{center}
\includegraphics[width=4in]{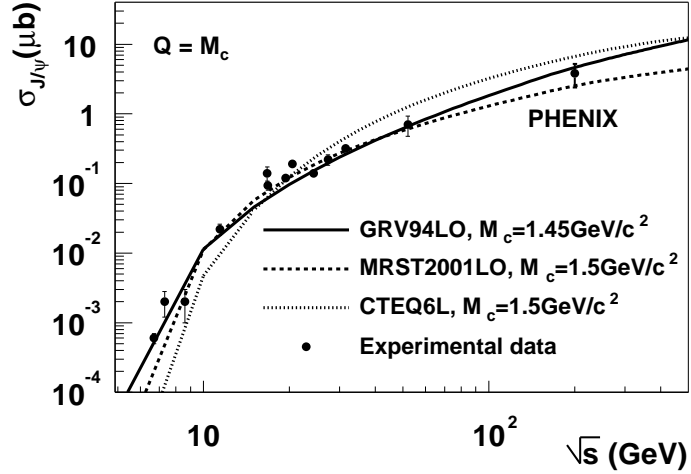}
\caption[$\sqrt{s}$ dependence of the production cross section
for $J/\psi$ in nucleon-nucleon collisions
($Q = M_{c}$ for the theoretical predictions)]
{Center of mass energy dependence of the production cross section
for $J/\psi$ in nucleon-nucleon collisions.
Lower energy results are summarized 
in Table~\ref{tab:jpsi_sigma_sqrts}.
Error bars include both statistical and systematic errors
which are added in quadrature.
The curves are for QCD predictions with different
PDF sets with a scale $Q$ = $M_{c}$.
}
\label{fig:jpsi_sigma_sqrts_scale1}
\end{center}
\end{figure}
\begin{figure}[htbp]
\begin{center}
\includegraphics[width=4in]{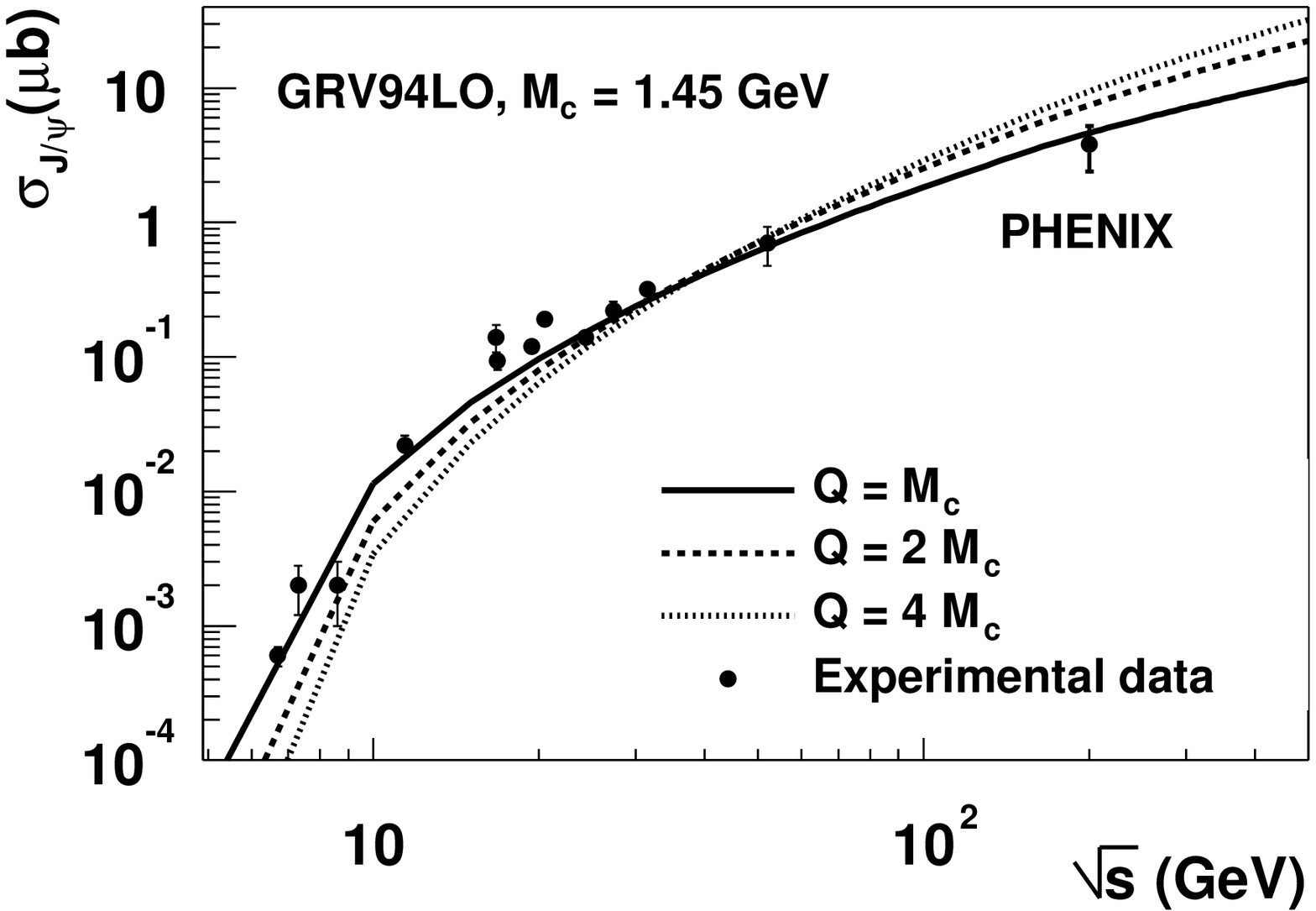}
\caption[
$\sqrt{s}$ dependence of the production cross section
for $J/\psi$ in nucleon-nucleon collisions with different
PDF scales]
{
Center of mass energy dependence of the production cross section
for $J/\psi$ in nucleon-nucleon collisions
with different PDF scales ($Q$).
GRV94LO PDFs and charm quark mass $M_{c} = 1.45$ GeV/$c^{2}$ are used.
}
\label{fig:jpsi_sigma_sqrts_scaledep}
\end{center}
\end{figure}

Figure~\ref{fig:jpsi_sigma_sqrts_scale1} shows 
$\sqrt{s}$ dependence of $\sigma_{J/\psi}$
together with predictions 
using gluon densities of 
three different PDF sets with a scale of $Q = M_{c}$.
Lower-energy experimental results are summarized
in Table~\ref{tab:jpsi_sigma_sqrts}.
Our result of the total cross section is at the highest energy since 
higher energy ($\sqrt{s} \ge$ 630~GeV) experimental data on $J/\psi$'s 
suffer from a $p_{T}$ cut ($p_{T} >$ 4 or 5~GeV/$c$)
as described in section~\ref{sec:intro}.
The increase in $\sigma_{J/\psi}$ at 
higher $\sqrt{s}$ 
can be explained by the increase in
gluon density in lower $x$-region 
according to the relation~(\ref{eq:x1x2}).
For the absolute normalization for $\sigma_{J/\psi}$, 
color-octet matrix elements
are used, whose uncertainties will be described later.
Charm quark mass is adjusted to $M_{c}$ = 1.45~GeV/$c^{2}$
or $M_{c}$ = 1.50~GeV/$c^{2}$ depending on PDF sets
to reproduce the normalization best.
Parton distributions are sensitive to a scale $Q$.
Figure~\ref{fig:jpsi_sigma_sqrts_scaledep}
shows $\sqrt{s}$ dependence of $\sigma_{J/\psi}$
with different values of $Q$
using the GRV94LO PDFs and $M_{c}$ = 1.45 GeV/$c^{2}$.
For higher $Q$, distribution is steeper since
gluon density increases more in lower-$x$
than in higher-$x$ as $Q$ increases.
The experimental data prefer $Q = M_{c}$
for this case.
In spite of uncertainty from the choice of a PDF set and a scale,
$\sqrt{s}$ dependence of $\sigma_{J/\psi}$ 
including our new result is
consistent with gluon distribution in the proton.

\begin{table}[htbp]
\begin{center}
\begin{tabular}{|c|c|c|c|c|} \hline
Target & $\sqrt{s}$ (GeV) & $\sigma_{J/\psi}$ (nb) 
& $\langle p_{T}\rangle$ (GeV/$c$) & Reference \\ \hline
p & 6.7 &  $0.62 \pm 0.18$ & | & \cite{ref:exp_sqrt_6.8} \\ 
Be & 7.3 & $2^{+2.0}_{-1.0}$ & | & \cite{ref:obs_J} \\ 
p & 8.6 &  $ 2.4 \pm 1.2$ & | & \cite{ref:exp_sqrt_8.7} \\ 
Be & 11.5  & 22 $\pm$ 6 & | & \cite{ref:exp_sqrt_11.5}\\
Be & 16.8 & 138 $\pm$ 46 & 0.96 $\pm$ 0.12 & \cite{ref:exp_sqrt_16.8} \\
p &  16.8 & 94 $\pm$ 20 & | & \cite{ref:exp_sqrt_19.4} \\
p & 19.4 & 122 $\pm$ 22 & 0.95 $\pm$ 0.02 & \cite{ref:exp_sqrt_19.4} \\
C & 20.5 & 190 $\pm$ 26 & 0.98 $\pm$ 0.04 & \cite{ref:exp_sqrt_20.5} \\
Li & 23.8 & 162 $\pm$ 22 & | & \cite{ref:exp_sqrt_23.7} \\
p & 24.3 & 144 $\pm$ 19 & 1.1 $\pm$ 0.2 & \cite{ref:exp_sqrt_24.3} \\
Be & 27.4 & 220 $\pm$ 54 & | & \cite{ref:exp_sqrt_27.4_2} \\
p & 30 & | & 1.14 $\pm$ 0.12 & \cite{ref:exp_sqrt_30-63} \\
Be & 31.5 & 322 $\pm$ 70 & 1.08 $\pm$ 0.05 &\cite{ref:exp_sqrt_31.5_E672}\\
Be & 31.5 & | & 1.15 $\pm$ 0.02 &\cite{ref:E672-706}\\
Be & 38.8 & |            & 1.22 $\pm$ 0.01 & \cite{ref:E672-706}\\
p & 52 & 700 $\pm$ 320 & 1.0 $\pm$ 0.2 & \cite{ref:exp_sqrt_52.0} \\
p & 53 & | & 1.39 $\pm$ 0.05 & \cite{ref:exp_sqrt_30-63} \\
p & 63 & | & 1.29 $\pm$ 0.05 & \cite{ref:exp_sqrt_30-63} \\ \hline

\end{tabular}
\caption[Experimental results of total cross sections
and the average values of $p_{T}$]
{
The $J/\psi$ total cross sections $\sigma_{J/\psi}$ 
and the average values of $p_{T}$ ($\langle p_{T}\rangle$)
in proton-induced interactions at lower energies.
Target mass dependence is accounted for assuming $A^{0.9}$ \cite{ref:Vogt}
where $A$ is the atomic number of the target.
}
\label{tab:jpsi_sigma_sqrts}
\end{center}
\end{table}

The absolute normalization for $\sigma_{J/\psi}$ 
depends on the production models.

A consequence of the CEM is the constant ratio,
that is $\sqrt{s}$ and process independent,
of $\sigma_{J/\psi}$ to $\sigma_{D\bar{D}}$ where 
$\sigma_{D\bar{D}}$ is the total production cross section
for $D$ mesons.
It is determined from the photo-production data
to be about 0.06~\cite{ref:cem_update}.
To normalize $\sigma_{D\bar{D}}$ in hadro-production,
the phenomenological $K$-factors
are used to account for higher order corrections.
As a result, $\sigma_{J/\psi}$ of hadro-production data
are well reproduced including our measurement.

The CSM (color-singlet model) has been known to fail
to describe the total $J/\psi$ cross sections 
at lower energies by a factor of about 20 \cite{ref:COM_fixed_target}. 
Despite large theoretical uncertainties
from the choice of scale parameters and charm quark
mass, this large discrepancy 
can be hardly explained.

Color-octet contributions have been considered
\cite{ref:COM_fixed_target,ref:Beneke_96,ref:WKT_96}
in 2 $\rightarrow$ 1 subprocesses (dominantly $g+g \rightarrow J/\psi$)
which is the largest contribution to the low $p_{T}$ 
(or $p_{T}$-integrated) yield.
Cross section for the direct $J/\psi$ production with
gluon-gluon fusion can be written as
\begin{eqnarray*}
  \sigma(gg \rightarrow J/\psi) &=& 
 \frac{5\pi^3 \alpha_s^2}{12(2M_c)^3s} \delta (x_1 x_2 - 4M_c^2/s) \\
 &\times&
 \Bigl[\langle O_{8}^{J/\psi}({}^{1}S_{0})\rangle +
       \frac{3}{M_{c}^{2}}\langle O_{8}^{J/\psi}({}^{3}P_{0})\rangle +
       \frac{4}{5M_{c}^{2}}\langle O_{8}^{J/\psi}({}^{3}P_{2})\rangle
 \Bigr]
\end{eqnarray*}
where $\alpha_s$ is the strong coupling as a function of 
a renormalization scale ($\mu$).
The color-octet matrix elements 
$\langle O_{8}^{J/\psi}({}^{2s+1}L_{J})\rangle$,
each of which represents probability for
a color-octet
$c\bar{c}$ pair in a state ${}^{2s+1}L_{J}$ to form a $J/\psi$ meson,
have been extracted from experimental data \cite{ref:Beneke_96}.
There are also decays from the color-octet production of $\chi_c$,
but their contribution is neglected,
since it is estimated to be small compared to the color-singlet
production of $\chi_c$~\cite{ref:Beneke_96}.
Including this color-octet contribution to
the direct $J/\psi$ production, total cross section
agrees with the data as shown
in Figs.~\ref{fig:jpsi_sigma_sqrts_scale1} and
\ref{fig:jpsi_sigma_sqrts_scaledep}
using reasonable charm quark mass $M_{c}$ 
and a renormalization scale $\mu$.
The uncertainty of $\sigma_{J/\psi}$ from $M_{c}$,
which is not constrained well yet~\cite{ref:RPP},
has been
studied by changing it by about 10\% and 
variation of $\sigma_{J/\psi}$ is found to be about 
a factor of two.
The effect of the uncertainty of 
$\mu$ is also studied by making it double or half
and $\sigma_{J/\psi}$ changes by a factor
of two to three.
The normalization for $\sigma_{J/\psi}$ with the COM
is consistent with the experimental data within 
large theoretical uncertainties inherent in perturbative QCD.

\subsection{Average $p_{T}$}

The average value of $p_{T}$ of $J/\psi$'s, 
$\langle p_{T} \rangle$ is 
sensitive to the intrinsic parton transverse-momentum
($k_{T}$) as well as initial momentum distribution of partons
(mainly gluons),
but not to the production mechanism much.
It is calculated as 
$\langle p_{T} \rangle = 
C_{eff} C_{mom-offset}\langle p_{T} \rangle_{uncorrected}$
where each factor has been obtained 
in section~\ref{sec:meanpt}.
Table~\ref{tab:meanpt_errors} summarizes their statistical and systematic
errors.
As a result,
\[ \langle p_{T} \rangle = 1.66 \pm 0.18\,(\mbox{stat.}) 
 \pm 0.12\,(\mbox{syst.})\, \mbox{GeV}/c 
\]
is obtained with cuts $0 < p_{T} < 5$ GeV/$c$ and $1.2 < y < 2.2$. 
Correction factors on $\langle p_{T} \rangle$
to extrapolate it to the entire kinematical region
are estimated in the next, followed by 
the comparison of $\langle p_{T} \rangle$ with lower-energy results.

\begin{table}[htbp]
\begin{center}
\begin{tabular}{|c|c|}\hline 
 Statistical & $\pm$ 11\% \\ \hline \hline
 Efficiency corrections & $\pm$ 5\% \\ \hline
 Momentum offset & $\pm$ 5\% \\ \hline
 Syst. total & $\pm$ 7\% \\ \hline
\end{tabular}
\end{center}
\caption {Summary of statistical and systematic errors of $\langle p_{T} \rangle$.}
\label{tab:meanpt_errors}
\end{table}

\subsubsection*{High-$p_{T}$ correction}

Since the statistical limit of the data is $p_{T}<$ 5 GeV/$c$,
the effect on $\langle p_{T} \rangle$ from missing
high-$p_{T}$ events is estimated
assuming the function form
(\ref{eq:pt_spectrum}) which fits 
the FNAL-E672/E706 data~\cite{ref:E672-706} 
(Fig.~\ref{fig:dsigma_dpt_sqrts40})
the FNAL-E789 data~\cite{ref:exp_sqrt_31.5_E789}
well.
As a result,
$\langle p_{T}\,(0<p_{T}<5\,\mbox{GeV}/c) \rangle$ = 1.82 GeV/$c$
and
$\langle p_{T}\,(0<p_{T}<\infty\,\mbox{GeV}/c) \rangle$ = 1.88 GeV/$c$
are obtained.
Therefore a correction factor from high-$p_{T}$ events
is expected to be small (3\%).

\begin{figure}[htbp]
  \begin{center}
  \includegraphics[width=4in]{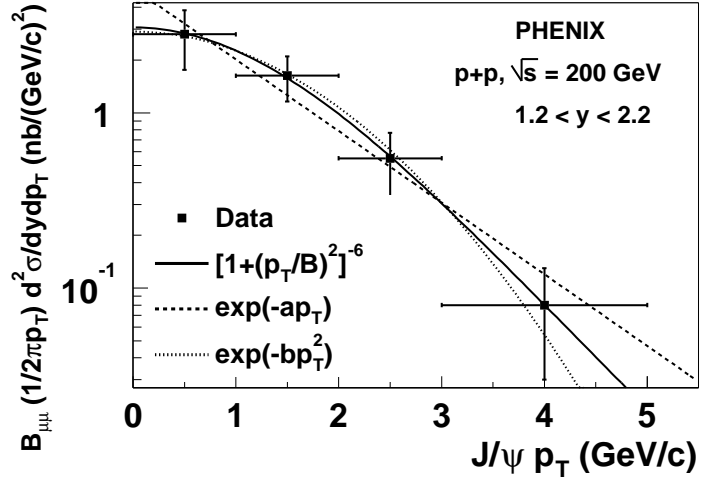}
  \caption{
    Invariant cross sections for the $J/\psi$ production
    as a function of $p_{T}$
    together with three different function forms.}
  \label{fig:dsigma_dpt_fit}
  \end{center}
\end{figure}

Other two empirical functions forms which have been used
for other experiments,
\begin{equation} 
  \frac{d\sigma}{dp_{T}^{2}} \propto \exp( -a p_{T} )
  \label{eq:pt_spectrum_2}
\end{equation}
and
\begin{equation} 
  \frac{d\sigma}{dp_{T}^{2}} \propto \exp( -b p_{T}^{2} )
  \label{eq:pt_spectrum_3}
\end{equation}
are tried where $a$ and $b$ are fit parameters.
Their fits to our data is shown in Fig.~\ref{fig:dsigma_dpt_fit}.
For (\ref{eq:pt_spectrum_2}), $p_{T}$ slope is slightly
harder than other functions but still consistent with
our results.
The correction factor of 11\% is obtained.
For (\ref{eq:pt_spectrum_3}), much smaller
a correction factor, 0.04\% is obtained.
The reduced chi-squares for the fit of (\ref{eq:pt_spectrum_2}) to 
the FNAL-E789 data ($\chi^{2}$/NDF = 6.6)
is much worse than others 
($\chi^{2}$/NDF = 0.5 for (\ref{eq:pt_spectrum}) 
and 1.3 for (\ref{eq:pt_spectrum_3})).

Consequently, high-$p_{T}$ correction ($p_{T}>$ 5 GeV/c)
is expected to be small ($<$ 3\%) 
assuming the function form for the $p_{T}$ spectrum which 
fits our results as well as
the FNAL-E789 and FNAL-E672/706 results
at $\sqrt{s}$ = 38.8 GeV with good chi-squares.

\subsubsection*{Rapidity correction}

\begin{figure}[htbp]
\begin{center}
\includegraphics[width=4in]{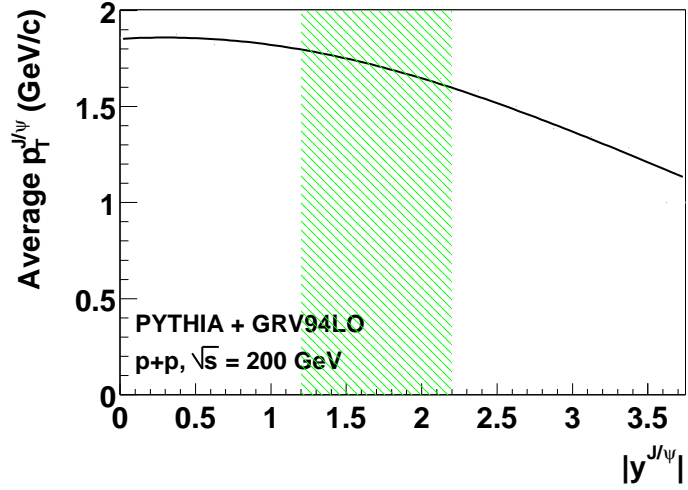}
\caption
[Rapidity dependence of $\langle p_{T}\rangle$ with PYTHIA]
{Rapidity dependence of $\langle p_{T}\rangle$ in
$\sqrt{s}$ = 200 GeV p+p collisions obtained with PYTHIA and the GRV94LO
PDFs. The hatched area shows the South Muon Arm acceptance.
}
\label{fig:pythia_jpsi_pt_rap}
\end{center}
\end{figure}

Figure~\ref{fig:pythia_jpsi_pt_rap} shows rapidity
dependence of $\langle p_{T}\rangle$ using PYTHIA with the GRV94LO
PDFs and other parameters which reproduce
$p_{T}$ and rapidity distributions
of both lower energy experiments and our result as already shown.
Using this dependence $\langle p_{T}\rangle_{1.2<y<2.2}/
\langle p_{T}\rangle_{-\infty < y < \infty}$ is found to be 0.99 which
is close to unity.

\subsubsection*{Comparison with lower energy results}

\begin{figure}[htbp]
\begin{center}
\includegraphics[width=4in]{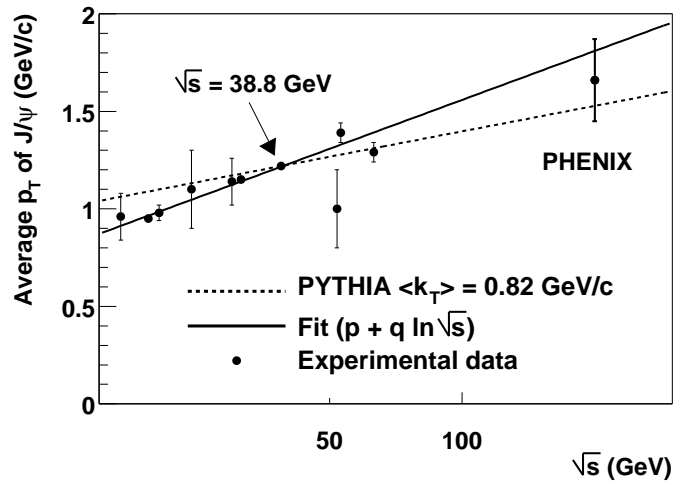}
\caption[$\sqrt{s}$ dependence of $\langle p_{T} \rangle$]
{
Center-of-mass energy dependence of 
average $p_{T}$
of $J/\psi$'s including the results of PHENIX and 
lower energy experiments.
The error bars include both statistical and systematic
errors which are added in quadrature.
The solid line shows a fit with a function 
$p + q \ln\sqrt{s}$ where $p$ and $q$ are fit parameters.
The dotted line shows the PYTHIA prediction using 
the GRV94LO PDFs and parton-transverse-motion $\langle
k_{T} \rangle$ = 0.82~GeV/$c$ which is 
tuned to reproduce the $p_{T}$ spectrum of 
the FNAL-E672/E706 data at $\sqrt{s}$ = 38.8 GeV.
}
\label{fig:meanPt}
\end{center}
\end{figure}

At higher energies, more energetic partons 
participate in interactions, thus
causing a slight increase in $\langle p_{T}\rangle$ of $J/\psi$'s.
Figure~\ref{fig:meanPt} shows $\sqrt{s}$ dependence of
$\langle p_{T} \rangle$ including the results of PHENIX and
the lower energy experiments
summarized in Table~\ref{tab:jpsi_sigma_sqrts}.
Neither high-$p_{T}$ correction nor 
rapidity correction is applied to the PHENIX data,
which are estimated to be small,
as well as the other experiments.
Our result, at the highest energy, is
slightly higher than the lower-energy results,
which is consistent with the expectation.
The experimental data are well fitted with 
a function $p + q\ln\sqrt{s}$
with a good reduced chi-square ($\chi^{2}$/NDF $\sim$ 1.1)
as shown in the same figure,
where $p$ and $q$ are fit parameters.
On the other hand, another empirical form
$p' + q'\sqrt{s}$ which fits the lower energy 
results well \cite{ref:exp_sqrt_31.5_E672}, 
does not fit including the PHENIX result,
indicating it is applicable only
in the limited $\sqrt{s}$ range.
Another curve in the figure shows the PYTHIA prediction
with a fixed $\langle k_{T} \rangle$ = 0.82 GeV/$c$,  
which reproduces
the $p_{T}$ spectrum of the FNAL-E672/E706 data
at $\sqrt{s}$ = 38.8 GeV
as shown in Fig.~\ref{fig:dsigma_dpt_sqrts40}.
The PYTHIA prediction is consistent with our data,
but slightly higher than the experimental
results at lower energy ($\sqrt{s} \leq $ 20~GeV), indicating 
$\langle k_{T}\rangle$ increasing with $\sqrt{s}$~\cite{ref:LiLai,ref:Cox_84}.

Such $\sqrt{s}$ dependence has also been observed in 
$\langle p_{T}\rangle$ of muon pairs from the Drell-Yan process,
which suffers less theoretical uncertainties 
such as next-to-leading-order corrections
thus has direct relation to $\langle k_{T}\rangle$ 
\cite{ref:Cox_84}.
Its magnitude is also consistent with that of $J/\psi$
at lower energies 
\cite{ref:exp_sqrt_31.5_E789,ref:E672-706,ref:kt,ref:CFS_81}.
This similarity supports that  
$\langle p_{T}\rangle$ of $J/\psi$ can be
understood by a perturbative-QCD based model (PYTHIA) with 
reasonable $\langle k_{T}\rangle$ values.

\subsection{Future measurements}

Our result on the total cross section for the $J/\psi$ production
in $\sqrt{s}$ = 200 GeV p+p collisions is consistent with 
both the color-evaporation model and color-octet model predictions.
In this section, we will discuss future measurements expected at RHIC
which will provide further critical tests for these models.

%

\subsubsection*{Polarization (Spin Alignment) of $J/\psi$}

One critical test for these models
is measurement of polarization $\lambda$ (or spin-alignment) of 
the $J/\psi$. The CEM predicts zero-polarization due to 
multiple gluon emissions while the CSM and COM
predict sizable polarization.
All the lower energy experiments show 
zero-polarization ($|\lambda|<0.1$)~\cite{ref:jpsi_pol_fixed_target} 
which is consistent with the CEM prediction
except at large $x_{F}$ ($\lambda \rightarrow -1$ with $x_{F} \rightarrow 1$).
The CDF experiment measures polarization
of $J/\psi$'s with $p_{T} >$ 4 GeV/$c$~\cite{ref:cdf_jpsi_pol}.
Medium-$p_{T}$ $J/\psi$'s (4 $< p_{T} < $ 12 GeV/$c$) are
transversely polarized 
($\lambda >$ 0) and consistent with the COM prediction.
However, at high $p_{T}$ ($p_{T} >$ 12 GeV/$c$), polarization 
decreases with $p_{T}$ unlike the COM prediction, causing
a sizable discrepancy at the highest $p_{T}$ bin (15 $< p_{T} <$ 20 GeV/$c$). 


Our measurement of $\lambda$ at $\sqrt{s}$ = 200 GeV~will serve as
another vital input for the production models.
Since $\lambda$ is obtained by fitting 
$\cos\theta^{*}$ 
distribution to the form
\[ \frac{d\sigma}{d\cos\theta^{*}} \propto 1 + \lambda
(\cos\theta^{*})^{2}, 
\vspace*{1mm}
\]

\noindent
sensitivity to higher $|\cos\theta^{*}|$ 
is important for the measurement of $\lambda$,
where $\theta^{*}$ represents the decay angle of a muon 
in the $J/\psi$ rest frame
with respect to the momentum direction of the $J/\psi$ in 
the laboratory frame
(Gottfried-Jackson frame).

\begin{figure}[htbp]
\begin{center}
\includegraphics[width=4in]{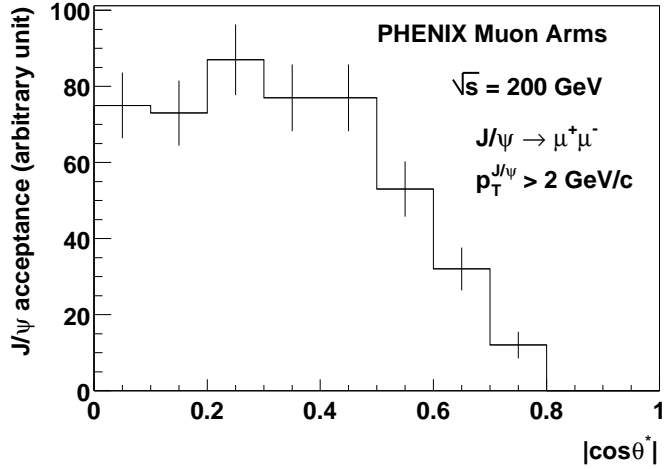}
\caption[Muon Arm acceptance as a function of $|\cos\theta^{*}|$]
{The Muon Arm acceptance for $J/\psi$'s with $p_{T} >$ 2~GeV/$c$
as a function of $|\cos\theta^{*}|$.}
\label{fig:jpsi_costh}
\end{center}
\end{figure}

Figure~\ref{fig:jpsi_costh}
shows the acceptance of the Muon Arms
for $J/\psi$'s with $p_{T} >$ 2~GeV/$c$,
as a function of 
$|\cos\theta^{*}|$.
Our acceptance is sensitive up to $|\cos\theta^{*}| \sim$ 0.8.
Acceptance for larger $|\cos\theta^{*}|$
increases with $p_{T}$ of $J/\psi$'s,
because the backward-going muon in the $J/\psi$ rest frame
is boosted toward the detector in the laboratory frame.

\subsubsection*{Polarized cross section for the $J/\psi$ production}

In Run-2, RHIC has successfully accelerated 
transversely polarized protons up to 100 GeV/$c$
momentum with about 20\%
polarization and collided them at $\sqrt{s}$ = 200 GeV.
Details of spin physics at RHIC
will be found in~\cite{ref:RHIC-spin}.

In Run-3 longitudinally polarized p+p collisions with about 50\% polarization
are expected at RHIC.
We plan to measure
double-longitudinal spin asymmetries
for $J/\psi$ production $\bigl(A_{LL}^{J/\psi}\bigr)$
defined as
\[ A_{LL}^{J/\psi} \equiv \frac{\sigma_{J/\psi}^{(++)} -
  \sigma_{J/\psi}^{(+-)}}{\sigma_{J/\psi}^{(++)} + \sigma_{J/\psi}^{(+-)}}, \]
where $\sigma_{J/\psi}^{(++)} \bigl( \sigma_{J/\psi}^{(+-)} \bigr)$
stands for 
production cross-section when incident protons have 
the same (opposite) signs of helicities.

\begin{figure}[htbp]
\begin{center}
\includegraphics[width=3.in]{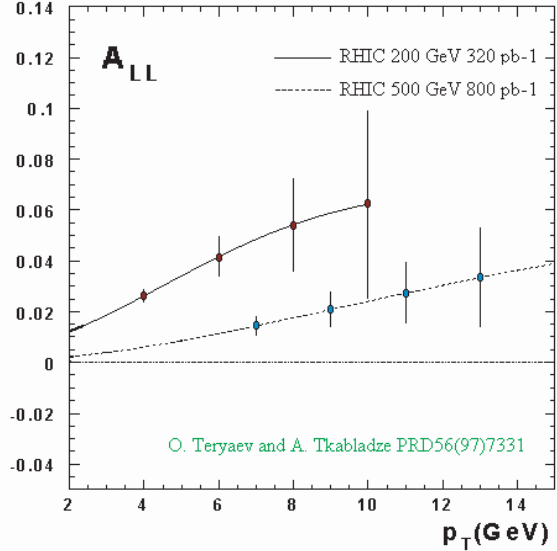}
\caption[Double-longitudinal spin asymmetries
for the $J/\psi$ production as a function of $p_{T}^{J/\psi}$]
{
Double-longitudinal spin asymmetries
for the $J/\psi$ production ($A_{LL}^{J/\psi}$)
as a function of $p_{T}^{J/\psi}$
with COM expectations together with statistical 
sensitivities of PHENIX with a RHIC one-year full luminosity.
}
\label{fig:jpsi_asym}
\end{center}
\end{figure}

It is sensitive to polarized gluon density
($\Delta g(x) = g^{+}(x) - g^{-}(x)$) in the proton
where $g^{+(-)}(x)$ denotes density of gluons
which are polarized in the same (opposite) direction as the proton
helicity and can be written as
\[ A_{LL}^{J/\psi}(x_1,x_2) = \frac{\Delta g(x_1)}{g(x_1)} 
\frac{\Delta g(x_2)}{g(x_2)}
 a_{LL}^{g+g \rightarrow J/\psi X} \]
where $x_{1}$ and $x_{2}$ are momentum fractions
of each gluon.
The partonic asymmetry $a_{LL}^{g+g \rightarrow J/\psi X}$
for the subprocess $g+g \rightarrow J/\psi X$
is calculated by each model in~\cite{ref:Gupta_jpsi_asym}:
\begin{align*}
a_{LL} &\sim +1 \,\, \mbox{with the CEM}, \\
a_{LL} &\sim -1 \,\, \mbox{with the CSM}\,\,\mbox{and} \\
a_{LL} &\sim \frac
{5M_{c}^{2}\tilde{\Theta} + 15B_{0}\langle O^{\chi_{c0}}_{1}
({}^{3}P_{0})\rangle - 4B_{2}\langle O^{\chi_{c2}}_{1}({}^{3}P_{2})\rangle}
{5M_{c}^{2}\Theta + 15B_{0}\langle O^{\chi_{c0}}_{1}
({}^{3}P_{0})\rangle + 4B_{2}\langle O^{\chi_{c2}}_{1}({}^{3}P_{2})\rangle}
\,\, \mbox{with the COM,} \\
\intertext{where}
\Theta &= \langle O_{8}^{J/\psi}({}^{1}S_{0})\rangle +
    \frac{7}{M_{c}^{2}} \langle O_{8}^{J/\psi}({}^{3}P_{0})\rangle, \\
\tilde{\Theta} &= \langle O_{8}^{J/\psi}({}^{1}S_{0})\rangle -
    \frac{1}{M_{c}^{2}} \langle O_{8}^{J/\psi}({}^{3}P_{0})\rangle \,\,\mbox{and} \\
B_{J} &= Br (\chi_{cJ} \rightarrow J/\psi \gamma). 
\end{align*}
Calculation with the COM is for the $2 \rightarrow 1$
subprocesses, which dominate the low-$p_{T}$ yield.
The mere determination of the sign
of $A_{LL}^{J/\psi}$ can choose either the CSM or CEM.
For the COM, $\tilde{\Theta}$ is currently unknown and expected to be
extracted from $A_{LL}^{J/\psi}$ of photo-production data
to be measured in polarized proton and electron collisions
at the Electron Ion Collider (EIC)~\cite{ref:EIC}.
Using $\Theta$ obtained from the unpolarized photo-production
data~\cite{ref:Cacciari_COM_matrix},
branching fractions~\cite{ref:RPP} and the color-singlet
matrix elements $\langle O_{1}^{\chi_{cJ}}({}^{3}P_{J})\rangle$
from the data of hadronic decays of $\chi_c$
states~\cite{ref:Mangano_CSM_matrix}, 
$a_{LL}$ with the color-octet model is expected to vary from $-$0.3 
to 0.7 assuming two extreme cases, 
$\langle O_{8}^{J/\psi}({}^{1}S_{0})\rangle$ = 0 and
$\langle O_{8}^{J/\psi}({}^{3}P_{0})\rangle$ = 0.
For the color-octet $2 \rightarrow 2$ subprocess, $A_{LL}^{J/\psi}$
has also been calculated in~\cite{ref:jpsi_asym_Teryaev}
assuming a function form of $\Delta g(x)$ found in~\cite{ref:GS95},
which is shown in Fig.~\ref{fig:jpsi_asym}
as a function of $p_{T}$ of $J/\psi$'s
together with statistical sensitivities of PHENIX
expected with one RHIC-year luminosity.\footnote{
In the enhanced mode. Also in this paper, 100\% detection
efficiency for $J/\psi$'s are assumed.}
Measurement of $A_{LL}^{J/\psi}$ is 
motivated not only by the direct determination
of polarized gluon distribution but also 
by another test for the production models for charmonia.

\vspace{1cm}

Table~\ref{tab:model_summary} shows 
the summary of model sensitivities
of each observable measured at RHIC.
Our measurement of $\sigma_{J/\psi}$ prefers
the CEM and COM rather than the CSM.
Measurements of rapidity and 
low-$p_{T}$ distributions ($\langle p_{T}\rangle$)
do not have sensitivities to the models.
Future measurements of the polarization of $J/\psi$
and $A_{LL}^{J/\psi}$ will provide
discrimination of the CEM from the COM.

\begin{table}[htbp]
\begin{center}
\begin{tabular}{|c||c|c|c||c|} \hline

                  & CSM    & CEM & COM & comments \\ \hline \hline
$\sigma_{J/\psi}$ & $\times$ &  \multicolumn{2}{|c||}{$\bigcirc$} & done \\ \hline
rapidity distribution &  \multicolumn{3}{|c||}{no sensitivity} & done \\ \hline
$\langle p_{T}\rangle$ &  \multicolumn{3}{|c||}{no sensitivity} & done \\ \hline
Polarization & $>0$ & 0 & $>0$ & needs high-$p_{T}$ events \\ \hline
$A_{LL}^{J/\psi}$ & $<0$  & $>0$  & vary & needs $\Delta g(x)$ \\ \hline

\end{tabular}
\end{center}
\caption[Summary of model sensitivities of each observables expected
at RHIC]
{Summary of model sensitivities of each observables 
regarding $J/\psi$ measured at RHIC.
Our measurement of $\sigma_{J/\psi}$ prefers
the CEM and COM rather than the CSM.
Future measurements of the polarization of $J/\psi$
and $A_{LL}^{J/\psi}$ will provide
discrimination of the CEM from the COM.
Expected ranges of values for those measurements
are shown with each model.
}
\label{tab:model_summary}
\end{table}

\section{Conclusion}

Measurements of $J/\psi$ mesons in p+p collisions at RHIC
provide useful information on both the perturbative
and non-perturbative aspects of QCD.
They also play crucial roles in both
QGP physics and spin physics expected at RHIC.

With the South Muon Arm in the PHENIX detector
covering $1.2 < y < 2.2$,
$J/\psi$ particles have been 
successfully triggered 
by the simple trigger system
and clearly identified with a small background
via $\mu^{+}\mu^{-}$ decays 
in the first p+p collisions at $\sqrt{s}$ = 200 GeV.
Total and differential cross sections for
inclusive $J/\psi$ production have been obtained.

Transverse momentum distribution is consistent with
the various function forms and PYTHIA which fit the other experimental data.
Average transverse momentum 
$\langle p_{T} \rangle$ = 1.66 $\pm$ 0.18 (stat.)
 $\pm$ 0.12 (syst.) GeV/$c$ is slightly higher than
the lower energy results
and consistent with the PYTHIA's expectation
with a reasonable value of $\langle k_{T}\rangle$.

Rapidity-differential cross section 
for inclusive $J/\psi$ production
times the branching fraction to a $\mu^{+}\mu^{-}$ pair
$B_{\mu\mu} d\sigma/dy|_{y=1.7}$ = 37 $\pm$ 7 (stat.) $\pm$
11 (syst.) nb has been obtained.
Including the measurement at mid
rapidity ($|y|<$ 0.35) using the electron decay channel, 
rapidity distribution is found to be consistent with
the gluon distribution function of some typical PDF sets
in the Bjorken-$x$ range $2 \times 10^{-3} < x < 0.2$.

The total cross section $\sigma_{J/\psi}$ = 
3.8 $\pm$ 0.6 (stat.) $\pm$ 1.3 (syst.) $\mu$b has been extracted
from the fit of the rapidity distribution,
which is related with the lower energy results
by a gluon distribution function with a reasonable scale.
Proper normalization for $\sigma_{J/\psi}$
is reproduced well by both
the color-evaporation model and the color-octet model
using reasonable gluon densities, scales and 
charm-quark mass.

Our results 
confirm the validity of the production mechanism of $J/\psi$
based on perturbative QCD.
The further studies of the production mechanism,
via measurements of such as the spin alignment 
of $J/\psi$ and $A_{LL}^{J/\psi}$,
are clearly needed especially to 
understand the non-perturbative aspects
involved in the production of $J/\psi$.
The results have also established the baseline
necessary for the QGP search in Au+Au collisions
as well as gluon polarization measurements in 
polarized p+p collisions at RHIC.


\appendix

\section{Estimation of various contributions to the inclusive $J/\psi$ yield}
\label{chap:feeddown}

$J/\psi$ mesons can be produced either directly or
via decays from higher state charmonia ($\chi_{c}$ and $\psi'$)
as well as from $b$-quarks.
Our measurement is inclusive and it is not possible to 
separate each contribution from others.
We will estimate fractions of 
each contribution taking into account
results of both higher and lower energy experiments.

\subsubsection*{$b$-quarks}

Contribution from $b$-quark decays is energy-dependent.
Production cross section for a $b\bar{b}$ pair times the branching fraction
for a $b$ (or $\bar{b}$)-quark decaying into a $J/\psi$ meson
in p (or $\bar{\mbox{p}}$)+N collisions  
$\sigma (\mbox{p}(\bar{\mbox{p}})\mbox{N}
 \rightarrow b\bar{b} X) Br (b\bar{b} \rightarrow
J/\psi X)$ has been measured in p+Au collisions 
at $\sqrt{s}$ = 38.8 GeV by the FNAL-E789 experiment 
\cite{ref:b_jpsi_E789} and the result is 
148 $\pm$ 34 (stat.) $\pm$ 28 (syst.)
pb/nucleon. 
It has been also measured in p+$\bar{\mbox{p}}$ collisions at 
$\sqrt{s}$ = 630 GeV by the CERN-UA1 experiment
\cite{ref:ua1_jpsi} 
and 32.7 $\pm$ 2.6 (stat.) $\pm$ 13.4 (syst.) nb
has been obtained
with $p_{T}^{J/\psi} > $ 5 GeV/$c$ and $|y^{J/\psi}| < $2.0 cuts.
The ratio of those results is consistent with 
a perturbative QCD prediction using PYTHIA.
It is extrapolated to $\sqrt{s}$ = 200 GeV and
$\sigma (\mbox{p+p}
 \rightarrow b\bar{b} X) Br (b\bar{b} \rightarrow
J/\psi X)$ = 
 40 nb is obtained
which is about 1\% of our inclusive result.

\subsubsection*{$\psi'$}

Production cross sections for $\psi'$, or $\psi$ (2S) mesons,
have been also measured
by the FNAL-E789 \cite{ref:b_jpsi_E789} experiment.
The relative production cross section times the branching fraction
to a muon pair
for a $\psi'$ to that for a $J/\psi$, 
$R  \equiv Br (\psi' \rightarrow \mu^{+} \mu^{-})
\sigma_{\psi'} / Br (J/\psi \rightarrow \mu^{+} \mu^{-}) \sigma_{J/\psi}$
= 0.018 $\pm$ 0.001 (stat.) $\pm$ 0.002 (syst.) has been obtained.
This $\psi$' fraction can contribute to our $J/\psi$ measurement
because of the poor mass-resolution (200 to 300 MeV/$c^{2}$)
currently observed.
Using the branching fractions $Br (\psi' \rightarrow \mu^{+}\mu^{-}) = 
(7.0 \pm 0.9) \times 10^{-3}$ and 
$Br (J/\psi \rightarrow \mu^{+}\mu^{-}) = 5.88 \pm 0.10$ \%
\cite{ref:RPP}, 
$\sigma_{\psi'} / \sigma_{J/\psi}$
is determined to be 0.15 $\pm$ 0.03. 
Contribution of $\psi' \rightarrow J/\psi \rightarrow \mu^{+}\mu^{-}$
to inclusive $J/\psi$ yield
is therefore expected to 15\% $\times$ 0.55 ($Br(\psi' \rightarrow J/\psi X)$)
$\sim$ 8\%.
The color evaporation model expects energy and process independence
of the fractions of $\psi'$ and $\chi_{c}$
to the inclusive $J/\psi$ yield.
Slightly higher or consistent $R$ values have been obtained
by the UA1
($R$ = 0.029 $\pm$ 0.010 (stat.) $\pm$ 0.007 (syst.) 
with $p_{T} >$ 5 GeV/$c$ and $|y| < $ 2.0 cuts)
\cite{ref:ua1_jpsi}
and CDF
($R$ = 0.033 $\pm$ 0.002 (stat.) $^{+0.007}_{-0.008}$ (syst.) nb
with $p_{T} >$ 5 GeV/$c$ and $|\eta| < $ 0.6 cuts)
\cite{ref:cdf_jpsi} experiments.

\subsubsection*{$\chi_{c}$}

The fraction of the inclusive $J/\psi$ yield from
radiative $\chi$ decays has been determined to be
(30 $\pm$ 4) \% by the FNAL-E705 experiment in p+Li collisions at
$\sqrt{s}$ = 23.8 GeV \cite{ref:E705_chi}.
A consistent value, (29.7 $\pm$ 1.7 (stat.) $\pm$ 5.7 (syst.))\% 
was obtained in p+$\bar{\mbox{p}}$ collisions at $\sqrt{s}$ = 1.8 TeV
by the CDF experiment at Fermilab \cite{ref:cdf_jpsi_chi}.

\vspace{1cm}

In summary, the fractions of $b$-quark, $\psi$' and 
$\chi$ contributions to inclusive $J/\psi$ yield
at RHIC energy are expected to be 1\%, 10-15\% and 30\%
respectively. The remains are ascribed to the direct production.
The fractions of $\psi'$ and $\chi_{c}$ contributions
do not agree with
the color-singlet model prediction. Including 
color-octet contributions, they are successfully explained
\cite{ref:Beneke_96}.
The color-evaporation model can not predict their fractions.
However, small $\sqrt{s}$ dependence of them
in nucleon-nucleon collisions
is consistent with its assumption that
they are process-independent.



\section*{Acknowledgment}

First of all, I would like to thank Prof.~K.~Imai for being my supervisor.
Prof.~N.~Saito initiated my interest in spin physics  
and provided me a lot of scientific advice throughout my research
at RHIC.

I wish to thank staff of the Collider-Accelerator Department (C-AD) at
Brookhaven National Laboratory (BNL)
for making the experiment possible.
In particular, help of Dr.~A.~Drees was crucial to understand
the beam-related background for the muon trigger.

The measurement of $J/\psi$ 
would not have been carried out without
coherent work of each collaborator of the PHENIX experiment.
I would like to give special thanks to 
members of the Muon detector group, BBC detector group,
and online and offline groups.
For the $J/\psi$ analysis, discussions with Dr.~V.~Cianciolo and 
J.~Nagle were helpful. 
I am grateful to
Dr.~Y.~Akiba for providing details 
of the electron channel measurement.
I wish thank Dr.~E.~M.~Gregores for affording me
the prediction of the color-evaporation model.

I am indebted to RIKEN and RIKEN BNL Research Center (RBRC) staff
for supporting my life at BNL, especially to
the former and current chiefs of the RIKEN Radiation Laboratory,
Prof.~M.~Ishihara and Dr.~H.~En'yo.
The latter was also my supervisor
and gave me a great deal of advice and encouragement.
A large number of discussions with scientific staff of RIKEN and RBRC
were useful for this work and spin physics
including Dr.~G.~Bunce, M.~G.~Perdekamp, B.~Fox, Y.~Goto and A.~Taketani.

My research was partially supported by the
RIKEN Junior Research Associate program (FY 1998-2000)
and JSPS Research Fellowship for Young Scientists
(FY 2001-2002).

\section{Attached figures' captions}

\begin{itemize}

\item figure 1: The RHIC accelerator complex.

figure 2: The PHENIX experiment overview.

figure 3: Magnetic field lines inside the Central and Muon Magnets.

figure 4: A photograph of a Beam-Beam Counter (BBC).
Dimensions are 29 cm in outer-diameter and 25 cm in length.

figure 5: A schematic view of a Muon Arm.
It consists of a conical magnet (Muon Magnet) with tracking
chambers inside at three stations (Muon Tracker, or MuTr) and an array
of chamber planes interleaved with absorbers (Muon Identifier, or
MuID).

figure 6: Cross section of a MuTr station.
One MuTr station consists of two or three gaps,
each of which has one anode-wire and two cathode-strip planes.

figure 7: Octant and half-octant structures of a MuTr gap.
The beam line passes at the center of this figure.
The bold lines show the boundaries for octants
and this lines for half-octants.
Octant and half-octant numbers are also shown.
For the South Arm, the left-hand side
in the figure (octant-5 side) corresponds to
the east direction.

figure 8: A photograph of a MuTr station-2 octant.
A cathode plane is made of etched 25-$\mu$m copper coated mylar foils.

figure 9: Directions of cathode strips and anode wires in
station-1, octant-2, half-octant-0 planes.
Bold lines show cathode strips
in non-stereo-angle planes.
Dotted lines show cathode strips
in stereo-angle planes with angles
shown in Table~\ref{tab:cath_angle}.
Angles in the figure are not exact.

figure 10: The optical alignment system in which
light from an optical fiber is projected
from station 1 through station~2 to a CCD
mounted on station~3.  The relative chamber
positions are monitored to $\pm$25~$\mu$m.

figure 11: A schematic diagram for the MuTr Front-End Electronics
(FEE).
Raw chamber signals are continuously amplified with
CPAs (Charge Pre-Amps) and stored in
AMUs (Analog Memory Units).
Upon receipt of a level-1 trigger bit from
a GTM (Granule Timing Module),
stored samples from all channels are
digitized by ADCs (Analog to Digital Converters) and the results are
sent to a DCM (Data Collection Module.

figure 12: Measurement of position resolution obtained in cosmic-ray
tests of a station-2 octant.  The composite chamber plus projection
error (position resolution for a track obtained
with the other five planes)
was about 131 $\mu$m, consistent with the
100 $\mu$m specification for the chambers and readout alone.

Figure 13: Setup for the MuID beam test carried out at KEK-PS.
Four scintillation counters (ST1 to ST4)
define the beam. Three gaseous $\check{\mbox{C}}$erenkov counters
(GC1 to GC3) identify pions and muons.

Figure 14: Pion mis-identification rate as a function of pion momentum
for both the South and North Muon Arms
obtained with the beam test at KEK.
The solid (dotted) lines show simulation results
including (excluding) $\pi^{\pm} \rightarrow \mu^{\pm}\nu$ decays.

Figure 15: Schematic diagram of the MuID NIM-logic trigger.
Each ``Quadrant'' MLU takes 16 (4 for each gap) output signals
from the MuID FEE (Front-End Electronics). The ``Decision'' MLU takes
output signals of all the ``shallow'' and ``deep'' quadrant MLUs
and generates final trigger bits,
which are 1S (single shallow),
2S (double shallow),
1D (single deep),
1D1S (deep-shallow) and 2D (double deep),
as PHENIX Global-Level-1 input.

figure 16: An accepted-pattern example of the MuID NIM-logic trigger.
Capital letters represent positions of each segment.
Gap 2 was not used because of the limitation of
input channels of the trigger logic.

Figure 17: The integrated luminosity RHIC has delivered
to the PHENIX and STAR experiments
as a function of day starting from December 21, 2001.

Figure 18; BBC efficiency for p+p $\rightarrow$ $\pi^{0}X$ events
as a function of $p_{T}$ of $\pi^{0}$
measured in the Central Arms ($|\eta|<$ 0.35)
\cite{ref:PHENIX_pi0_run2}.

\end{itemize}



\end{document}